\documentclass[11pt,aps,prd,nofootinbib,tightenlines]{revtex4-1}
\usepackage{dcolumn}% Align table columns on decimal point
\usepackage{graphicx}
\usepackage{color}
\usepackage{amsmath}
\usepackage{amssymb}
\usepackage{latexsym}
\usepackage{bm}
\usepackage{multirow}
\usepackage{hyperref}
\usepackage{tabularx}
\usepackage{subcaption}
\def\beq{\begin{equation}}
\def\eeq{\end{equation}}
\def\be{\begin{eqnarray}}
\def\ed{\end{eqnarray}}
\def\non{\nonumber}

\newcommand*{\bea}{\begin{eqnarray}}
\newcommand*{\eea}{\end{eqnarray}}

\def\be{\begin{eqnarray}}
\def\ed{\end{eqnarray}}
\def\non{\nonumber}

\def\lam{\lambda}
\def\beq{\begin{eqnarray}}
\def\eeq{\end{eqnarray}}
\def\beq{\begin{equation}}
\def\eeq{\end{equation}}
\def\be{\begin{eqnarray}}
\def\ed{\end{eqnarray}}
\def\non{\nonumber}

\newcommand{\ba}{\begin{array}}
\newcommand{\ea}{\end{array}}
\begin{document}
%%%%%%%%%%%%%%%%%%%%%%%%%%%%%%%%%%%%%%%%%%%%%%%%%%%%%%
\vspace*{2cm}
\title{{Signal to background interference in  $pp\to t H^-\to t W^- b\bar b$ at the LHC Run-II}}

\author{Abdesslam Arhrib$^{1}$, Rachid Benbrik$^{2,3,4}$, Stefano Moretti$^{5}$, Rui Santos$^{6,7,8}$ and Pankaj Sharma$^{9}$}
%\email[]{aarhrib@gmail.com}
\affiliation{ 
$^{1}$Facult\'e des Sciences et Techniques, Abdelmalek Essaadi University, B.P. 416, Tangier, Morocco \\
%\affiliation{\small Physics Division, National Center for Theoretical Sciences, Hsinchu, Taiwan}
%\author{R. Benbrik}
%\email[]{r.benbrik@uca.ac.ma}
%\affiliation{\small 
$^{2}$  LPHEA Facult\'e Semlalia Marrakech and MSISM Team, Facult\'e Polydisciplinaire de Safi, Sidi Bouzid, BP 4162,  Safi, Morocco \\
$^{3}$School of Physics Sciences, University of Chinese Academy of Sciences, Beijing 100039, China \\
$^{4}$Center for future high energy physics, Chinese Academy of Sciences, Beijing 100039, China \\
%
%\author{Stefano Moretti}
%\email[]{s.moretti@soton.ac.uk}
$^{5}$School of Physics and Astronomy, University of Southampton, Southampton, SO17 1BJ, United Kingdom \\
%
%\author{Rui Santos}
%\email[]{srasantos@fc.ul.pt}
%\affiliation{\small 
$^{6}$ISEL - Instituto Superior de Engenharia de Lisboa, \\
	Instituto Polit\'ecnico de Lisboa, 1959-007 Lisboa, Portugal\\
$^{7}$Centro de F\'{\i}sica Te\'{o}rica e Computacional, Faculdade de Ci\^{e}ncias, Universidade de Lisboa, \\
	Campo Grande, Edif\'{\i}cio C8, 1749-016 Lisboa, Portugal\\
$^{8}$LIP, Departamento de F\'{\i}sica, Universidade do Minho, 4710-057 Braga, Portugal\\
%\author{Pankaj Sharma}
%\email[]{pankaj.sharma@adelaide.edu.au}
$^{9}$ARC Center of Excellence for Particle Physics at the Terascale,\\ Department of Physics, University of Adelaide, 5005 Adelaide, South Australia}

\date{\today}

\vspace*{-3cm}
\begin{abstract}
We investigate in the Large Hadron Collider (LHC) environment the possibility that sizeable  interference effects between a {\sl heavy}  charged Higgs boson signal produced  via  $bg\to tH^-$ (+ c.c.) scattering and decaying via $H^-\to W^- A\to W^- b\bar b$ (+ c.c.)  and the irreducible background given by $bg\to t W^- b\bar b$ topologies could spoil current search approaches where the former and latter channels are treated separately. The rationale for this comes from the fact that a heavy charged Higgs state can have a large width, and so can happen for the CP-odd neutral Higgs state emerging in the ensuing decays well, which in turn enables such interferences. We conclude that effects are very significant, both at the inclusive and exclusive level (i.e., both before and after $H^\pm$ selection cuts are enforced, respectively) and typically of a destructive nature. This therefore implies that currently established LHC reaches for heavy charged Higgs bosons require some level of rescaling. However, this is possible a posteriori, as the aforementioned $H^\pm$ selection cuts shape the interference contributions at the differential level in a way similar to that of the isolated $H^\pm$ signal, so there is no need to reassess the efficiency of the individual cuts. We show such effects quantitatively by borrowing benchmarks points from different Yukawa types of a 2-Higgs Doublet Model parameter space for $H^\pm$ values starting from around 200 GeV.    
\end{abstract} 
\maketitle

%%%%%%%%%%%%%%%%%%%%%%%%%%%%%%%%%%%%%%%%
\section{Introduction} 
As much as one would welcome the production of a light charged Higgs boson from top-quark decay at the LHC, as
the event rate would be plentiful, it must be recognised by now that the likelihood of this being a design of Nature is becoming slimmer and slimmer. This is because extensive searches have been carried out by the ATLAS and CMS Collaborations in this mode, assuming a variety of $H^\pm$ decay channels, none of which has been fruitful.  Hence,
it is becoming more likely than otherwise that, if such a state indeed exists in Nature, it will be heavier than the top quark,
i.e., $M_{H^\pm}>m_t$\footnote{We should note however that both the production and  decay modes used in all present
searches may have a strong dependence on the parameters of the model. In particular versions of the
2-Higgs Doublet Model (2HDM) \cite{Aoki:2011wd}, for example, a very large value of the parameter $\tan \beta$, the ratio
of the Vacuum Expectation Values (VEVs) of the two doublets, will render useless any search involving Yukawa couplings.
For these scenarios only processes involving the electromagnetic coupling of the charged Higgs would be able
to settle the issue of existence of light charged Higgs bosons.}. 

The call for establishing an $H^\pm$ signal comes from an intriguing theoretical consideration, that  
the discovery of a (singly) charged Higgs boson would signal the existence of a second Higgs doublet in addition to the Standard Model (SM)-like one already established through the discovery of the $W^\pm$ and $Z$ bosons at the S$p\bar p$S \cite{Arnison:1983rp, Bagnaia:1983zx} in the eighties and of a Higgs boson itself at the LHC only five years ago \cite{Aad:2012tfa, Chatrchyan:2012xdj}. Such a spinless field can naturally be accommodated in 2HDMs, which are the standard theoretical frameworks assumed in experimental analyses. Indeed, in their CP-conserving versions, 2HDMs present in their spectra, after spontaneous Electro-Weak Symmetry Breaking (EWSB), five physical Higgs states: the neutral pseudoscalar ($A$), the lightest ($h$) and heaviest ($H$) neutral scalars and two charged ones ($H^\pm$). 

Of all 2HDM Yukawa types (see~\cite{Branco} for a review), we concentrate here on the 2HDM Type II, Flipped
and Type III ones (to be defined later). This is because  such Yukawa types of 2HDMs have a preference for heavy charged Higgs bosons. In the 2HDM Type II and Flipped, constraints from $b\to s\gamma$ decays put a lower limit on the $H^\pm$ mass at about 580 GeV, rather independently of $\tan\beta$~\cite{Misiak:2015xwa,Misiak:2017bgg}. In the 2HDM Type III, such constraint is relaxed, yet the combination of all available experimental data  places a lower limit on $M_{H^\pm}$ at about 200 GeV or possibly even less~\cite{Arhrib:2017yby}. Hence, both such 2HDM scenarios provide parameter spaces that are suitable to benchmark experimental searches for heavy charged Higgs bosons. 

Such a heavy mass region is very difficult to access because of the large reducible and irreducible backgrounds associated with the main decay mode $H^+\to t\bar b$, following the dominant production channel $bg\to t H^-$ \cite{bg}. (Notice that the production rate of the latter exceeds by far that of other possible production modes, like those identified in~\cite{bq, BBK, ioekosuke, Aoki:2011wd}, thus rendering it the only accessible production channel at the CERN machine in the heavy mass region.) The analysis of the $H^+\to t\bar b$ signature has been the subject of many early phenomenological studies  \cite{roger}--\cite{roy1}, their conclusion being that  the LHC discovery potential might be satisfactory, so long that $\tan\beta$ is small ($\leq 1.5$) or large ($\geq30$) enough and the charged Higgs boson mass is below 600 GeV or so. Such rather positive prospects have recently been revived by an ATLAS analysis of the full Run-I sample \cite{ATLAS}, which searched precisely for the aforementioned $H^\pm$ production and decay modes, by exploring the mass interval from 300 to 600 GeV. In fact, an excess with respect to the SM predictions was observed for  $M_{H^\pm}$ hypotheses in the heavy mass region.  While CMS does not confirm such an excess \cite{CMS}, the increased sensitivity that the two experiments are accruing with current Run-II data calls for a renewed interest in the search for such elusive Higgs states.

In this spirit, and recognising that the $H^+\to t\bar b$ decay channel eventually produces a $W^+b\bar b$ signature, Ref.~\cite{Uppsala} attempted to extend the reach afforded by this channel by exploiting the companion signature $H^+\to h_{\rm SM}W^+$ $\to$ $b\bar b W^+$, where $h_{\rm SM}$ is the SM-like Higgs boson discovered at CERN in 2012 (which is either the $h$ or $H$ state of  2HDMs). The knowledge of its mass now provides in fact an additional handle in the kinematic analysis when reconstructing a Breit-Wigner resonance in the $h_{\rm SM}\to b\bar b$  decay channel, thereby significantly improving the signal-to-background ratio afforded by pre-Higgs-discovery analyses \cite{whroy,me}. Such a study found  that significant portions of the parameter spaces of several 2HDMs are testable at Run-II. 

Spurred by the aforementioned experimental results and building upon Ref.~\cite{Uppsala}, some of us studied in Ref.~\cite{previous} all  intermediate decay channels of a heavy $H^\pm$ state also yielding a $W^\pm b\bar b$ signature, i.e., $H^+\to t\bar b$, $hW^\pm, HW^\pm$ and $AW^\pm$, starting from the production mode $bg\to tH^-$ (+ c.c.) (see also \cite{Akeroyd:2016ymd}). In doing so, we also took into account interference effects between these four channels, in the calculation of the total $H^\pm$ width as well as of the total yield in the cumulative $W^\pm b\bar b$ final state (wherein the $W^\pm$ decays leptonically), with the aim of maximising the experimental sensitivity of ATLAS and CMS. The outcome of this analysis was that somewhat more inclusive search strategies (historically geared towards extracting the prevalent $H^+\to t\bar b$ signature) ought to be deployed, that also capture $H^+\to W^+$~Higgs $\to W^+b\bar b$ channels. The exercise was performed specifically for a 2HDM Type II, but results therein can easily be extrapolated to other Yukawa types.

In~\cite{previous}, only interferences between the four 2HDM channels yielding $H^+\to W^+b\bar b$ decays were taken into account though, i.e., those between the different signal modes. While clearly all of these decay rates cannot be large at the same time, the important role of interferences amongst these decay modes was clearly established. However,
in that analysis, the role of interference effects between any of these signals and the irreducible background was not discussed, as illustrative examples of the $H^\pm$ production and decay phenomenology were chosen so as to nullify their impact. Unfortunately, this condition can only be realised in specific regions of the 2HDM parameter space considered, whichever the Yukawa type, not everywhere. It is the purpose of this paper to address this issue, i.e., to assess the impact of interference effects between signal and irreducible background in the $H^+\to W^+ b\bar b$ channel 
on current phenomenological approaches to extract the latter. We will show that such effects are indeed very large for
heavy $H^\pm$ masses over certain region of the 2HDM parameter space considered, both at the inclusive and exclusive level, i.e., before and after a selection is enforced, respectively. We will give some quantitative examples of this for the case of the specific $H^+\to W^+A\to W^+ b\bar b$ signal mode in three different Yukawa types of 2HDM (namely Type II, Flipped
and Type III), for several $M_{H^\pm}$ choices. 

The plan of this paper is as follows. In the next section we introduce the 2HDM types considered and define their available parameter spaces based on current experimental and theoretical constraints in the following one. Then we proceed to describe what are the relevant diagrams entering both signal and (irreducible) background as well as illustrate how we computed these. Sect. V is our numerical signal-to-background analysis. Finally, we draw our conclusions based on the results obtained in the last section of the paper.

%%%%%%%%%%%%%%%%%%%%%%%%%%%%%%%%%%%%%%%%
\section{Theoretical framework of 2HDMs}
\label{sec:formalism}
In this section we define  the scalar potential and the Yukawa sector of 
the 2HDM Type II, Flipped and Type III. The most general scalar potential which is 
 $SU(2)_L\otimes U(1)_Y$ invariant is given by \cite{Gunion:2002zf,Branco}
%%%%%%%%%%%%%%%
%{\color{red} 
\be
V(\Phi_1,\Phi_2)
&=& m^2_1 \Phi^{\dagger}_1\Phi_1+m^2_2 \Phi^{\dagger}_2\Phi_2 -(m^2_{12}
\Phi^{\dagger}_1\Phi_2+{\rm h.c}) +\frac{1}{2} \lam_1 (\Phi^{\dagger}_1\Phi_1)^2 
\nonumber \\ &+& \frac{1}{2} \lam_2
(\Phi^{\dagger}_2\Phi_2)^2 +
\lam_3 (\Phi^{\dagger}_1\Phi_1)(\Phi^{\dagger}_2\Phi_2) + \lam_4
(\Phi^{\dagger}_1\Phi_2)(\Phi^{\dagger}_1\Phi_2)  \non \\
&+& 
 \left[\frac{\lam_5}{2}(\Phi^{\dagger}_1\Phi_2)^2 +  \left(\lam_6 \Phi^\dagger_1 \Phi_1 + \lam_7 \Phi^\dagger_2 \Phi_2 \right) \Phi^\dagger_1 \Phi_2+{\rm h.c.} \right].
\label{higgspot}
\ed
%%%%%%%%%%%%
The scalar doublets $\Phi_i$ ($i=1,2$) can be parametrised as
\begin{align}
\Phi_i(x) = \begin{pmatrix}
\phi_i^+(x) \\ 
\frac{1}{\sqrt{2}}\left[v_1+\rho_1(x)+i \eta_1(x)\right]
\end{pmatrix}, 
\end{align}
with $v_{1,2}\geq 0$ being the VEVs satisfying $v=\sqrt{v_1^2+v_2^2}$, 
with $v=246.22$~GeV~\cite{Olive:2016xmw}.
Hermiticity of the potential forces $\lambda_{1,2,3,4}$ 
to be real while $\lambda_{5,6,7}$ and $m^2_{12}$ can be complex.
In this work we choose to work in a CP-conserving
potential where both VEVs are real 
and $\lambda_{5,6,7}$ and $m^2_{12}$ are also real.

After EWSB, three of the eight degrees
of freedom in 2HDMs are the Goldstone
bosons ($G^\pm$, $G^0$) and the remaining five degrees of freedom become the aforementioned 
physical Higgs bosons. After using the minimisation conditions for the potential
together with the $W^\pm$ boson mass requirement, we end up with nine independent parameters
which will be taken as:
%$(\lambda_i)_{i=1,\ldots,7}$, $m^2_{12}$, and  
%Equivalently, one can use instead, the following set of independant parameters:
\begin{equation}
\{ m_{h}\,, m_{H}\,, m_{A}\,, m_{H^\pm}\,, \alpha\,, \beta\,,  m^2_{12}, \lambda_6, \lambda_7 \}\,, 
\label{parameters} 
\end{equation}
where $\tan\beta \equiv v_2/v_1$ and $\beta$ is also the angle that diagonalises
the mass matrices of both the CP-odd  and  charged Higgs sector while the angle
$\alpha$ does so in the CP-even Higgs sector. 

The most commonly used version of a CP-conserving 2HDM is the one
where the terms proportional to $\lambda_6$ and $ \lambda_7 $ are absent.
This can be achieved by imposing a discrete $Z_2$ symmetry on the model
that usually takes the form $\Phi_i \to (-1)^{i+1} \Phi_i \quad i=1,2$. Such a symmetry
would also require $m^2_{12} = 0$, unless we allow a soft violation of this
discrete symmetry by the dimension two term $m^2_{12}$. 
When this $Z_2$ symmetry is extended to the Yukawa sector we end up with 
four possibilities regarding the Higgs bosons couplings to the fermions.
The two $Z_2$ symmetric models we will use in the work 
are the Type II model - where the symmetry
is extended in such a way that only $\Phi_1$ couples to up-type quarks while
only $\Phi_2$ couples to down-type quarks and leptons - 
and the Flipped model - where $\Phi_1$ couples to to up-type quarks and leptons while 
$\Phi_2$ couples to the down-type quarks. Besides the Type II and Flipped scenarios 
we will also study a version of the more general case of Type III, to be discussed below,
where neither the potential nor the Yukawa Lagrangian is $Z_2$ symmetric.
Therefore, for this particular case, $\lambda_6 \neq 0$ and $\lambda_7 \neq 0$.
Still in this work we will consider the limit $\lambda_6 \approx \lambda_7  \approx 0$.
The reason is basically that of simplicity and it is justified by the fact
that: a) the study does not depend on those parameters as there are no
Higgs self-coupling present in our analysis; b) it is a tree-level study 
and $\lambda_6 \approx \lambda_7  \approx 0$ is a tree-level condition; c)
the only possible effect on our study would be to enlarge the allowed
values of the parameter ranges which  would not change our conclusions.  

In the most general version of the 2HDM, the Yukawa sector 
is built such that both Higgs doublets couple to quarks and leptons.
The model is known as 2HDM Type III~\cite{Cheng:1987rs,Atwood:1996vj} and 
the Yukawa Lagrangian can be written as
\begin{align}
- \mathcal L_Y=
\bar Q_L( Y_1^d\Phi_1+Y_2^d\Phi_2) d_R 
+\bar Q_L({Y}^u_1 \tilde\Phi_1+{Y}^u_2 \tilde\Phi_2)u_R +
\bar L_L( Y_1^l\Phi_1+Y_2^l\Phi_2) l_R
+\text{h.c.},
\end{align}
where $Q^T_L = (u_L, d_L)$ and $L^T_L=(l_L, l_L)$ are 
 the left-handed quark doublet and lepton doublet, respectively, $Y^f_k$ 
 ($k=1,2$ and $f= u,d,l$) denote the $3\times 3$  Yukawa matrices and 
$\tilde\Phi_k = i\sigma_2 \Phi^*_k$, $k=1,2$. 
Since the mass matrices of the quarks and leptons are 
a linear combination of  $Y_{1}^f$ 
and $Y_2^f$,  $Y_{1,2}^{d,l}$ and  $Y_{1,2}^u$ 
 cannot be diagonalised simultaneously in general\footnote{Since we are interested
 in the couplings of a charged Higgs boson to quarks we just consider
 that the lepton flavour violating couplings are small enough not
 to show any effect in the measured processes involving leptons.}. 
Therefore, neutral Higgs Yukawa couplings with flavour violation appear 
at tree-level and lead to  a tree-level contribution to  $\Delta M_{K, B, D}$ as well as to 
$B_{d,s} \to \mu^+ \mu^-$ mediated by neutral Higgs exchange.
This is an important distinction with respect to  $Z_2$ symmetric models
 and can have important repercussions
for many different physical quantities.
Note that also the charged Higgs coupling to a pair of fermions
is modified, which will in turn induce changes in the contribution of the
charged Higgs loop in $b \to s \gamma$ at the one loop-level.   
In order to get naturally small Flavour Changing Neutral Currents (FCNCs), we will use the 
Cheng-Sher ansatz by taking $Y_k^{i,j} \propto \sqrt{m_i m_j}/v$ 
\cite{Cheng:1987rs,Atwood:1996vj}. 

After EWSB, the Yukawa Lagrangian can be expressed in the mass-eigenstate basis as 
\cite{GomezBock:2005hc,Arhrib:2017yby}:
\begin{eqnarray}
{\mathcal L}_Y &=& -\sum_{f=u,d,\ell} \frac{m_f}{v} \left(\xi_h^f \bar f fh + \xi_H^f \bar f fH - i \xi_A^f \bar f \gamma_5 f A \right) \nonumber \\
 && - \Big(\frac{\sqrt 2 V_{ud}}{v} \bar u \left (m_u \xi_A^u P_L +  m_d\xi_A^d P_R \right )d H^+ 
+\text{h.c.}\Big),\label{Eq:Yukawa}
\end{eqnarray}
where the couplings $\xi_\Phi^f$ are given in table~\ref{Tab:MixFactor} for Type II, Flipped and
 Type III. We stress that the parameters $\eta_{ij}^f$ are related to the Yukawa
couplings through the relations: $\eta_{ij}^u=U_L^u Y_1^u U_R^{u\dagger}/m_j$ and 
$\eta_{ij}^d=U_L^d Y_2^d U_R^{d\dagger}/m_j$, where $U_{L,R}^f$ are unitary
matrices that diagonalise the fermions mass matrices. 
Using the Cheng-Sher ansatz, we assume that $\eta_{ij}^f= \sqrt{m_i/m_j} \chi_{ij}^f/v$ where $\chi_{ij}^f$ is a free
parameter that will be taken in the range $[-1,1]$.
%As it can be seen from table-I, the Yukawa textures are different
% in 2HDM type II and type III, and this
% could lead to different phenomenological aspects.
%
\begin{table}[!t]
\small
\begin{center}
\renewcommand{\arraystretch}{1.5}
\begin{tabular}{|c|c|c|c|c|c|c|} \hline
\ \ $\Phi$ \ \  &  \multicolumn{2}{c|}{$\xi^u_{\Phi}$}  &  \multicolumn{2}{c|}{$\xi^d_{\Phi}$}  &  \multicolumn{2}{c|}{$\xi^\ell_{\Phi}$}   \\  \hline
& Type II & Type III & Type II & Type III & Type II & Type III \\   
\hline
$h$  & \ $\;   \frac{c_\alpha}{s_\beta}   \; $ \ 
        & \ $\;   \frac{c_\alpha}{s_\beta}\delta_{ij} - \eta_{ij}^f \frac{c_{\beta-\alpha}}{\sqrt{2}s_\beta}   \; $ \ 
        & \ $ \; -\frac{s_\alpha}{c_\beta}   \; $ \ 
        & \ $ \; -\frac{s_\alpha}{c_\beta}\delta_{ij} + \eta_{ij}^f \frac{c_{\beta-\alpha}}{\sqrt{2}c_\beta}  \; $ \ 
        & \ $ \; -\frac{s_\alpha}{c_\beta}   \; $ \ 
        & \ $ \; -\frac{s_\alpha}{c_\beta}\delta_{ij} + \eta_{ij}^f \frac{c_{\beta-\alpha}}{\sqrt{2}c_\beta}  \; $ \ \\
$H$  & \ $\;  \frac{s_\alpha}{s_\beta}  \; $ \ 
        & \ $\;   \frac{s_\alpha}{s_\beta}\delta_{ij} + \eta_{ij}^f \frac{s_{\beta-\alpha}}{\sqrt{2}s_\beta}  \; $ \ 
        & \ $ \;  \frac{c_\alpha}{c_\beta}    \; $ \ 
        & \ $ \;  \frac{c_\alpha}{c_\beta}\delta_{ij} - \eta_{ij}^f \frac{s_{\beta-\alpha}}{\sqrt{2}c_\beta}   \; $ \ 
        & \ $ \;  \frac{c_\alpha}{c_\beta}        \; $ \ 
        & \ $ \;  \frac{c_\alpha}{c_\beta}\delta_{ij} - \eta_{ij}^f \frac{s_{\beta-\alpha}}{\sqrt{2}c_\beta}   \; $ \ \\
$A$  & \ $\;   \frac{1}{t_\beta}   \; $ \ 
        & \ $\;    \frac{1}{t_\beta}\delta_{ij} - \eta_{ij}^f \frac{1}{\sqrt{2}s_\beta} \; $ \ 
        & \ $ \;   t_\beta                \; $ \ 
        & \ $ \;   t_\beta \delta_{ij} - \eta_{ij}^f \frac{1}{\sqrt{2}c_\beta}              \; $ \ 
        & \ $ \;   t_\beta                \; $ \ 
        & \ $ \;   t_\beta \delta_{ij} - \eta_{ij}^f \frac{1}{\sqrt{2}c_\beta}              \; $ \ \\ 
\hline
\end{tabular}
\end{center}
\caption{Neutral Higgs Yukawa couplings in Type II and Type III relative to the SM Higgs Yukawa couplings with 
$\eta_{ij}^f = \sqrt{m_i/m_j}\chi_{ij}^f/v $. The Yukawa couplings for the Flipped 
model are easily obtained from the Type II ones with the replacements: $\xi^{u, d, \ell}_{\Phi}$ (Flipped) = $\xi^{u, d, u}_{\Phi}$ (Type II).}
\label{Tab:MixFactor}
\end{table}
As can be seen from table~\ref{Tab:MixFactor}, if the $\chi_{ij}^f$'s are of ${\cal{O}}(1)$, 
the new effects are dominated by heavy fermions and comparable with those in the 2HDM Type II
and Flipped models. 
The effect of the $\chi_{ij}^f$'s can modify significantly the limit on the charged
Higgs boson mass coming from $b\to s\gamma$. As recently discussed in~\cite{Misiak:2017bgg}, 
the mass of the charged Higgs boson is bounded to be heavier than about 580 GeV for any value of
$\tan\beta$ in both  the Type II and Flipped models. As shown in~\cite{Arhrib:2017yby}, though, 
this bound can be weakened to about 200 GeV by judiciously tuning the $\chi_{ij}^f$'s  
together with the other 2HDM Type III parameters.  

The couplings of $h$ and $H$ to gauge bosons $V=W,Z$ are proportional to 
$\sin(\beta-\alpha)$ and  $\cos(\beta-\alpha)$, respectively.
Since these are gauge couplings, they are the same for all Yukawa types.
As we are considering the scenario where the neutral lightest Higgs state  is the 125 GeV scalar,
the SM-like Higgs boson $h$ is recovered 
when $\cos(\beta-\alpha)\approx 0$. For the Type II and Flipped models this is also
the limit where the Yukawa couplings of the discovered Higgs boson  become SM-like.
The limit $\cos(\beta-\alpha)\approx 0$ seems to be favoured by LHC data, 
except for the possibility of a wrong sign limit \cite{Ferreira:2014naa, Ferreira:2014dya} 
where the couplings to down-type quarks can have a relative sign
to the gauge bosons opposite to that of the SM. Our benchmarks will focus
on the SM-like limit where indeed $\cos(\beta-\alpha)\approx 0$ and
consequently the effect of the $\chi_{ij}^f$'s  in $hf\bar f$ and $Hf\bar f$
coupling is suppressed by the $\cos(\beta-\alpha)$ factor.

%%%%%%%%%%%%%%%%%%%%%%%%%%%%%%%%%%%%%%%%%%%%%%%%%%
\section{Theoretical and experimental constraints}
%%%%%%%%%%%%%%%%%%%%%%%%%%%%%%%%%%%%%%%%%%%%%%%%%%

In order to perform a systematic scan on the versions of the 2HDM Type II, Flipped and  Type III, 
we use the following theoretical and experimental constraints. 
\begin{itemize}
\item \textbf{\underline{Vacuum stability}}:
To ensure that the scalar potential is bounded from below, 
the quartic couplings should satisfy the relations~\cite{Deshpande:1977rw}
\begin{equation}
	\lambda_{1,2}>0,\qquad \lambda_3>-(\lambda_1 \lambda_2)^{1/2},\qquad \mathrm{and} \qquad \lambda_3+\lambda_4- \vert \lambda_5 \vert  >-(\lambda_1 \lambda_2)^{1/2}.
\end{equation}		
We impose that the potential has a minimum that is compatible with EWSB.
If this minimum is CP-conserving, any other possible charged or CP-violating stationary points
will be a saddle point above the minimum~\cite{Ferreira:2004yd}. However, there is still the possibility
of having two coexisting CP-conserving minima. In order to force the minimum compatible with
EWSB, one can impose the simple condition~\cite{Barroso:2013awa}:
\begin{equation}
m_{12}^2 \left(m_{11}^2-m_{22}^2 \sqrt{\lambda_1/\lambda_2} \right) \left( \tan \beta - \sqrt[4]{\lambda_1/\lambda_2}\right) >0.
\end{equation}

Writing the minimum conditions as
\begin{align}
m_{11}^2+\dfrac{\lambda_1 v_1^2}{2}+\dfrac{\lambda_3 v_2^2}{2} &= \frac{v_2}{v_1} \left[ m_{12}^2 - (\lambda_4+\lambda_5)\dfrac{v_1 v_2}{2}\right],\\
m_{22}^2+\dfrac{\lambda_2 v_2^2}{2}+\dfrac{\lambda_3 v_1^2}{2} &= \frac{v_1}{v_2} \left[ m_{12}^2 - (\lambda_4+\lambda_5)\dfrac{v_1 v_2}{2}\right],
\end{align} 
allows us to express $m_{11}^2$ and $m_{22}^2$ in terms of the soft
$Z_2$ breaking term $m_{12}^2$ and the quartic couplings
$\lambda_{1-5}$.
\item \textbf{\underline{Perturbative unitarity}}:
Another important theoretical constraint on the scalar sector of  2HDMs  stems 
from the perturbative unitarity requirement of the $S$-wave component 
of the various scalar scattering amplitudes. 
That condition implies a set of constraints that have to 
be fulfilled and are given by~\cite{Kanemura:1993hm}
\begin{equation}
|a_\pm|, |b_\pm|, |c_\pm|, |f_\pm|, |e_{1,2}|, |f_1|, |p_1| < 8 \pi,
\end{equation}
where
\begin{align}
\begin{split}
a_\pm &= \dfrac{3}{2}(\lambda_1+\lambda_2)\pm \sqrt{\dfrac{9}{4}(\lambda_1-\lambda_2)^2+(2\lambda_3+\lambda_4)^2},\\
b_\pm &= \dfrac{1}{2}(\lambda_1+\lambda_2)\pm \dfrac{1}{2} \sqrt{(\lambda_1-\lambda_2)^2+4\lambda_4^2},\\
c_\pm &= \dfrac{1}{2}(\lambda_1+\lambda_2)\pm \dfrac{1}{2} \sqrt{(\lambda_1-\lambda_2)^2+4\lambda_5^2},\\
e_1 &= \lambda_3 + 2 \lambda_4 -3\lambda_5,\hspace*{3cm}
e_2 = \lambda_3-\lambda_5,\\
f_+ &= \lambda_3+2 \lambda_4+3\lambda_5, \hspace*{2.9cm} f_- =\lambda_3+\lambda_5,\\
f_1 &= \lambda_3+\lambda_4, \hspace*{4.3cm}p_1 = \lambda_3-\lambda_4.
\end{split}
\end{align}
\item \textbf{\underline{EW Precision Tests}}:	
The additional neutral and charged 
scalars contribute to the gauge boson vacuum
polarisation through their coupling to gauge bosons. 
As a result, the updated EW precision data provide important
constraints on new physic models. In particular, the universal parameters
$S, T$ and $U$ provides constraint on the mass splitting 
between the heavy states $m_H$, $m_{H^\pm}$ and $m_A$
in the scenario in which $h$ is identified with the SM-like Higgs state. 
%(see for instance~\cite{Becirevic:2015fmu}). 
The general expressions for the parameters $S$, $T$ and $U$ in 2HDMs  
can be found in~\cite{Barbieri:2006bg}. 
To derive constraints on the scalar spectrum we consider the following 
updated values for $S, T$ and $U$:
\begin{equation}
     \Delta S = 0.05\pm 0.11,\quad \quad  \Delta T = 0.09\pm 0.13, \quad \quad 
     \Delta U = 0.01\pm 0.11, \\
\end{equation}
and use the corresponding covariance matrix given in \cite{Baak:2014ora}.
The $\chi^2$ function is then expressed as
\begin{equation}
  \chi^2_{ST}= \sum_{i,j}(X_i - X_i^{\rm SM})(\sigma^2)_{ij}^{-1}(X_j - X_j^{\rm SM}),
\end{equation}
with correlation factor  +0.91. 
\item \textbf{\underline{LHC constraints}}:
Moreover, we take into account the new 
experimental data at 13 TeV from the observed cross section 
 times Branching Ratio (BR) 
divided by the SM predictions, i.e., the so-called `signal strengths' of the Higgs 
boson defined by
\begin{eqnarray}
\mu^f_{i}&=&\frac{\sigma(i \rightarrow h)^{\rm 2HDM} {\rm BR}(h\rightarrow
  f)^{\rm 2HDM}}{\sigma(i\rightarrow h)^{\rm SM} {\rm BR}(h\rightarrow f)^{\rm SM}},
\quad\quad\quad i=1,2,
\end{eqnarray}
where $\sigma(i \rightarrow h)$ denotes the Higgs boson production cross 
section through channel $i$ and  ${\rm BR}(h\rightarrow f)$  the 
BR for the Higgs decay $h\rightarrow f$. 
Since several Higgs production channels are 
available at the LHC,  they are grouped to be 
$\mu^f_{1} =\mu^f_{{\rm ggF}+tth}$ and  $\mu^f_{2} = \mu^f_{{\rm{VBF}}+Vh}$,
containing gluon-gluon Fusion (ggF) plus associated Higgs production $t\bar{t}h$ as well as 
Vector Boson Fusion (VBF) plus Higgs-strahlung $Vh$ with $V=W^\pm,Z$.  
The values of the  observed signal strengths are 
shown with their correlation factor in table~\ref{tab:Higgs data}.
According to LHC results,  which appear to be in 
good agreement with the SM predictions~\cite{Corbett:2015ksa}, 
 data seems to favour a scenario with alignment limit where 
$\sin(\beta-\alpha)\approx 1$  where $h$ is the SM-like 
or $\cos(\beta-\alpha)\approx 1$  where $H$ is the SM-like.
As intimated, in our study, we identify the lightest CP-even state $h$ with the
SM-like scalar observed at the LHC with mass
$m_h=125.09(24)$~GeV~\cite{Olive:2016xmw} which, because we have discarded
the possibility of being in the wrong sign limit, in turn  implies that 
 $\sin(\beta-\alpha)\approx 1$.
 \end{itemize} 
%%%%%%%%%%%%%%%%%%%%%%%%%%%%%%%%%%%%%
   \begin{table}[hpbt]
 \begin{ruledtabular}
\begin{tabular}{ccccc} 
 $f$ & $\widehat{\mu}^{f}_{\rm{1}}$ & $\widehat{\mu}^{f}_{\rm{2}}$ & $\pm\,\,1\widehat{\sigma}_{\rm{1}}$ & $\pm\,\,1\widehat{\sigma}_{\rm{2}}$ \\ \hline
$\gamma\gamma$ & 1.09 &  1.14 &  0.23 &  0.25 \\ \hline
$ZZ^*$  &  1.31  & 1.25 & 0.24 &  0.28 \\ \hline
 $WW^*$ & 1.06 &  1.27& 0.18 &  0.21 \\ \hline
 $\tau^+\tau^-$ & 1.05 & 1.24 & 0.35 & 0.40  \\ \hline
$b\bar{b}$ & 3.9 & 3.7 &  2.8 & 2.4  \\ \hline
\end{tabular}
\caption{Combined best-fit signal strengths $\widehat{\mu}_{\rm{1}}$ and
$\widehat{\mu}_{\rm{2}}$ for corresponding Higgs decay mode from ~\cite{run2}.}
\label{tab:Higgs data}
\end{ruledtabular}
\end{table}
%%%%%%%%%%%%%%%%%%%%%%%%%%%%%%%%%%%%%%%%
\begin{itemize}
 %Moreover, we forbid the decay 
 %the dangerous decays
%$h\to A A$ which could over-saturate the total width of $h$ ($\simeq
%\Gamma_h^\mathrm{\rm SM}$) by assuming $m_A> m_h/2$. 

\item \textbf{\underline{Flavour physics constraints}}:
We take into account all the relevant flavour constraints which,
as previously discussed, force the the charged Higgs mass to be above about 580 GeV
from $b \to s \gamma$ at the $2\sigma$ level in Type II and Flipped~\cite{Misiak:2017bgg}.
However, we relax this condition to the $3\sigma$ level in order to obtain $H^\pm$  signal rates
that are more within the reach of the next run of the LHC. All other flavour
constraints were discussed recently for Type III in~\cite{Arhrib:2017yby} and
are also taken into account here. We again note that the tuning of the $\chi_{ij}^f$'s 
together with the other 2HDM Type III parameters allows us to relax the bound
on the charged Higgs boson mass significantly in this scenario.

\end{itemize}

%%%%%%%%%%%%%%%%%%%%%%%%%%%%%%%%%%%%%%%%
%\section{Charged Higgs production at the LHC}
%%%%%%%%%%%%%%%%%%%%%%%%%%%%%%%%%%%%%%%%

%%%%%%%%%%%%%%%%%%%%%%%%%%%%%%%%%%%%%%%%%%%%%%%%%%%%%%
\section{Numerical analysis}
The above mentioned constraints are then imposed onto a set of 
 randomly generated points in the ranges: 
\begin{eqnarray}
&& 200 \,{\rm GeV} \le m_{H^\pm} \le 1 \,{\rm TeV}, \quad 126 \,{\rm GeV} 
\le m_{H} \le 1 \,{\rm TeV}, 
\quad 100 \,{\rm GeV} \le m_{A} \le 1 \,{\rm TeV} \,, \non \\
&& -1 \le \sin\alpha \le 1,  \quad 2 \le \tan\beta \le 50,  
\quad -(1000\, {\rm GeV})^2 \le m^2_{12} \le (1000\, {\rm GeV})^2 \,.
\label{numbers}
\end{eqnarray}
We note again that we take the $\chi_{ij}^f$'s  in the range $[-1,1]$ and that 
all constraints are taken at the 2$\sigma$ level except the ones from
the $b\to s\gamma$ measurement where we allow compatibility at the 3$\sigma$ level 
which in Type II and Flipped mean a reduction in the bound from 580 GeV to 440 GeV. 
For Type III the scan starts at 200 GeV. 
%A scan of parameters consistent with the constraints listed above 
%favours the moderate and small values of $\tan\beta \in (2,50)$. 
%To see that the larger values of $\tan \beta$ cannot be discarded it 
%is sufficient to examine our BPs in the alignment limit. 
%%%%%%%%%%%%%%%%%%%%%%%%%%%%%%%%%%%%%%%%%%%%
\begin{figure}[h!]
\includegraphics[width=0.47\textwidth]{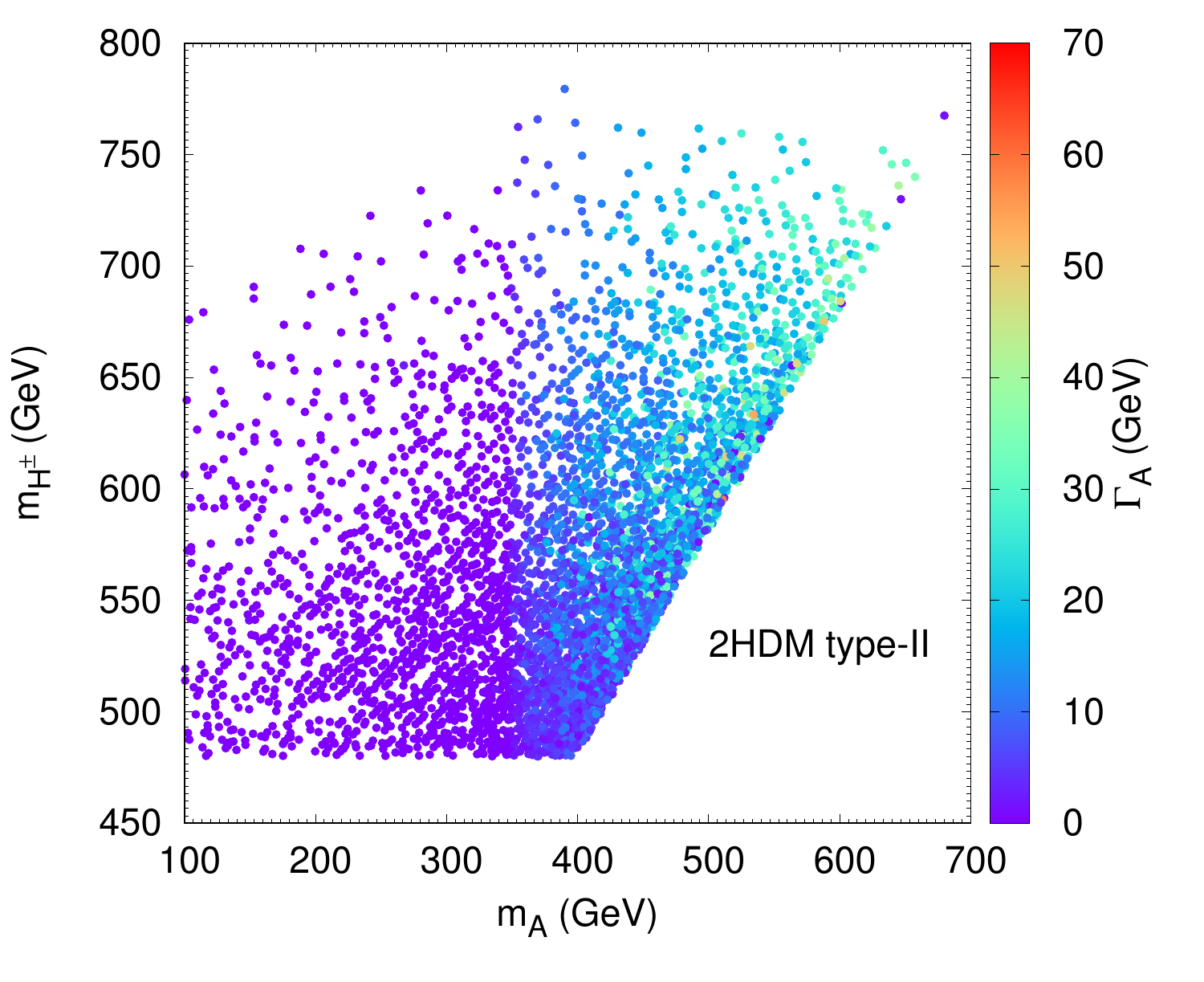}
\includegraphics[width=0.47\textwidth]{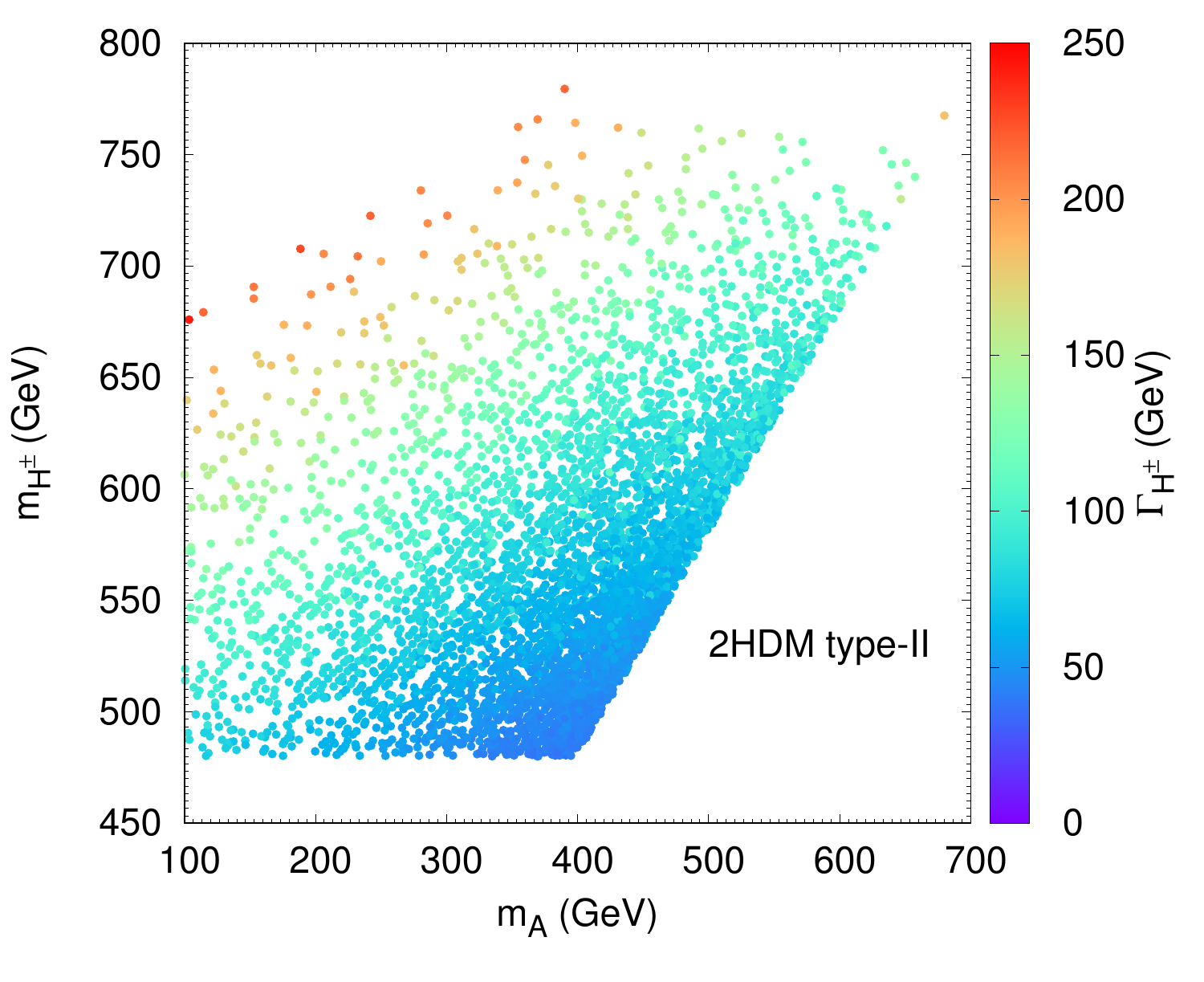}
\includegraphics[width=0.47\textwidth]{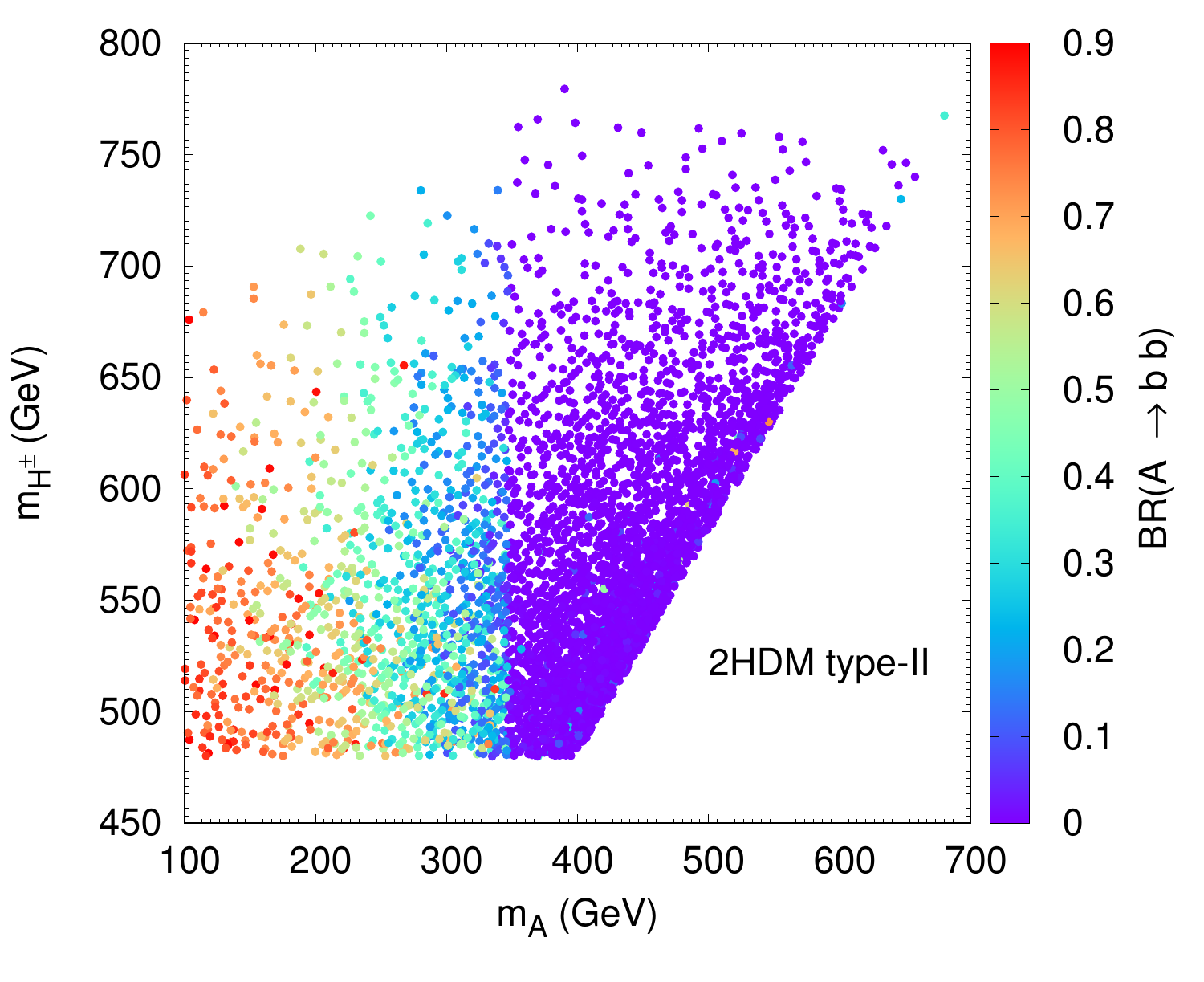}
\includegraphics[width=0.47\textwidth]{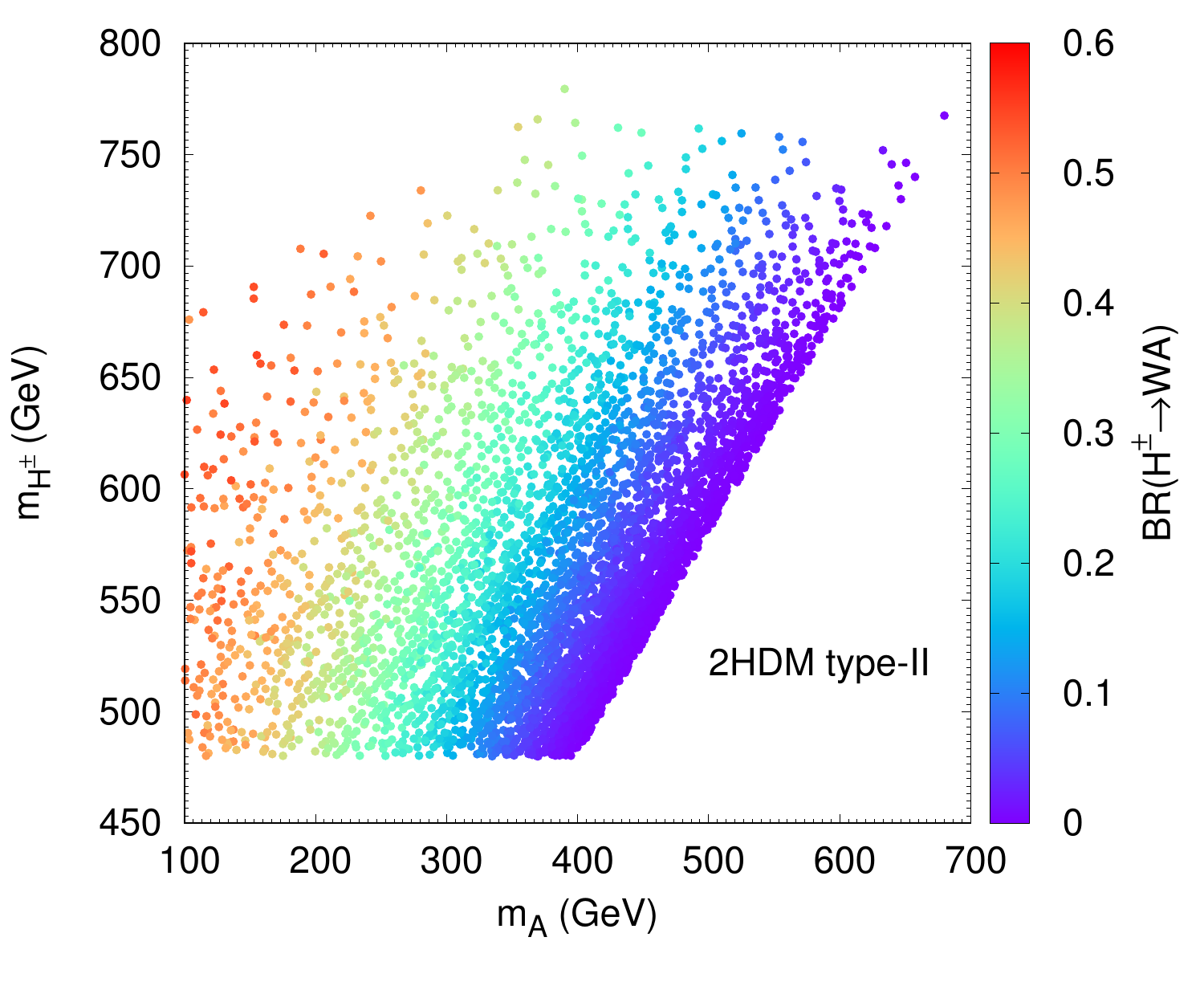}
\caption{Upper panels: Total decay widths (in GeV) 
of the CP-odd $A$ and charged Higgs $H^\pm$ bosons in the plane $(m_A,m_{H^\pm})$.
Lower panels: BR($A\to b\bar{b}$)  (left) and 
${\rm BR}(H^\pm \to W^\pm A)$ (right) in the plane $(m_A,m_{H\pm})$. All panels are
for Type II.}
\label{figure:ty2a}
\end{figure}
%%%%%%%%%%%%%%%%%%%%%%%%%%%%%%
\begin{figure}[h!]
\includegraphics[width=0.47\textwidth]{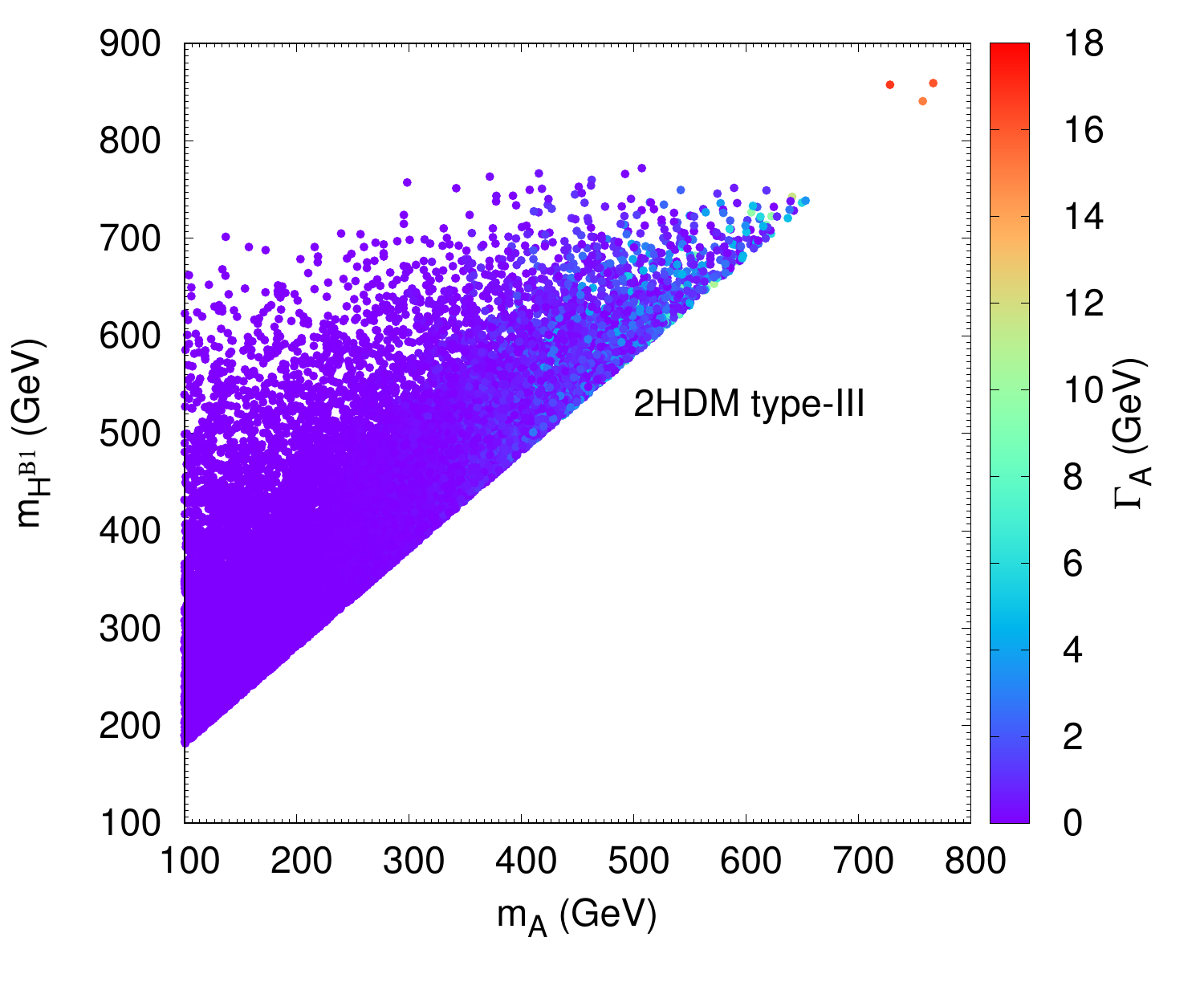}
\includegraphics[width=0.47\textwidth]{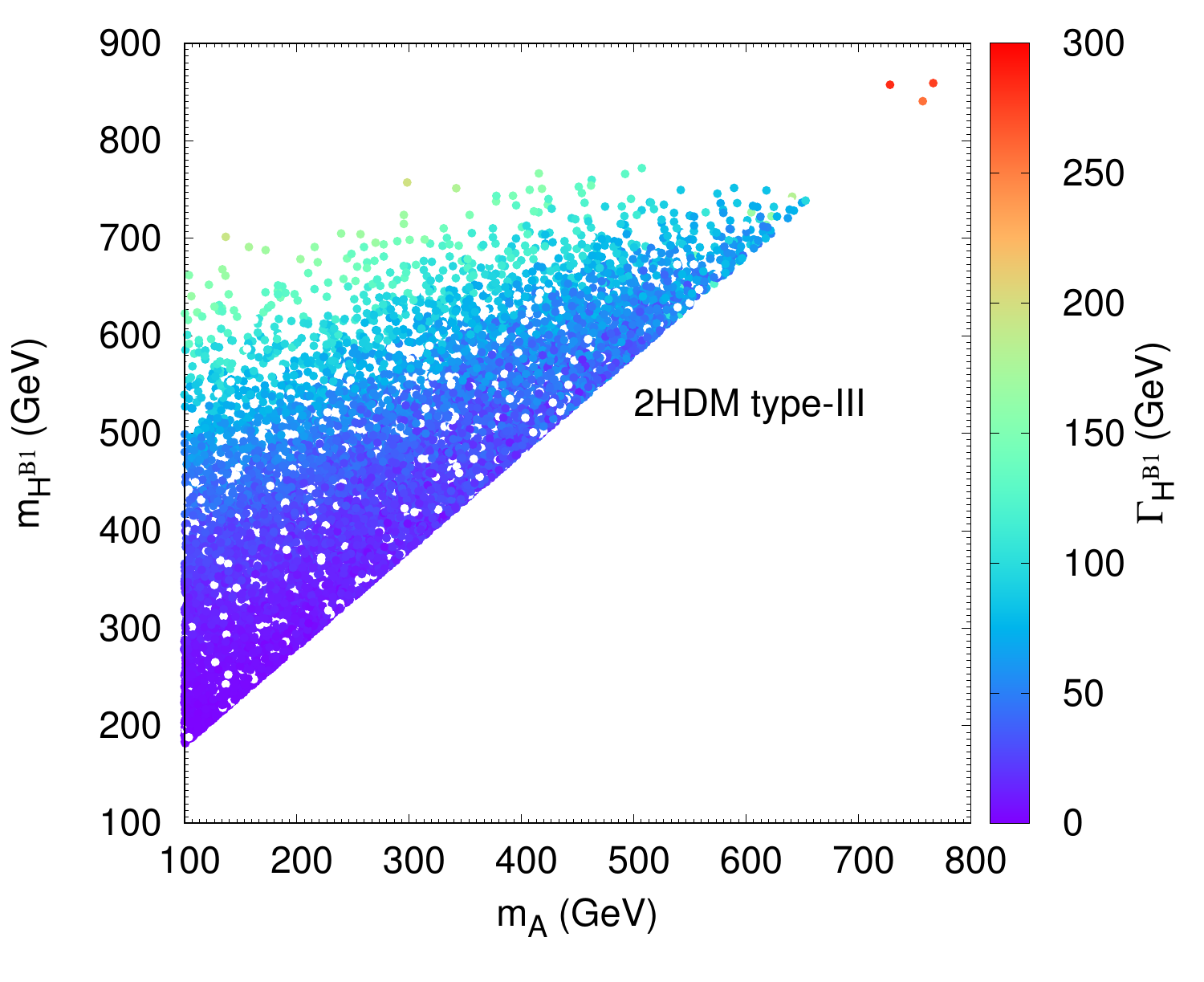}
\includegraphics[width=0.47\textwidth]{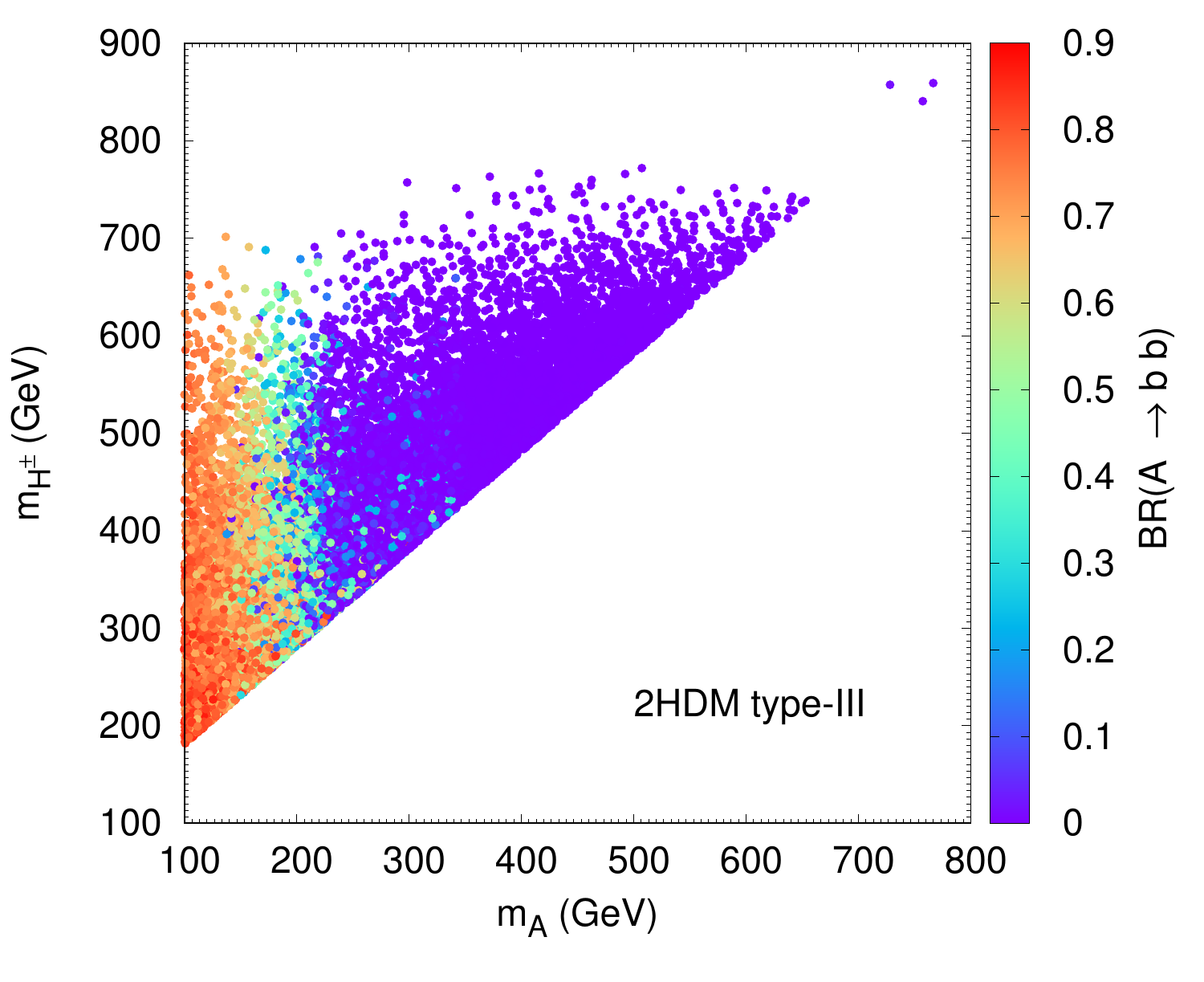}
\includegraphics[width=0.47\textwidth]{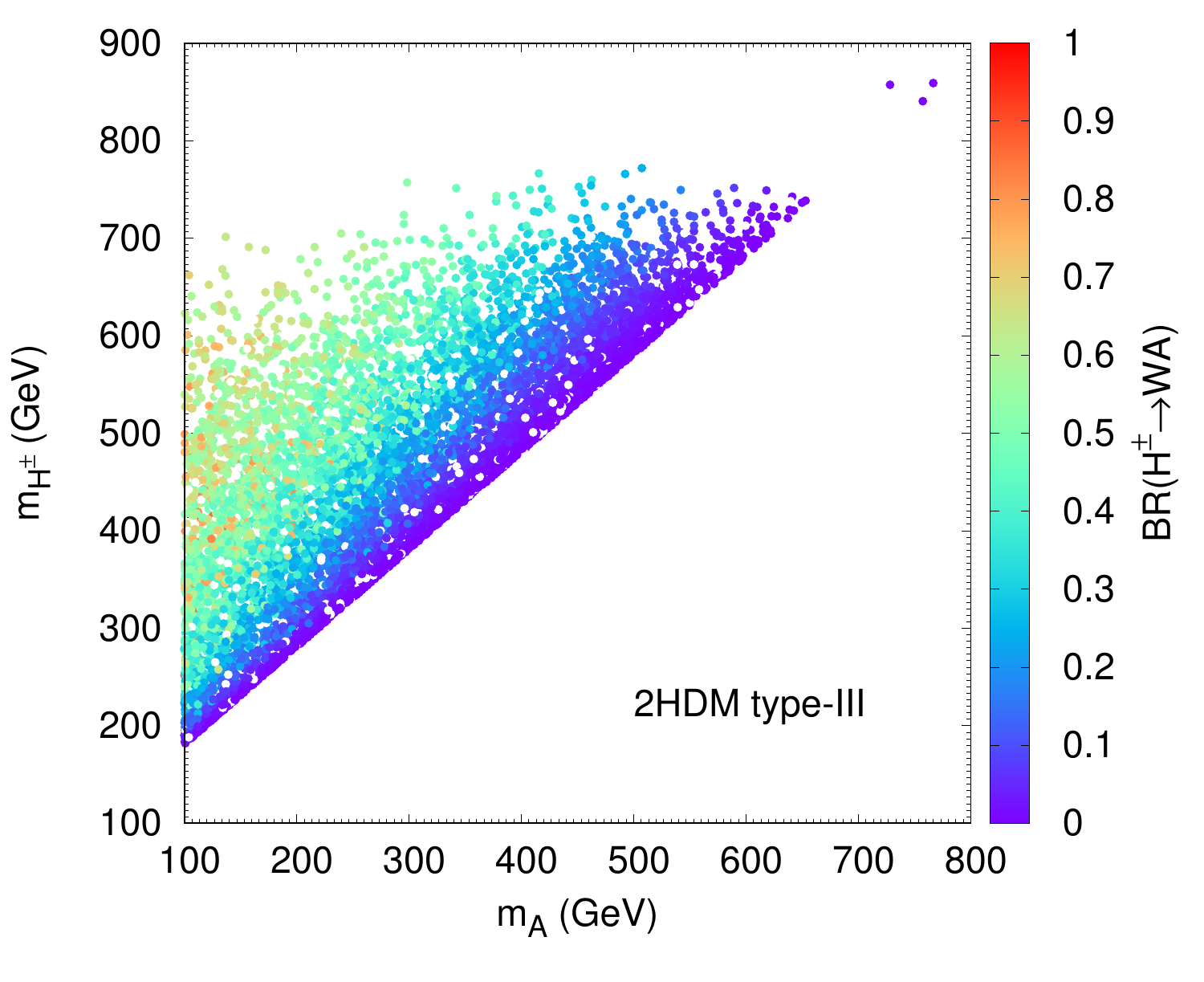}
\caption{Same as in figure~\ref{figure:ty2a} but  for Type III.}
\label{figure:ty3a}
\end{figure}
%%%%%%%%%%%%%%%%%%%%%%%%%%%%%%

In figure~\ref{figure:ty2a} (upper panels) we show the total widths of 
the CP-odd Higgs $A$ and the charged Higgs boson $H^\pm$ in the plane $(m_A,m_{H^\pm})$.
It is clear that the two widths can be simultaneously large.
In the case of the charged Higgs boson the total width is amplified by the 
opening of the bosonic decay $H^\pm \to W^\pm A$ for $m_A\leq 350$ GeV
while for the CP-odd Higgs the total width gets enhanced after the opening of
$A\to t\bar{t}$.
In the lower panels of figure~\ref{figure:ty2a} we present the 
BRs of $A\to b\bar{b}$ (left) and  $H^\pm \to W^\pm A$ (right).
One can see that the ${\rm BR}(A\to b\bar{b})$ could be sizeable and above
about 70\% below the $t\bar{t}$ threshold. 
In the case of the charged Higgs boson, since $g_{H^\pm W^\mp A}$ is a 
gauge coupling with no suppression factor, we expect    
${\rm BR}(H^\pm \to W^\pm A)$ to be large when kinematically allowed
and  able to compete with the $H^+ \to \bar{t}b$ and $H^\pm \to
 W^\pm H$ decays. The lighter the pseudoscalar is the larger 
the ${\rm BR}(H^\pm \to W^\pm A)$ can be, easily reaching 
values above 50\% as can be seen from the figure. Therefore, since we need
a charged Higgs boson with a large width, in order to compromise and to
obtain large ${\rm BR}(H^\pm \to W^\pm A)$ and large ${\rm BR}(A\to b\bar{b})$,
we need a heavy charged Higgs and a much lighter pseudoscalar. Still
the charged Higgs mass should not be too large so that the rate of
signal events is large enough to be seen at the LHC Run-II. We note 
that the plots for the Flipped model would be very similar and
therefore we will refrain from presenting them here.

In figure~\ref{figure:ty3a} we show the total width of the CP-odd 
$A$ and of the charged Higgs boson $H^\pm$ together with ${\rm BR}(A\to b\bar{b})$ and
${\rm BR}(H^\pm \to W^\pm A)$ in Type III. The picture is rather similar to the one for Type II
except that the charged Higgs mass for Type III is 
relaxed up to 200 GeV. The conclusions regarding the possible decays
are the same as for Type II and Flipped.

%%%%%%%%%%%%%%%%%%%%%%%%%%%%%%%%%%%%%%%% 
\section{Monte Carlo Analysis}
%%%%%%%%%%%%%%%%%%%%%%%%%%%%%%%%%%%%%%%% 

\begin{figure}
 \begin{subfigure}[t]{0.3\textwidth}
 \centering
 \includegraphics[scale=0.325]{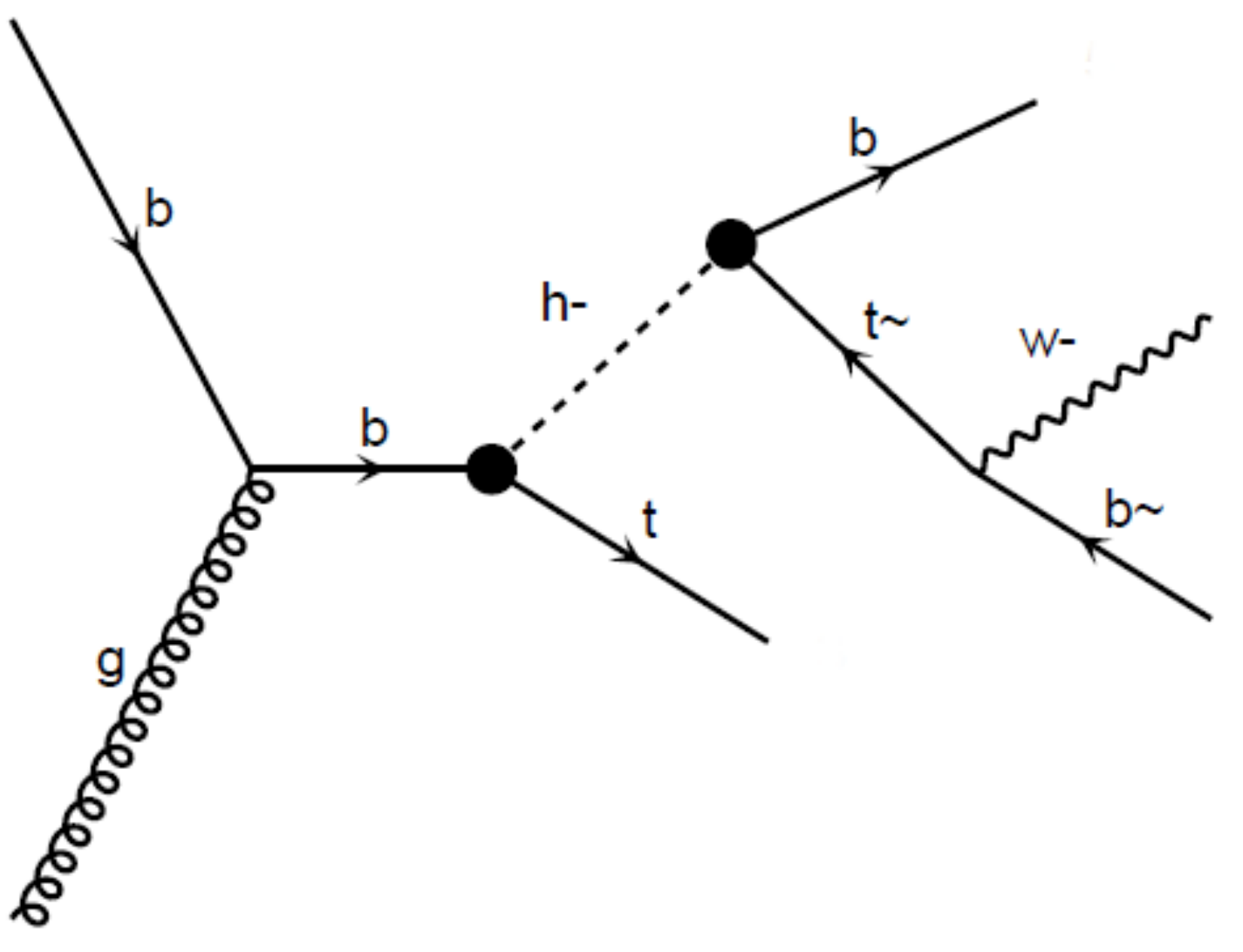} 
 \caption{}
 \end{subfigure}
 \begin{subfigure}[t]{0.3\textwidth}
 \centering
 \includegraphics[scale=0.325]{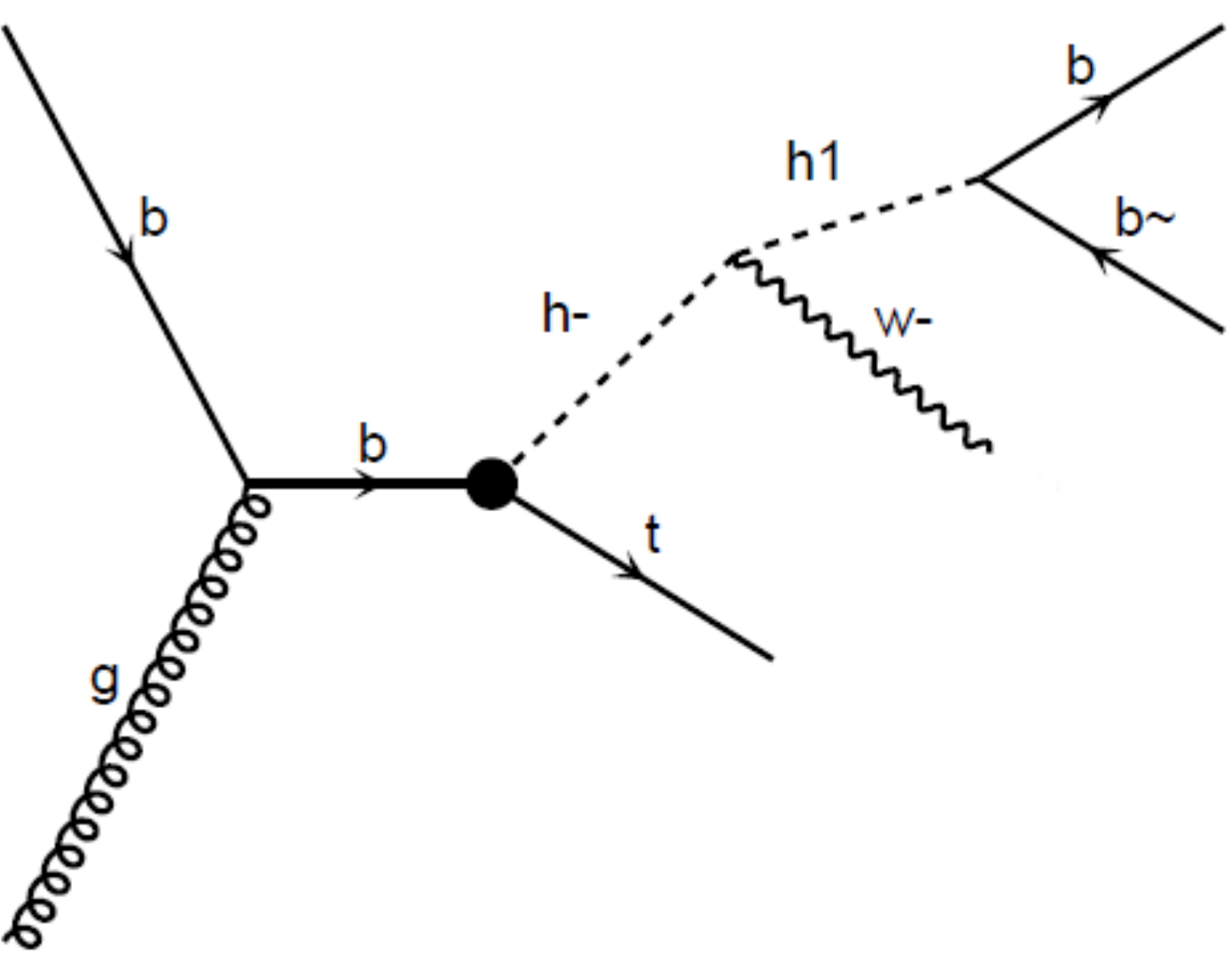} 
 \caption{}
 \end{subfigure}
 \begin{subfigure}[t]{0.3\textwidth}
 \centering
 \includegraphics[scale=0.325]{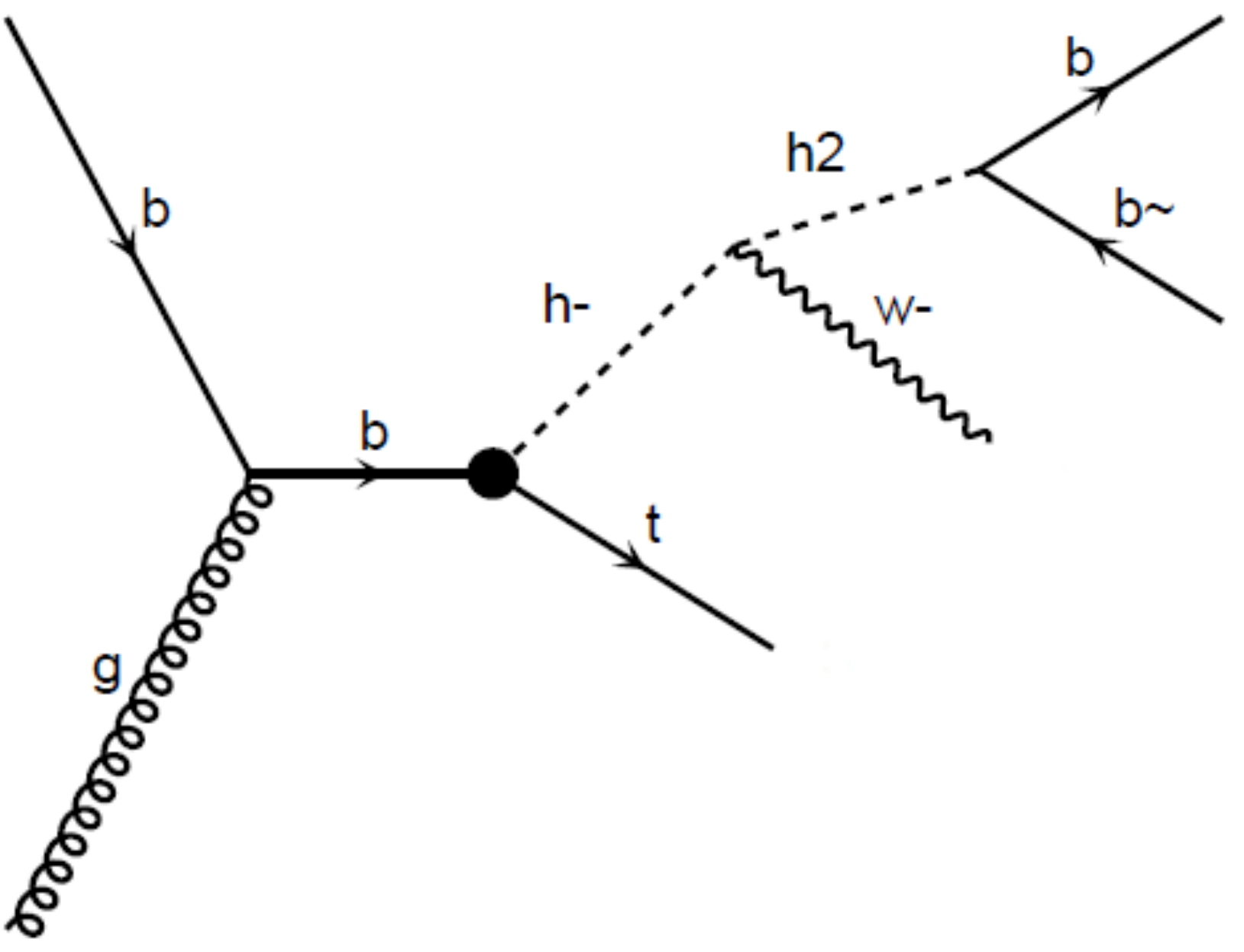} 
 \caption{}
 \end{subfigure}
 \begin{subfigure}[t]{0.3\textwidth}
 \centering
 \includegraphics[scale=0.325]{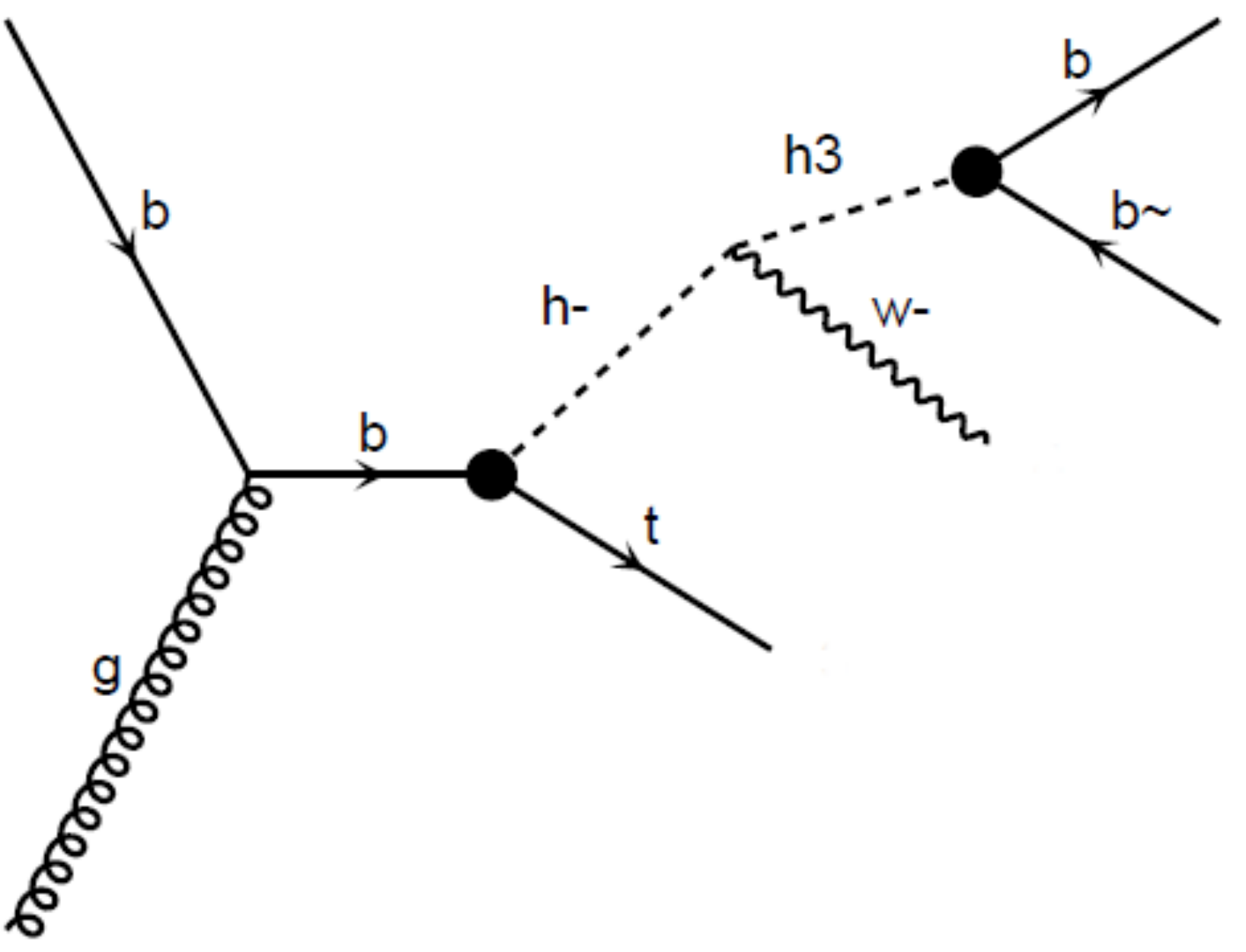} 
 \caption{}
 \end{subfigure}
 \begin{subfigure}[t]{0.3\textwidth}
 \centering
 \includegraphics[scale=0.325]{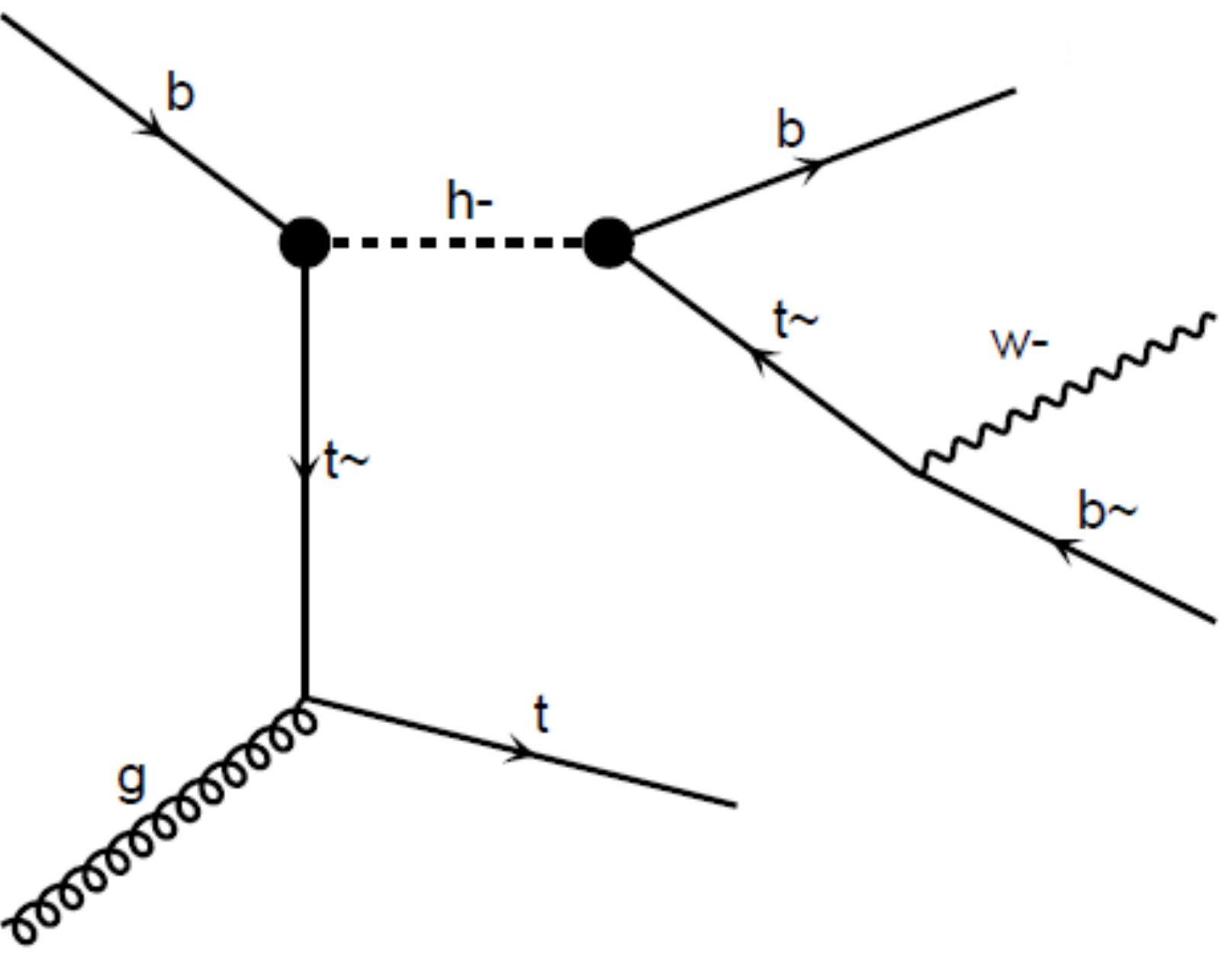} 
 \caption{}
 \end{subfigure}
 \begin{subfigure}[t]{0.3\textwidth}
 \centering
 \includegraphics[scale=0.325]{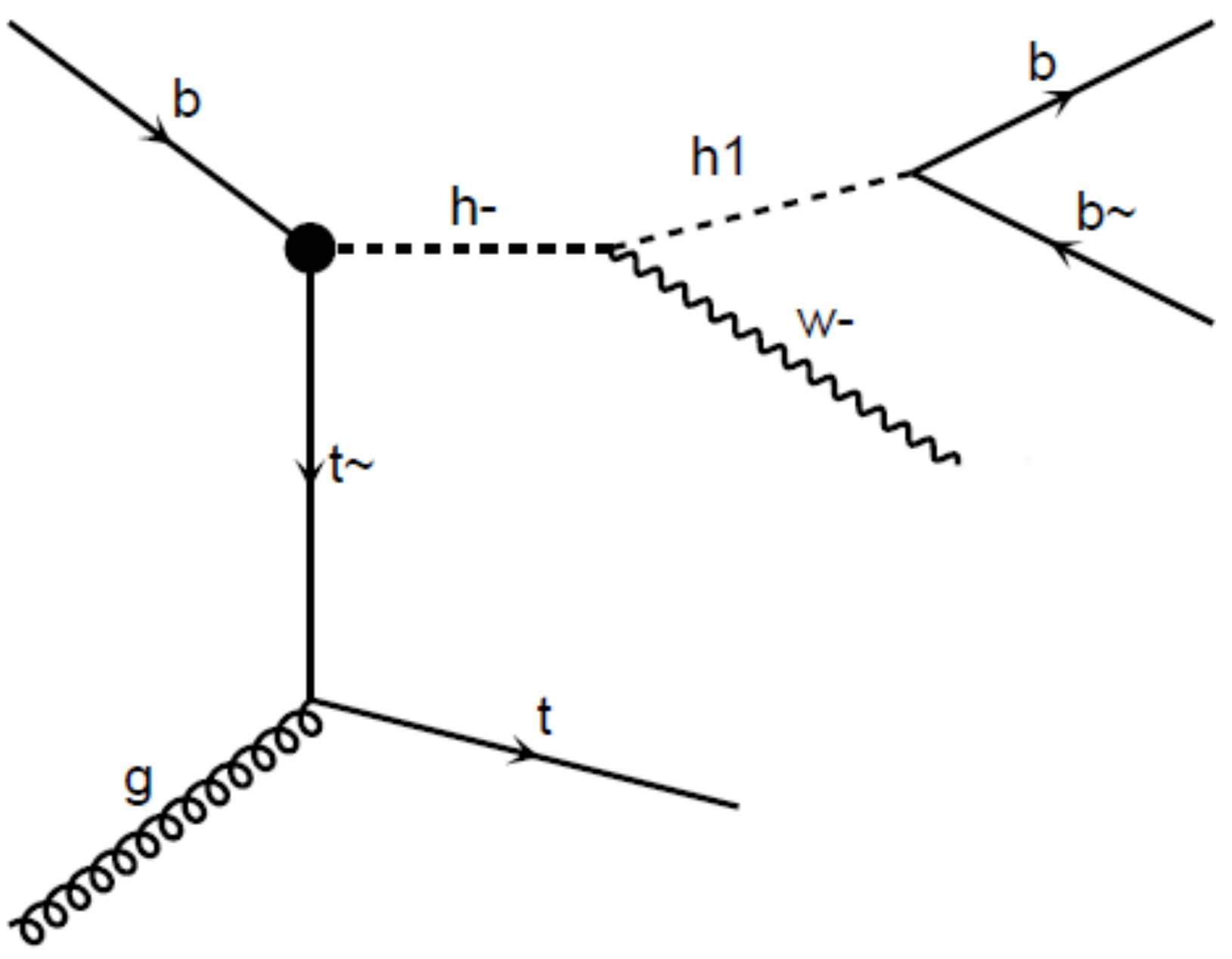} 
 \caption{}
 \end{subfigure}
 \begin{subfigure}[t]{0.3\textwidth}
 \centering
 \includegraphics[scale=0.325]{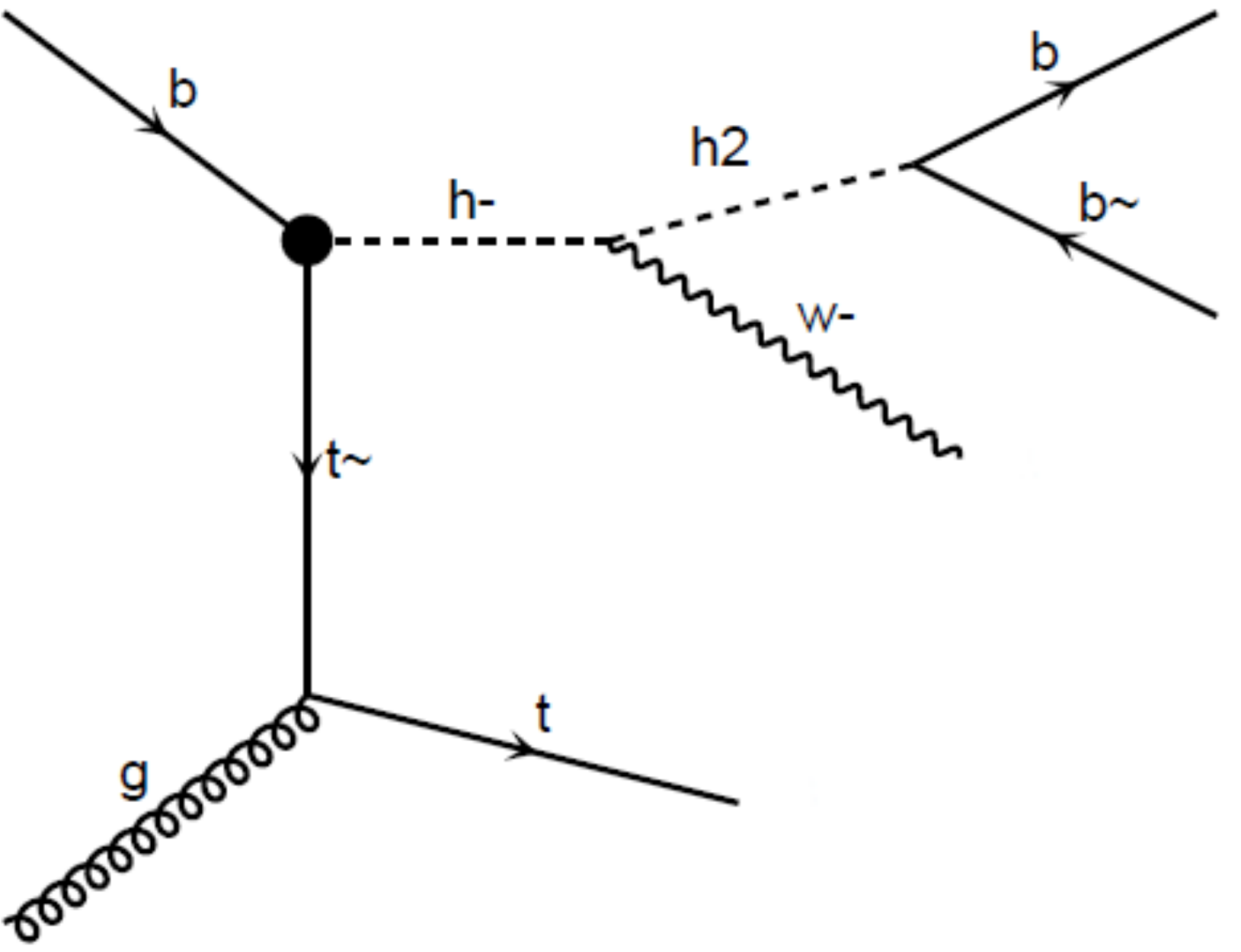} 
 \caption{}
 \end{subfigure}
 \begin{subfigure}[t]{0.3\textwidth}
 \centering
 \includegraphics[scale=0.325]{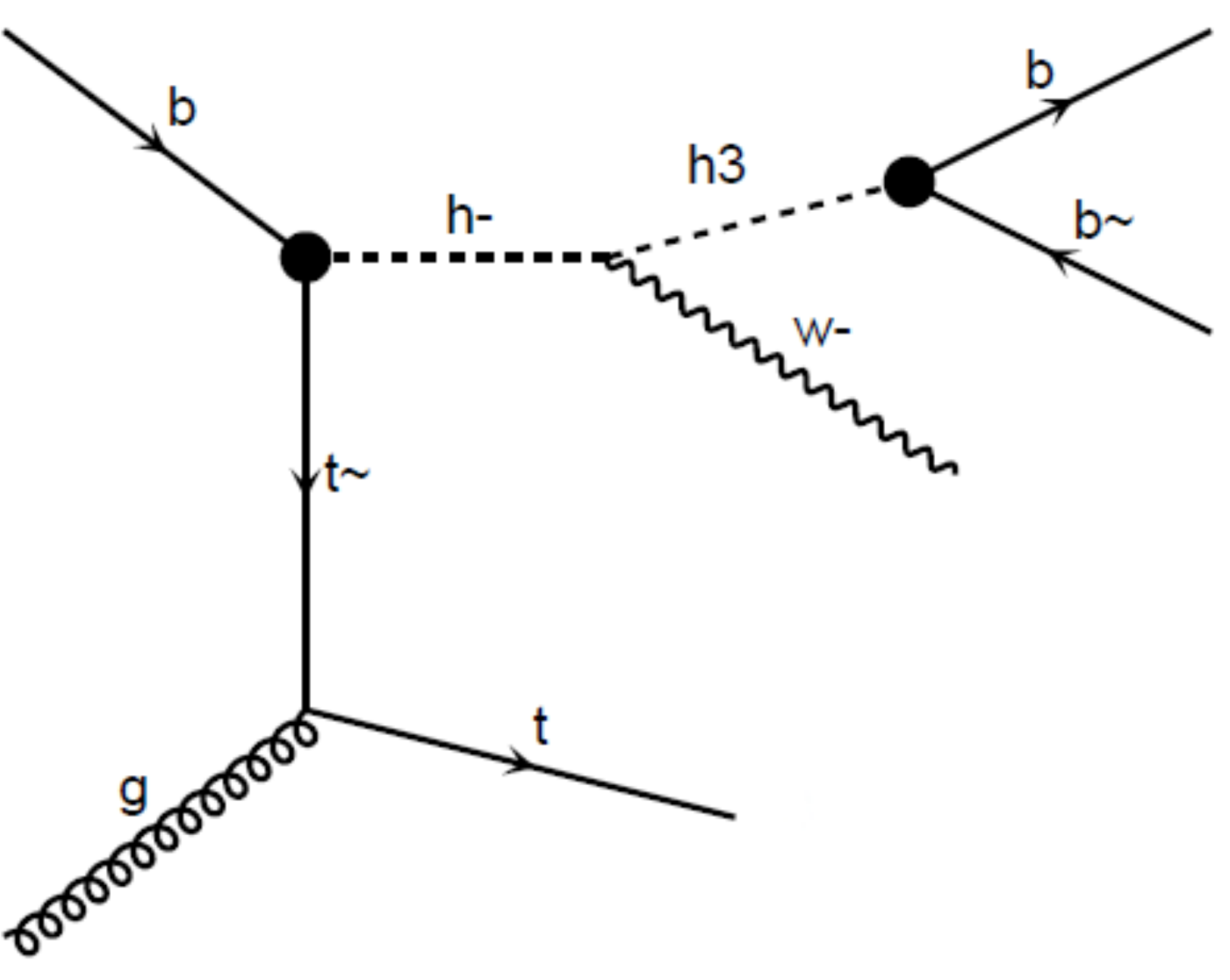} 
 \caption{}
 \end{subfigure}
\caption{\label{figs:sig} Feynman diagrams in a 2HDM contributing to resonant charged Higgs production and corresponding decays leading to the signal $pp\to tW^- b \bar b$ with h- $\equiv H^-$, h1 $\equiv h$, h2 $\equiv H$ and h3 $\equiv A$
 (as appropriate).}
\end{figure}

\begin{figure}
 \begin{subfigure}[t]{0.3\textwidth}
 \centering
 \includegraphics[scale=0.325]{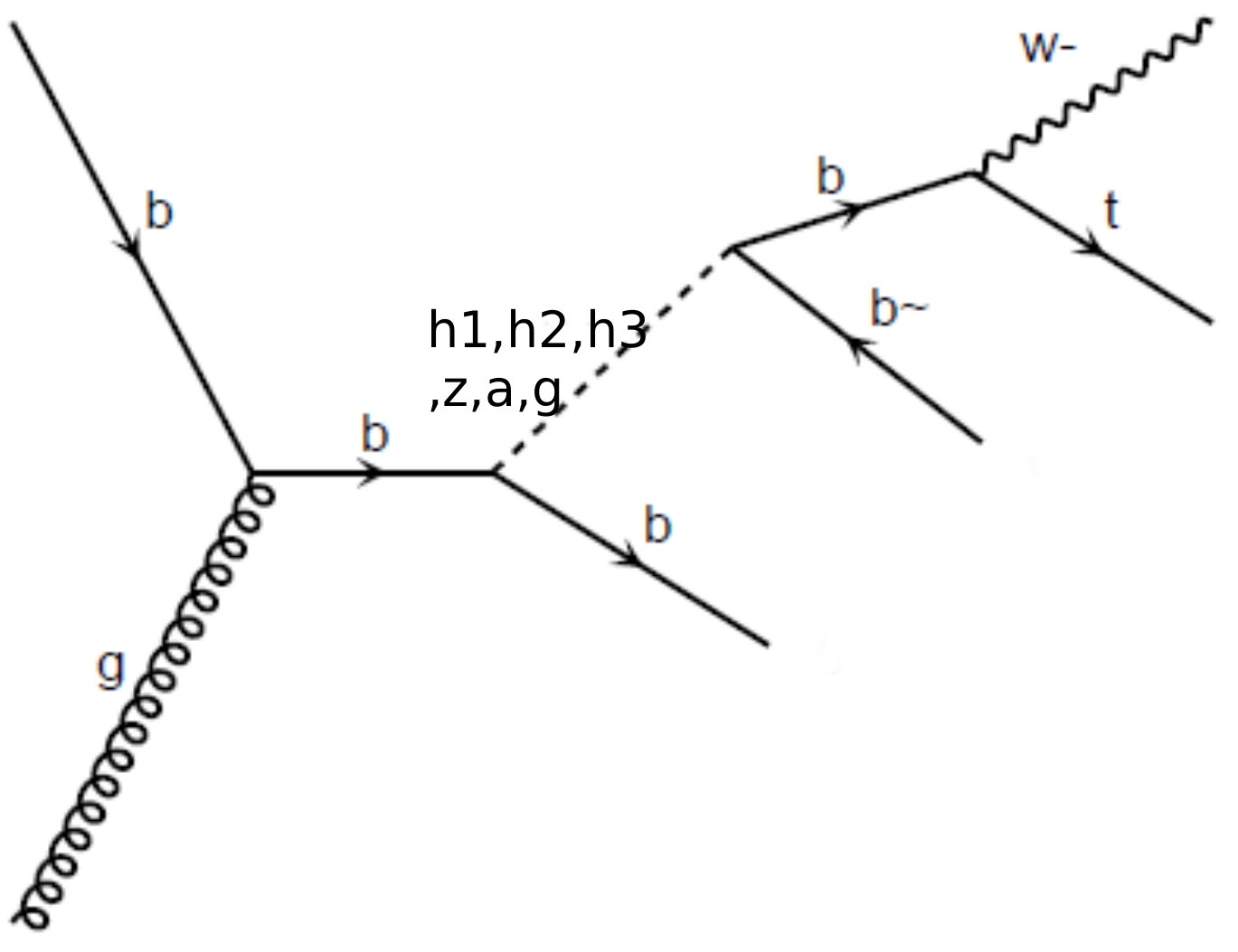} 
 \caption{}
 \end{subfigure}
 \begin{subfigure}[t]{0.3\textwidth}
 \centering
 \includegraphics[scale=0.325]{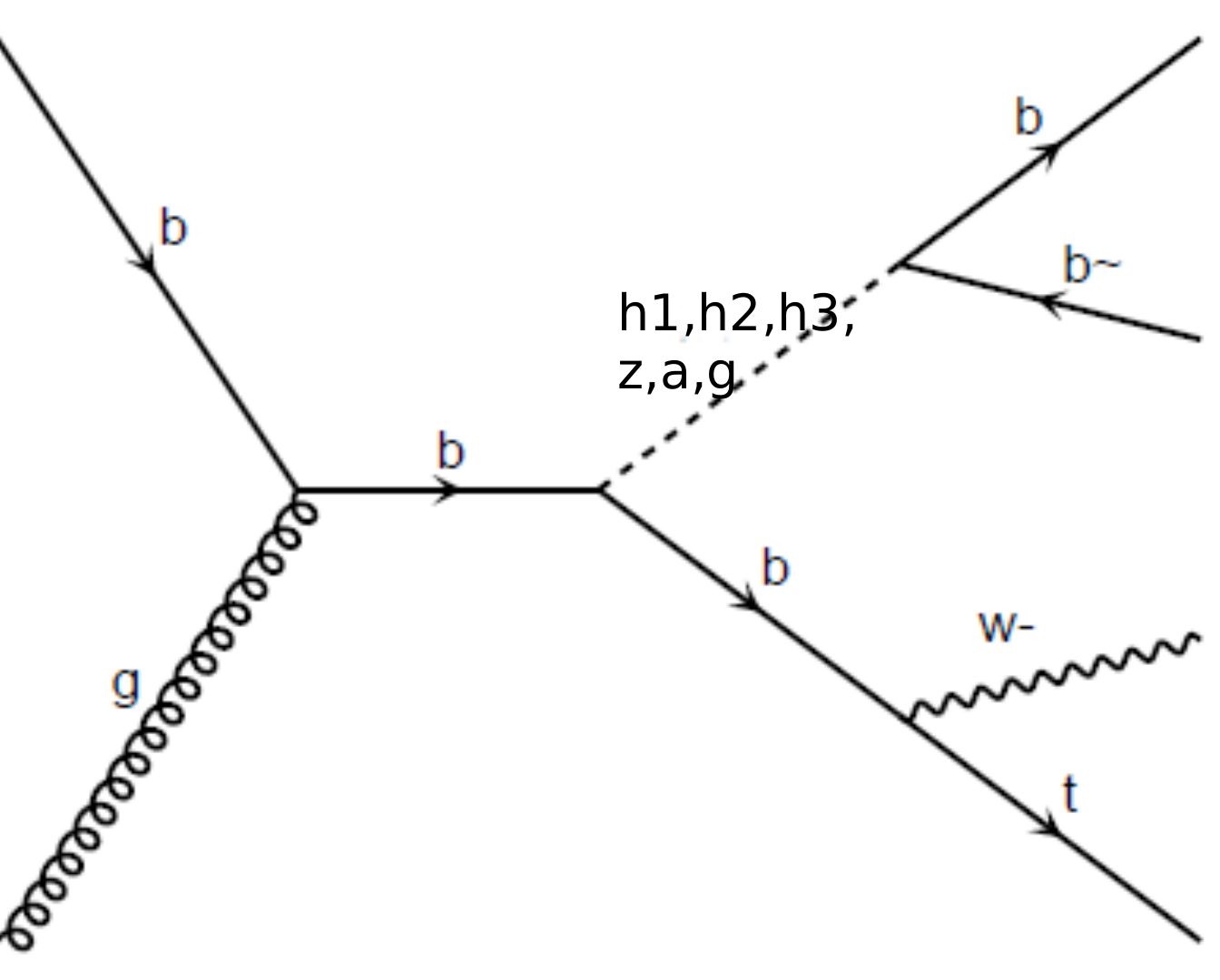} 
 \caption{}
 \end{subfigure}
 \begin{subfigure}[t]{0.3\textwidth}
 \centering
 \includegraphics[scale=0.325]{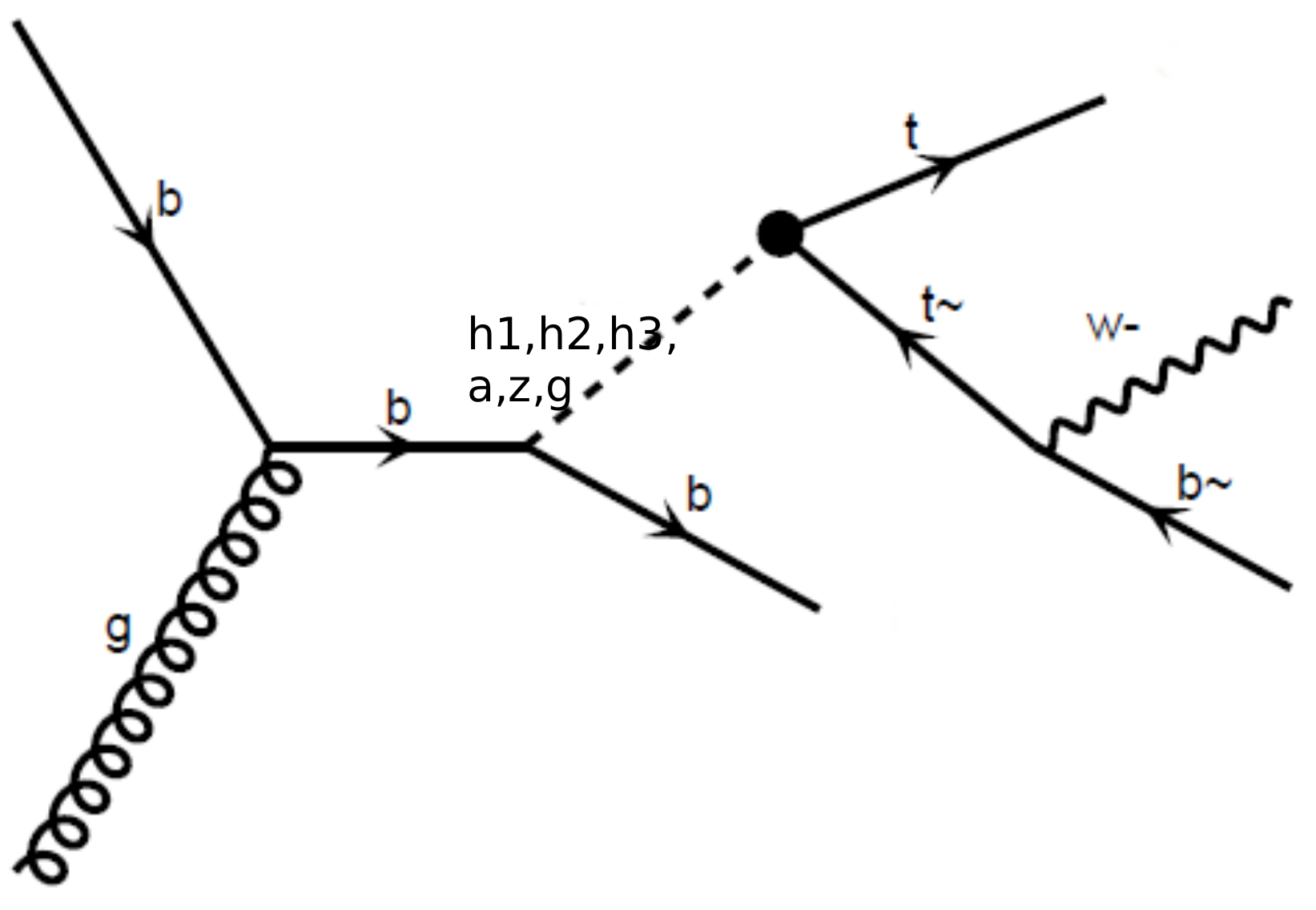} 
 \caption{}
 \end{subfigure}
 \begin{subfigure}[t]{0.3\textwidth}
 \centering
 \includegraphics[scale=0.325]{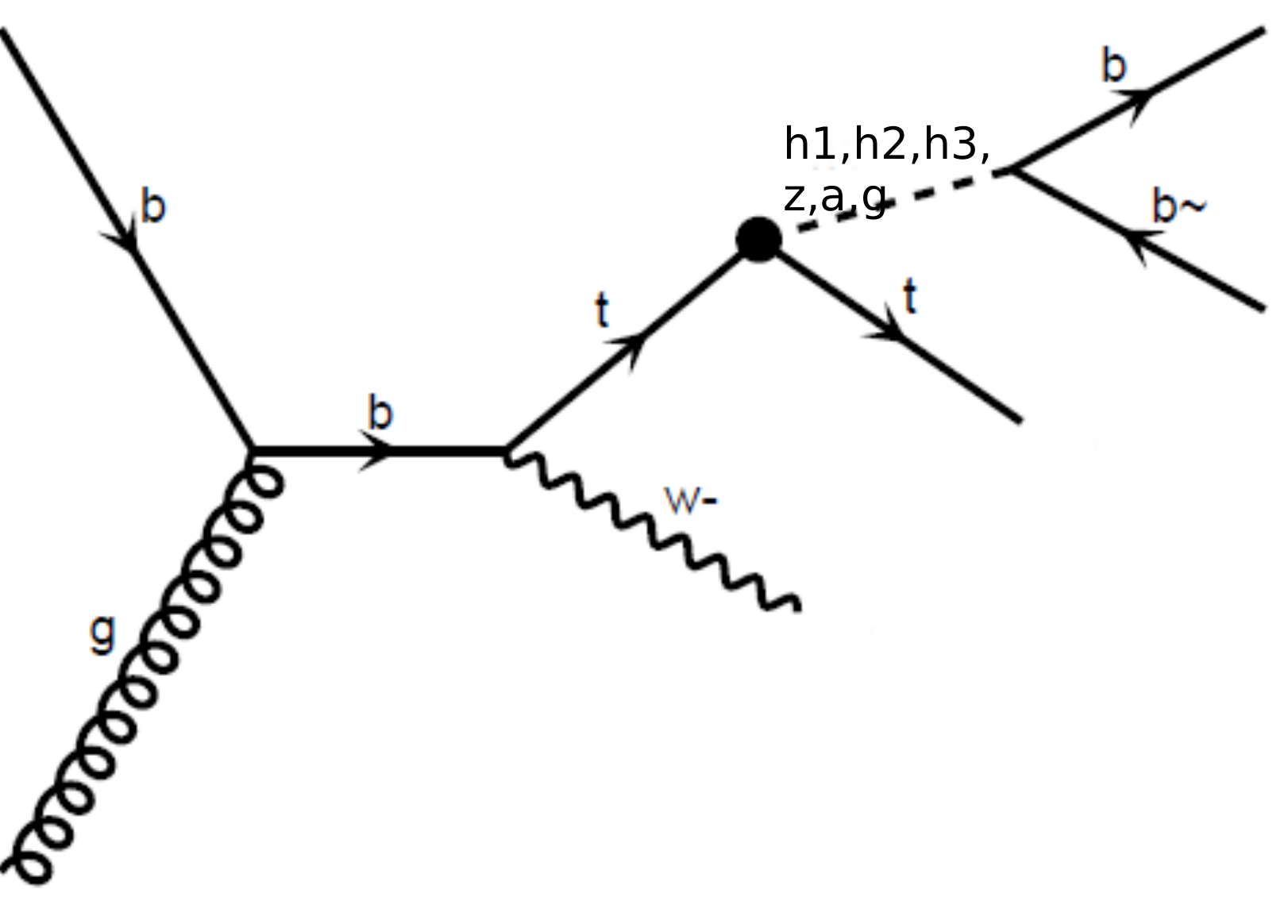} 
 \caption{}
 \end{subfigure}
 \begin{subfigure}[t]{0.3\textwidth}
 \centering
 \includegraphics[scale=0.325]{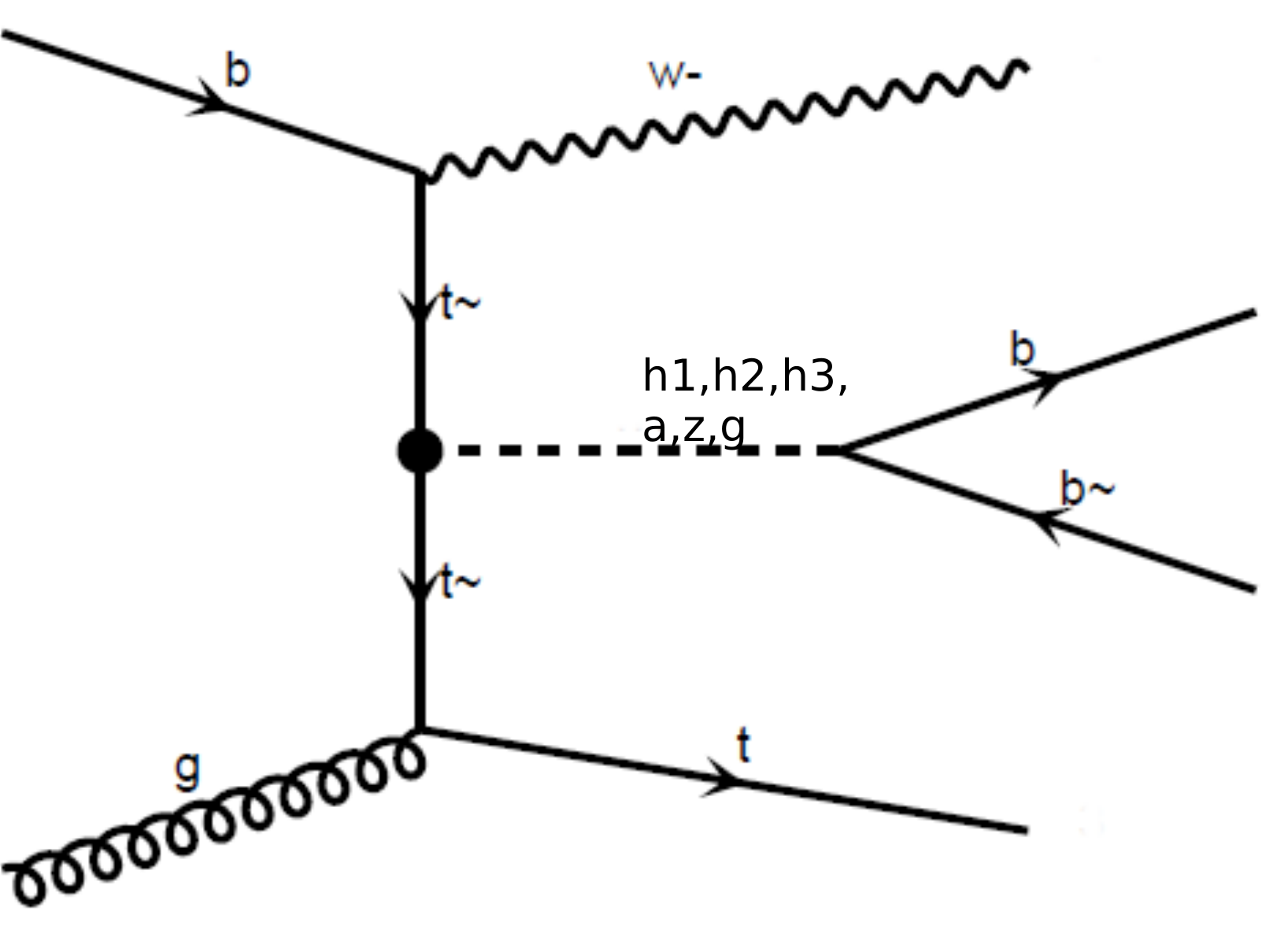} 
 \caption{}
 \end{subfigure} 
 \begin{subfigure}[t]{0.3\textwidth}
 \centering
 \includegraphics[scale=0.325]{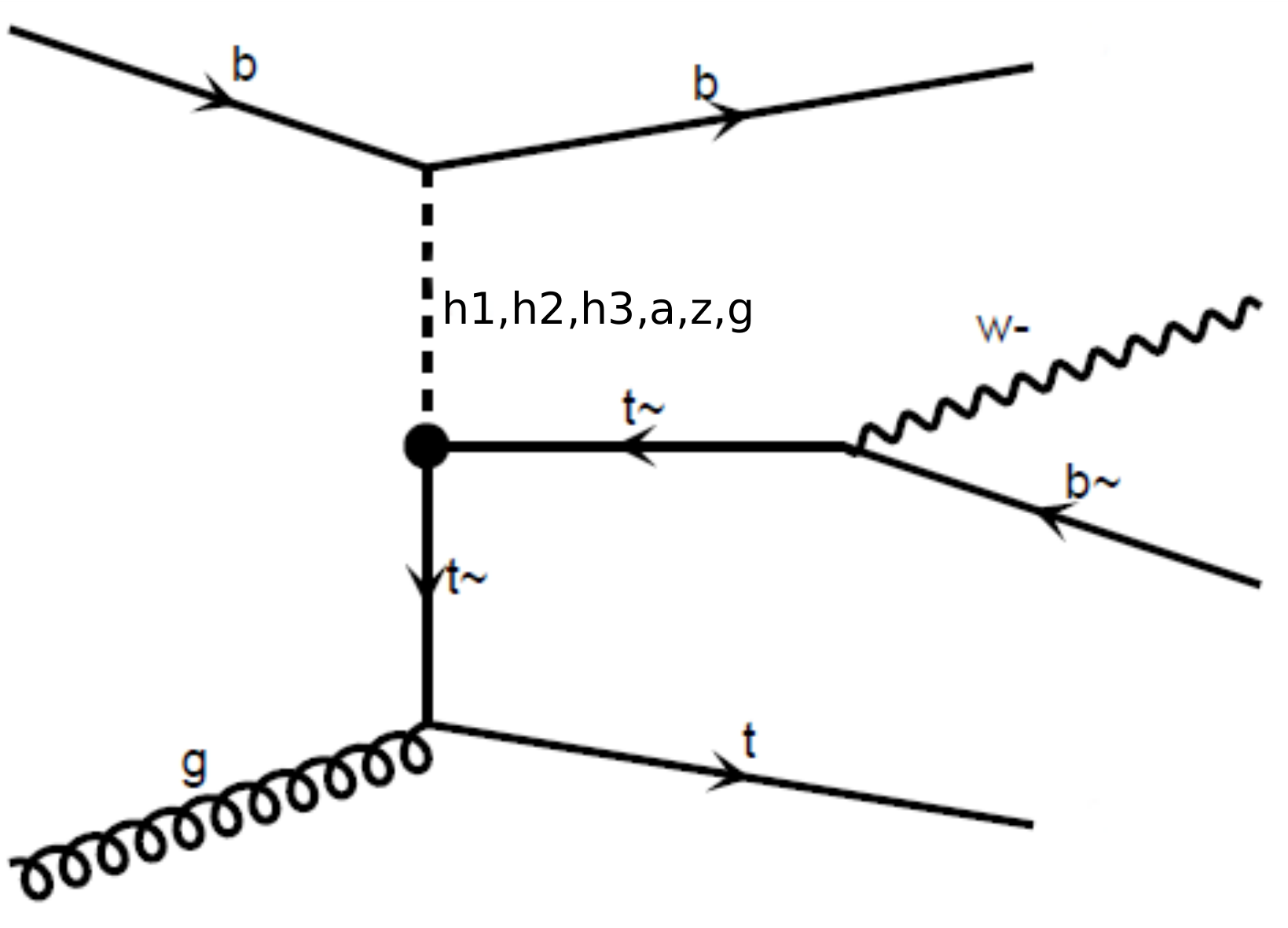} 
 \caption{}
 \end{subfigure}
 \begin{subfigure}[t]{0.3\textwidth}
 \centering
 \includegraphics[scale=0.325]{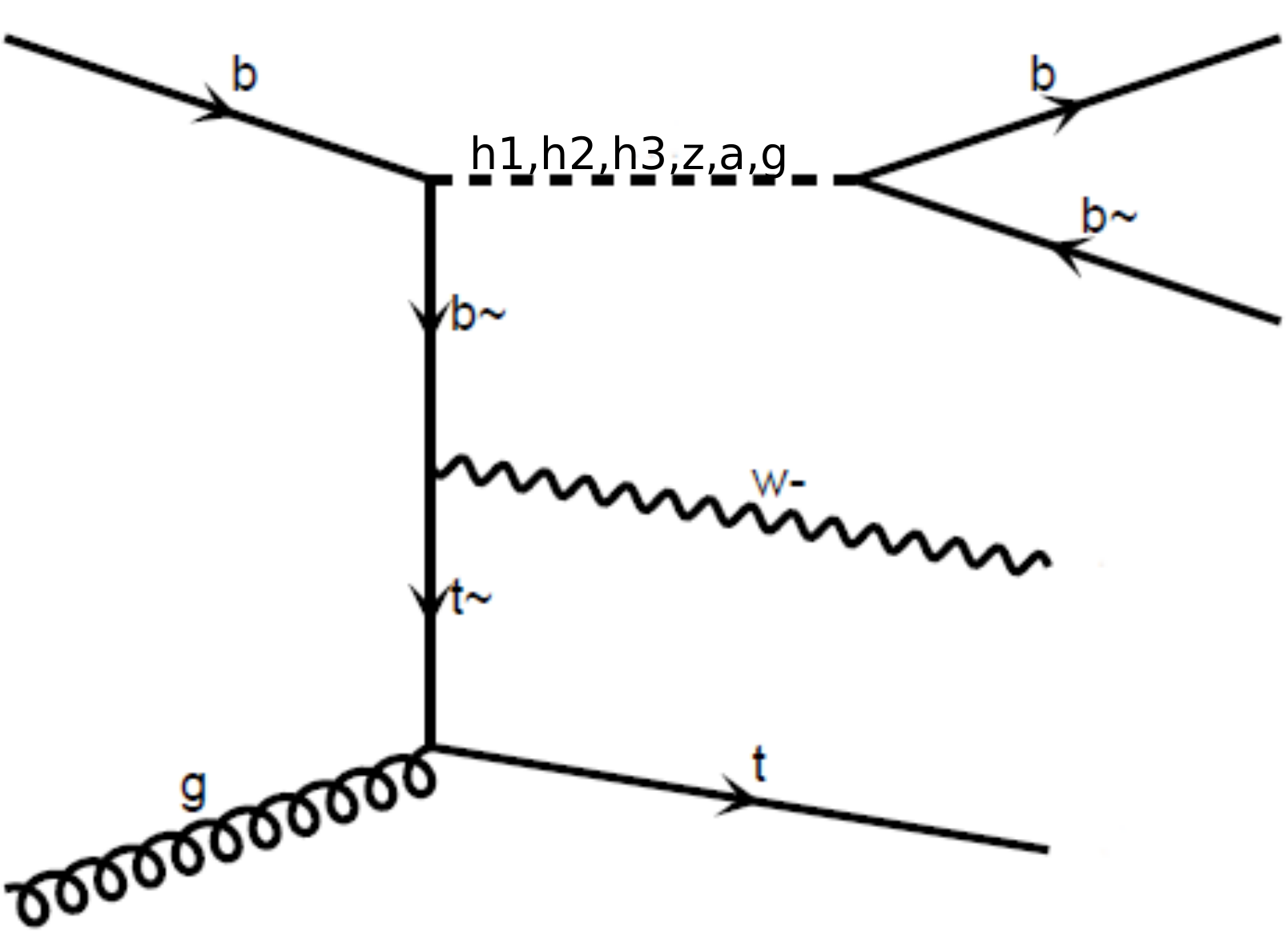} 
 \caption{}
 \end{subfigure}
 \begin{subfigure}[t]{0.3\textwidth}
 \centering
 \includegraphics[scale=0.325]{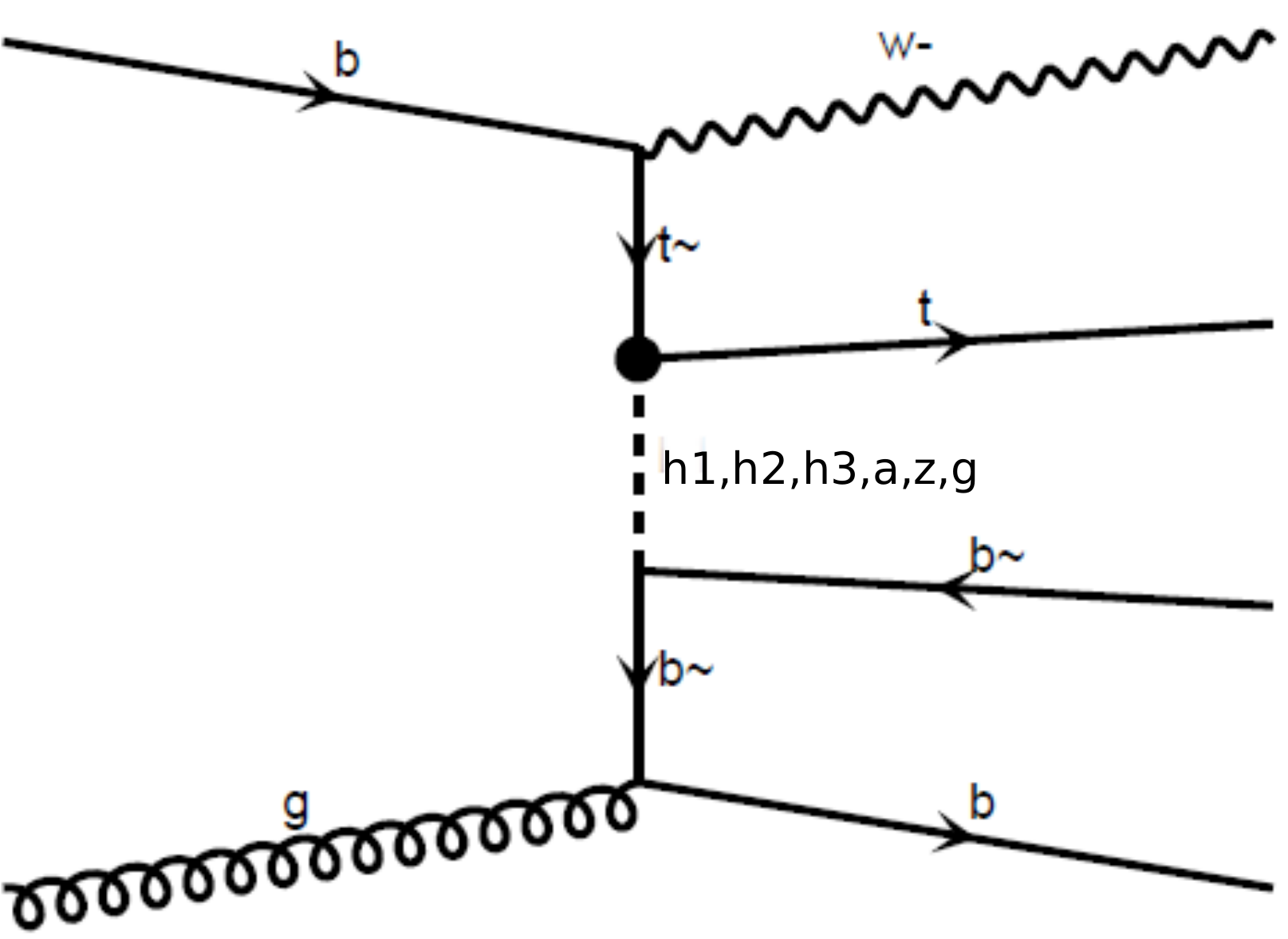} 
 \caption{}
 \end{subfigure}
 \begin{subfigure}[t]{0.3\textwidth}
 \centering
 \includegraphics[scale=0.325]{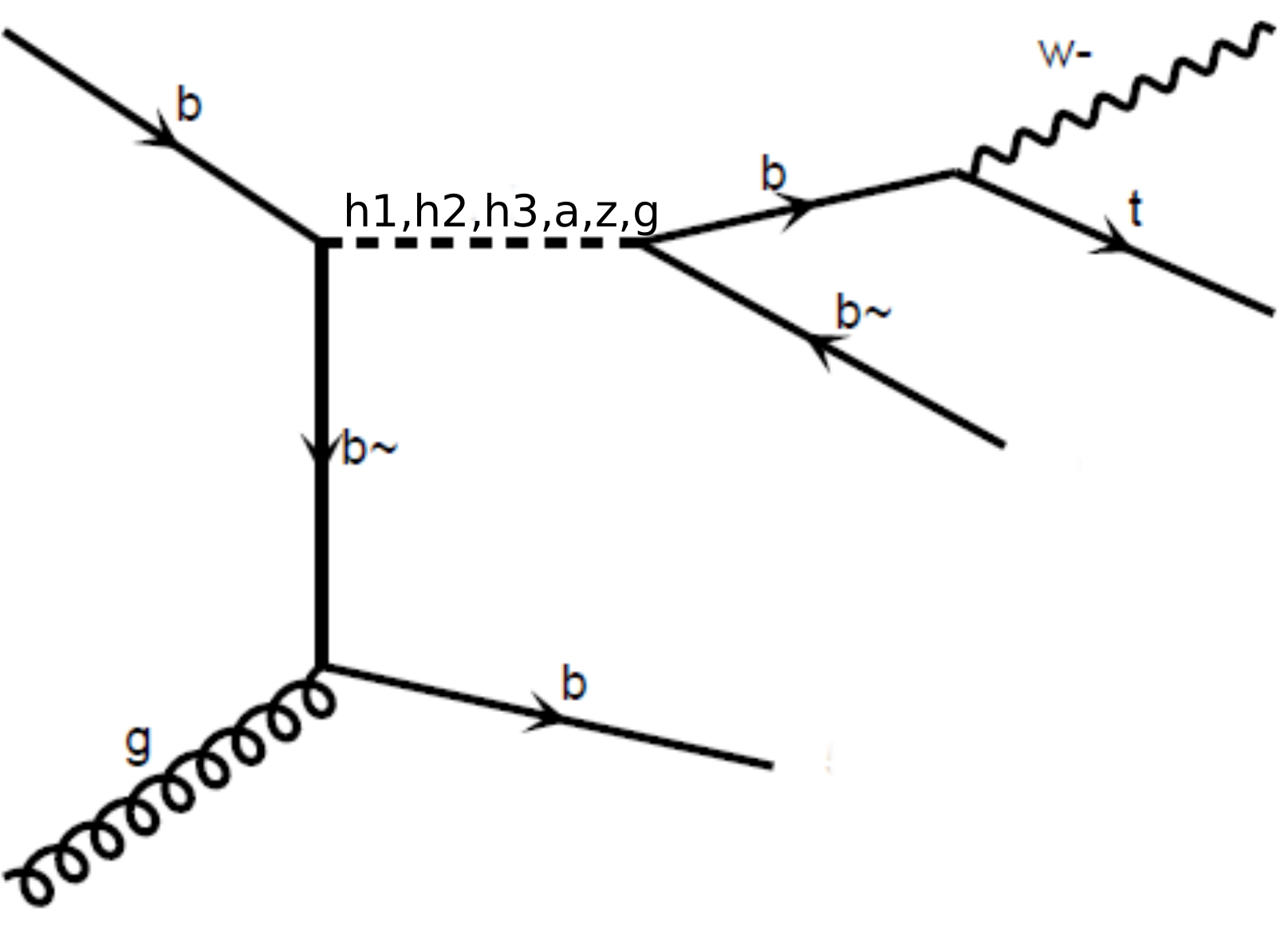} 
 \caption{}
 \end{subfigure}
 \begin{subfigure}[t]{0.3\textwidth}
 \centering
 \includegraphics[scale=0.325]{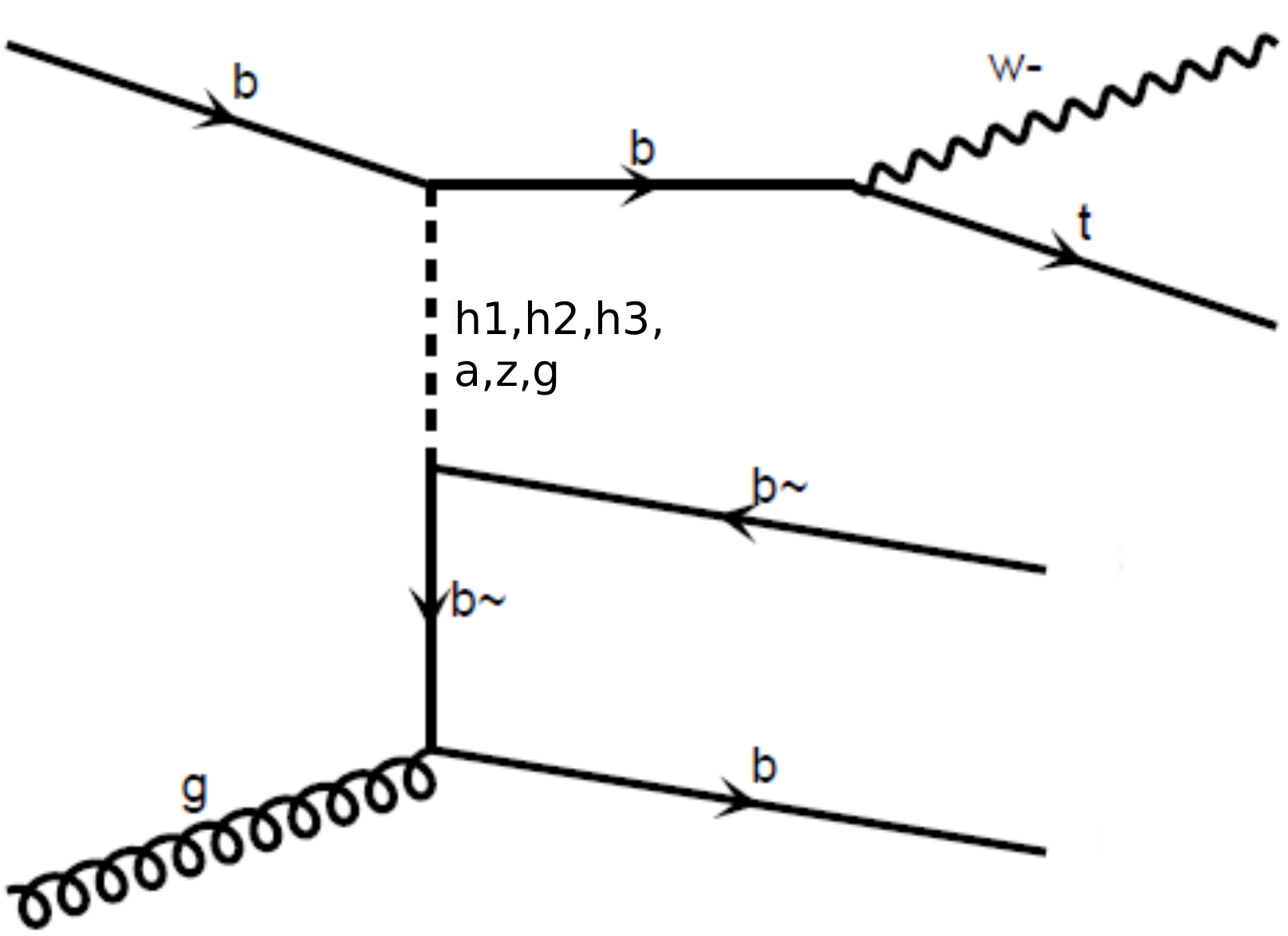} 
 \caption{}
 \end{subfigure}
 \begin{subfigure}[t]{0.3\textwidth}
 \centering
 \includegraphics[scale=0.325]{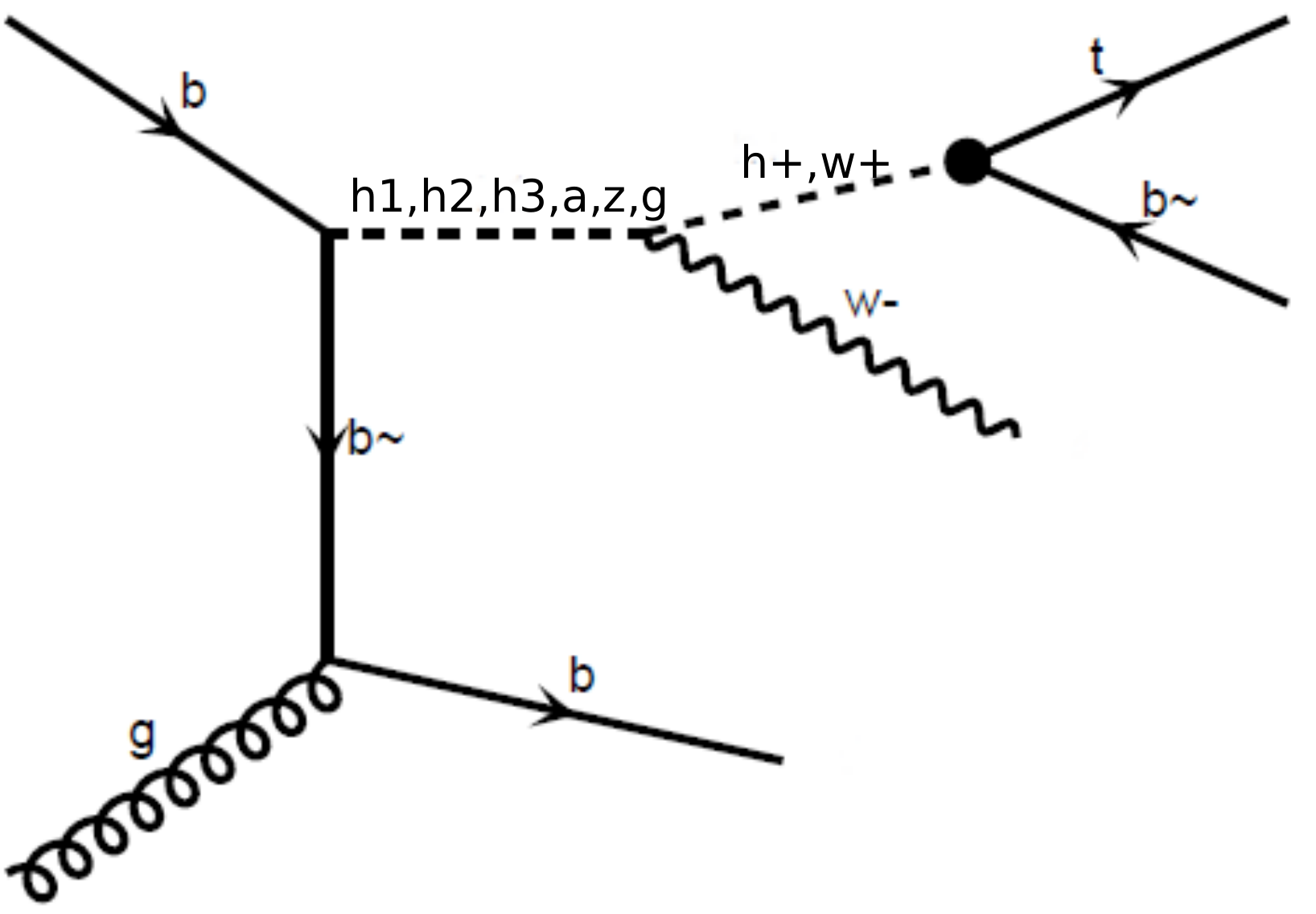} 
 \caption{}
 \end{subfigure}
 \begin{subfigure}[t]{0.3\textwidth}
 \centering
 \includegraphics[scale=0.325]{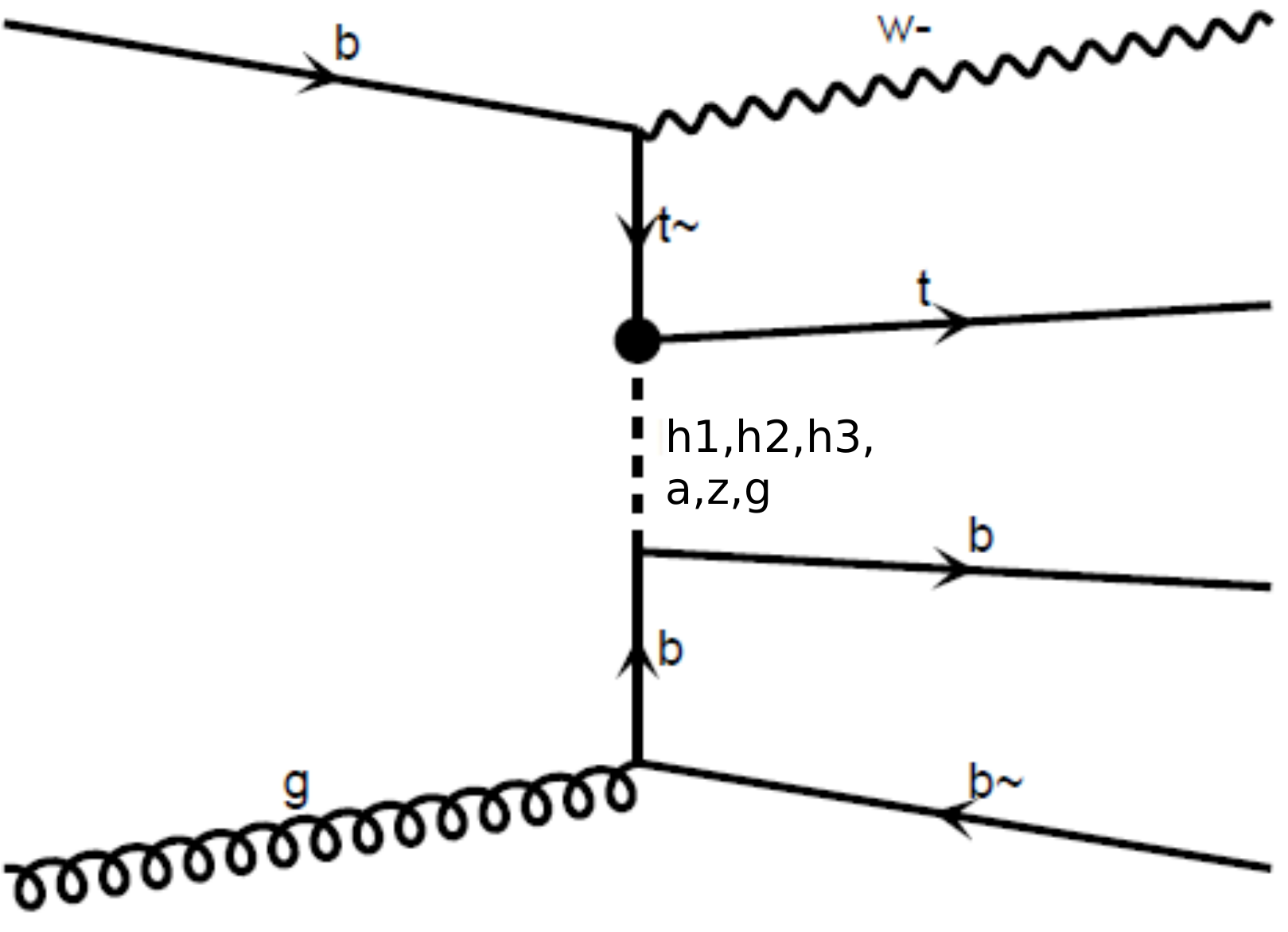} 
 \caption{}
 \end{subfigure} 
 \begin{subfigure}[t]{0.3\textwidth}
 \centering
 \includegraphics[scale=0.325]{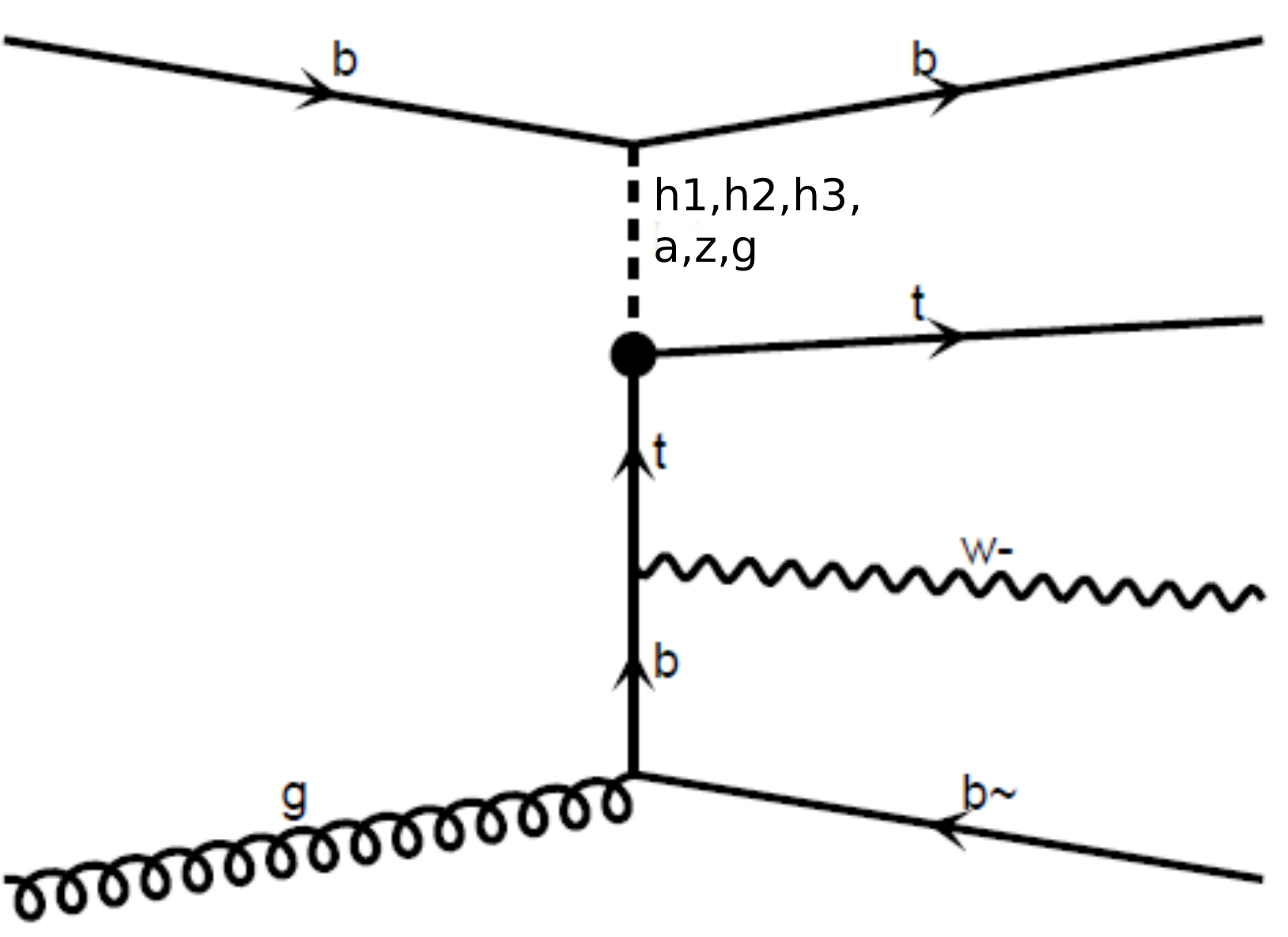} 
 \caption{}
 \end{subfigure}  
 \begin{subfigure}[t]{0.3\textwidth}
 \centering
 \includegraphics[scale=0.325]{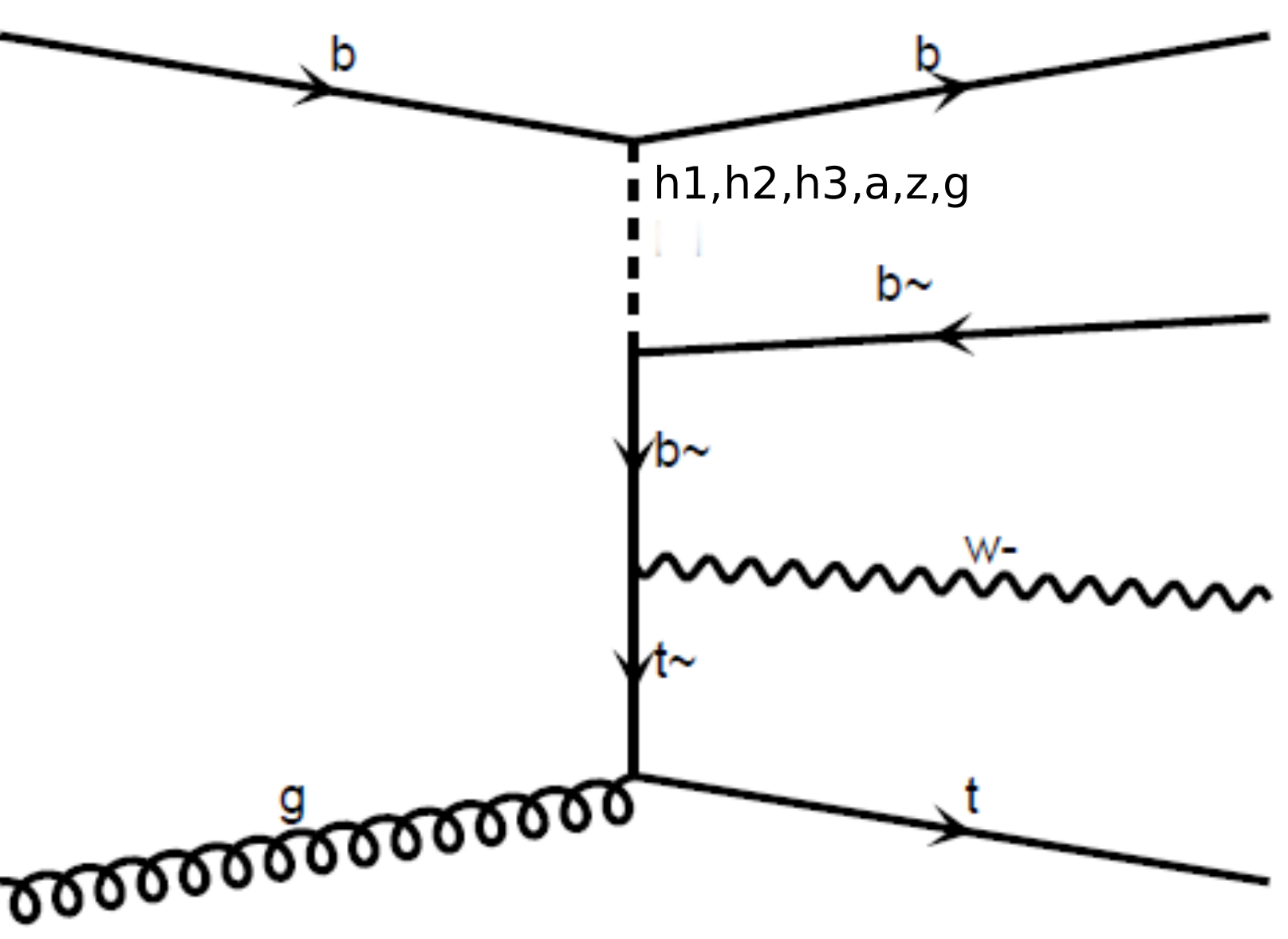} 
 \caption{}
 \end{subfigure}
 \begin{subfigure}[t]{0.3\textwidth}
 \centering
 \includegraphics[scale=0.325]{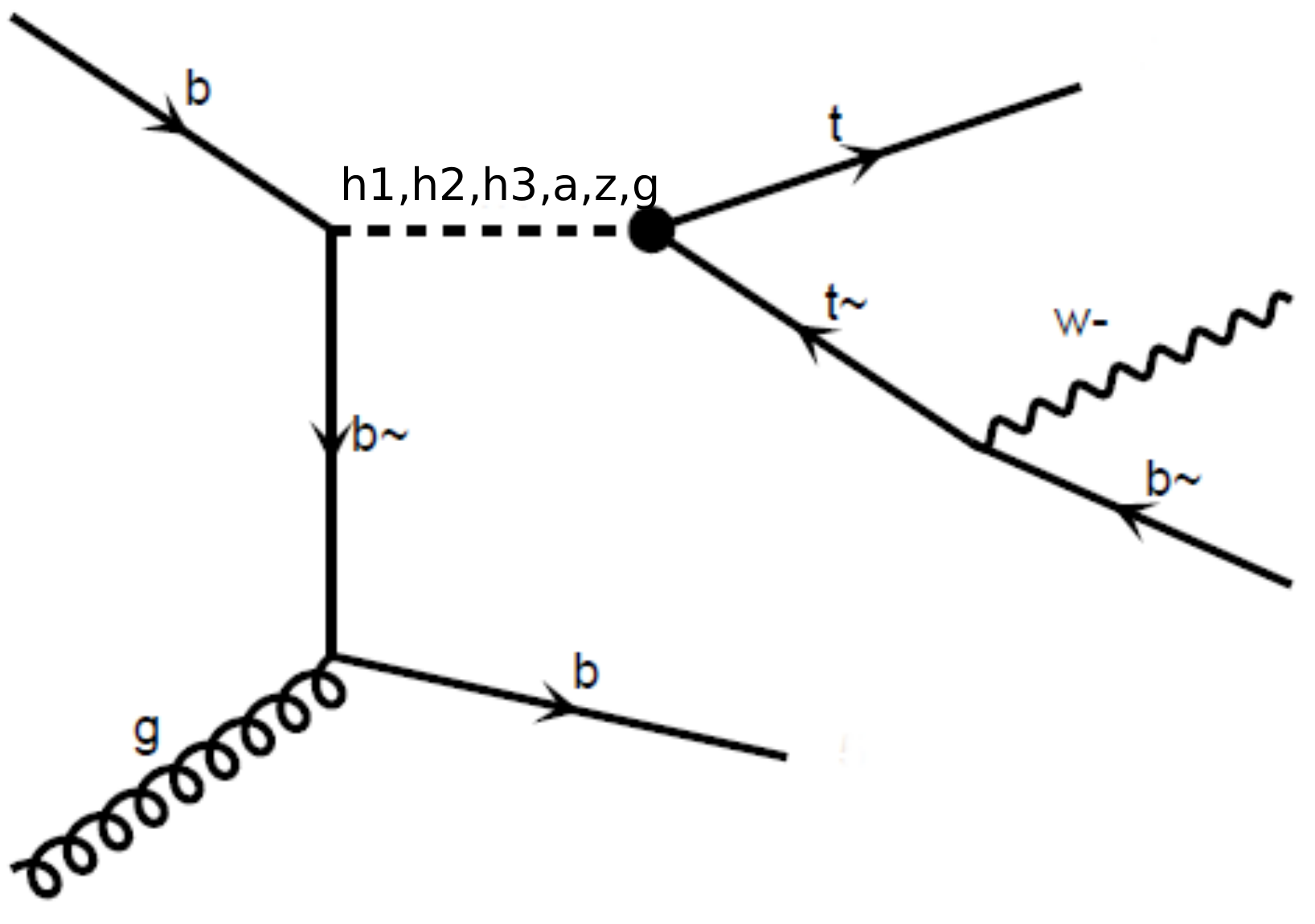} 
 \caption{}
 \end{subfigure} 
 \caption{\label{figs:bckg_a} Non-resonant Feynman diagrams contributing to the background for the process $pp\to tW^- b \bar b$ with h- $\equiv H^-$, h1 $\equiv h$, h2 $\equiv H$, h3 $\equiv A$ and a $\equiv \gamma$
 (as appropriate).}
\end{figure}

\begin{figure}
 \begin{subfigure}[t]{0.3\textwidth}
 \centering
 \includegraphics[scale=0.325]{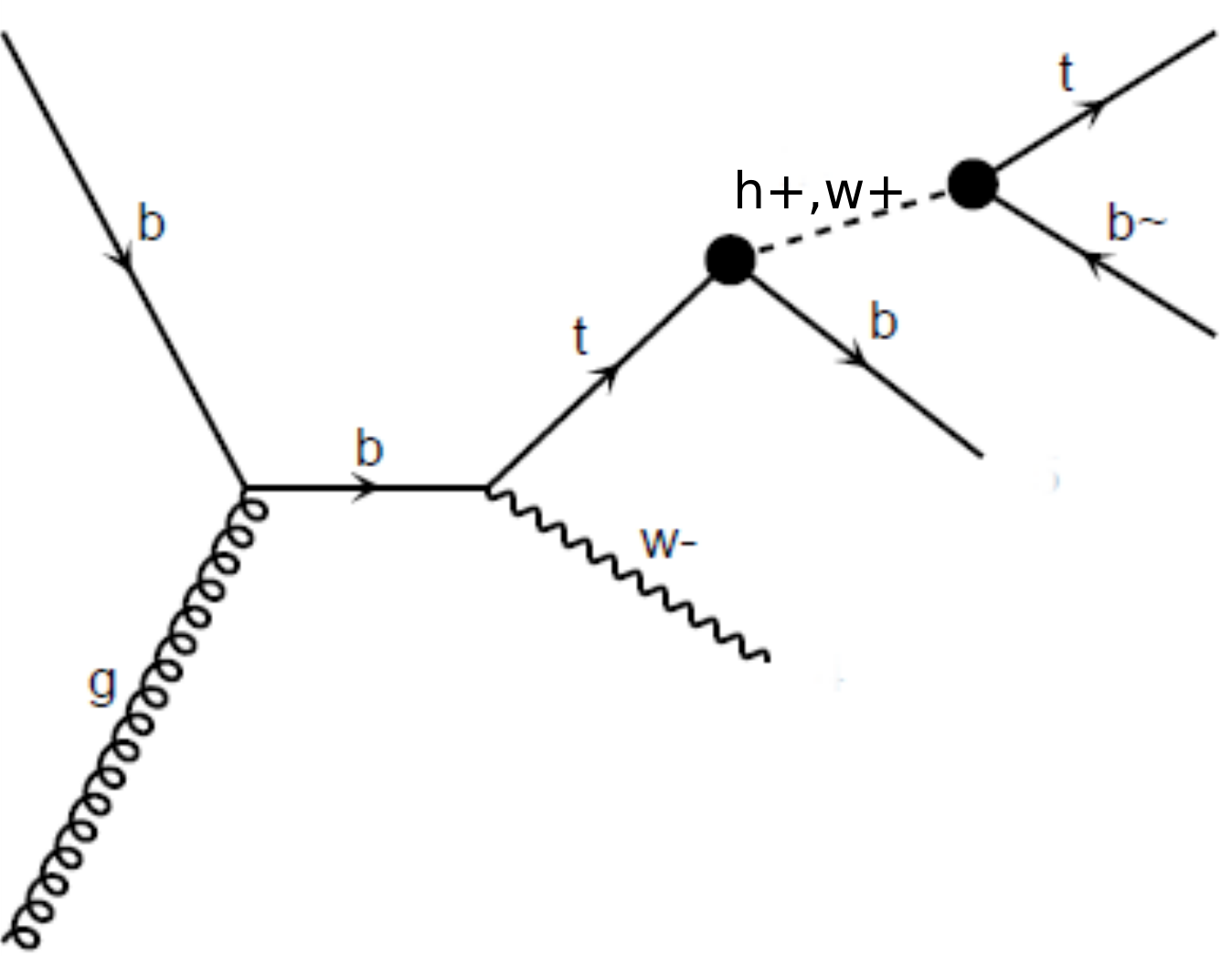} 
 \caption{}
 \end{subfigure}
 \begin{subfigure}[t]{0.3\textwidth}
 \centering
 \includegraphics[scale=0.325]{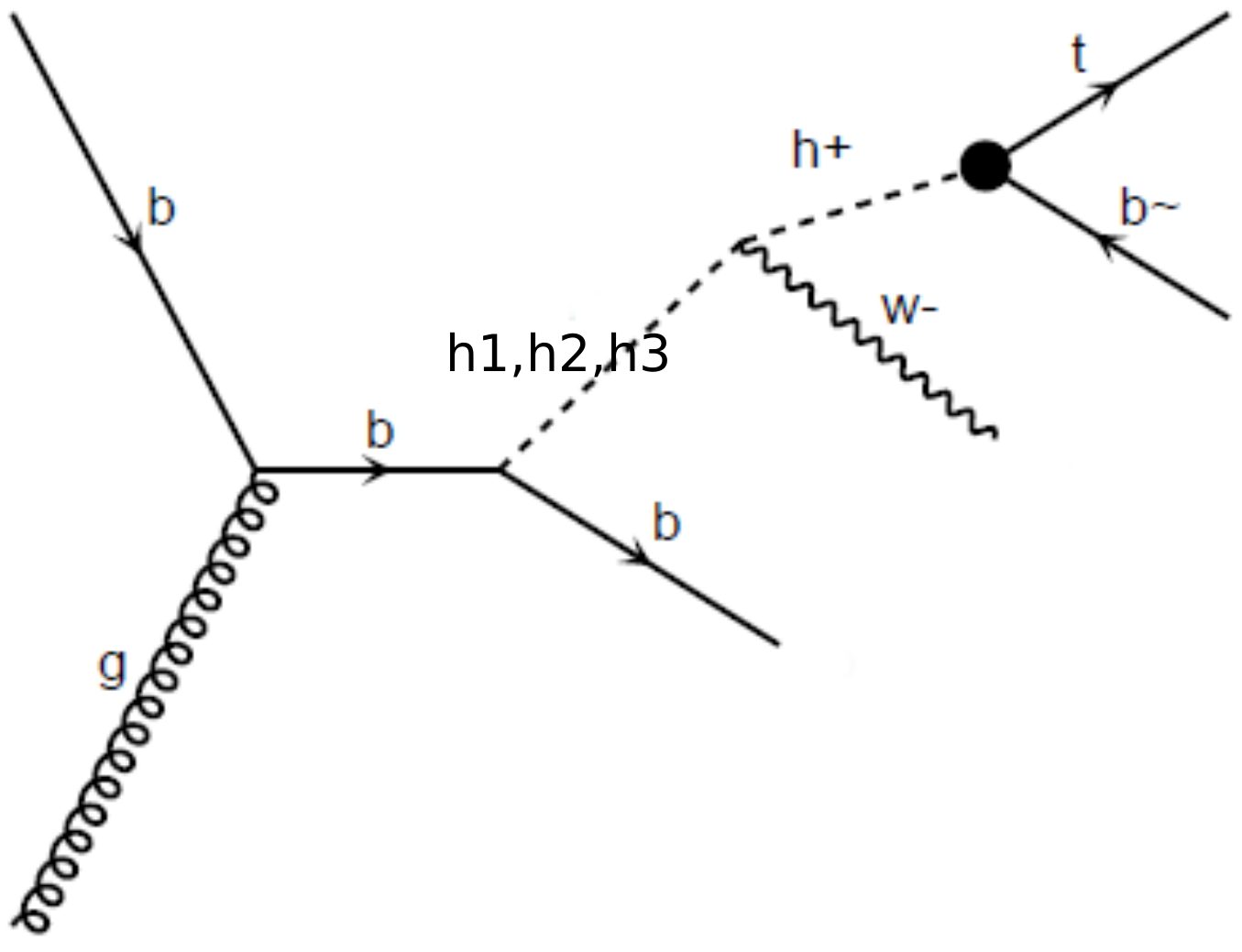} 
 \caption{}
 \end{subfigure}
 \begin{subfigure}[t]{0.3\textwidth}
 \centering
 \includegraphics[scale=0.325]{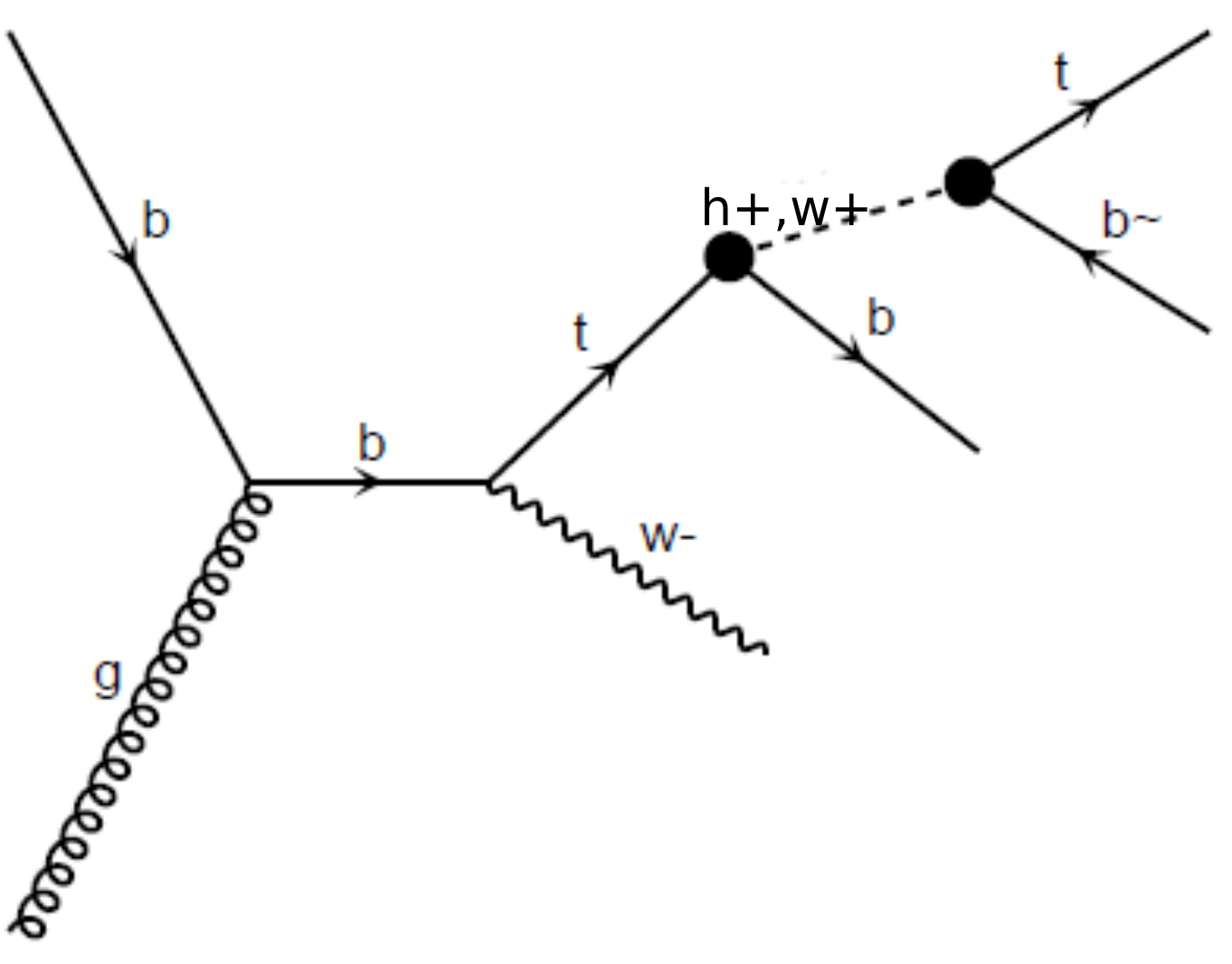} 
 \caption{}
 \end{subfigure}
 \begin{subfigure}[t]{0.3\textwidth}
 \centering
 \includegraphics[scale=0.325]{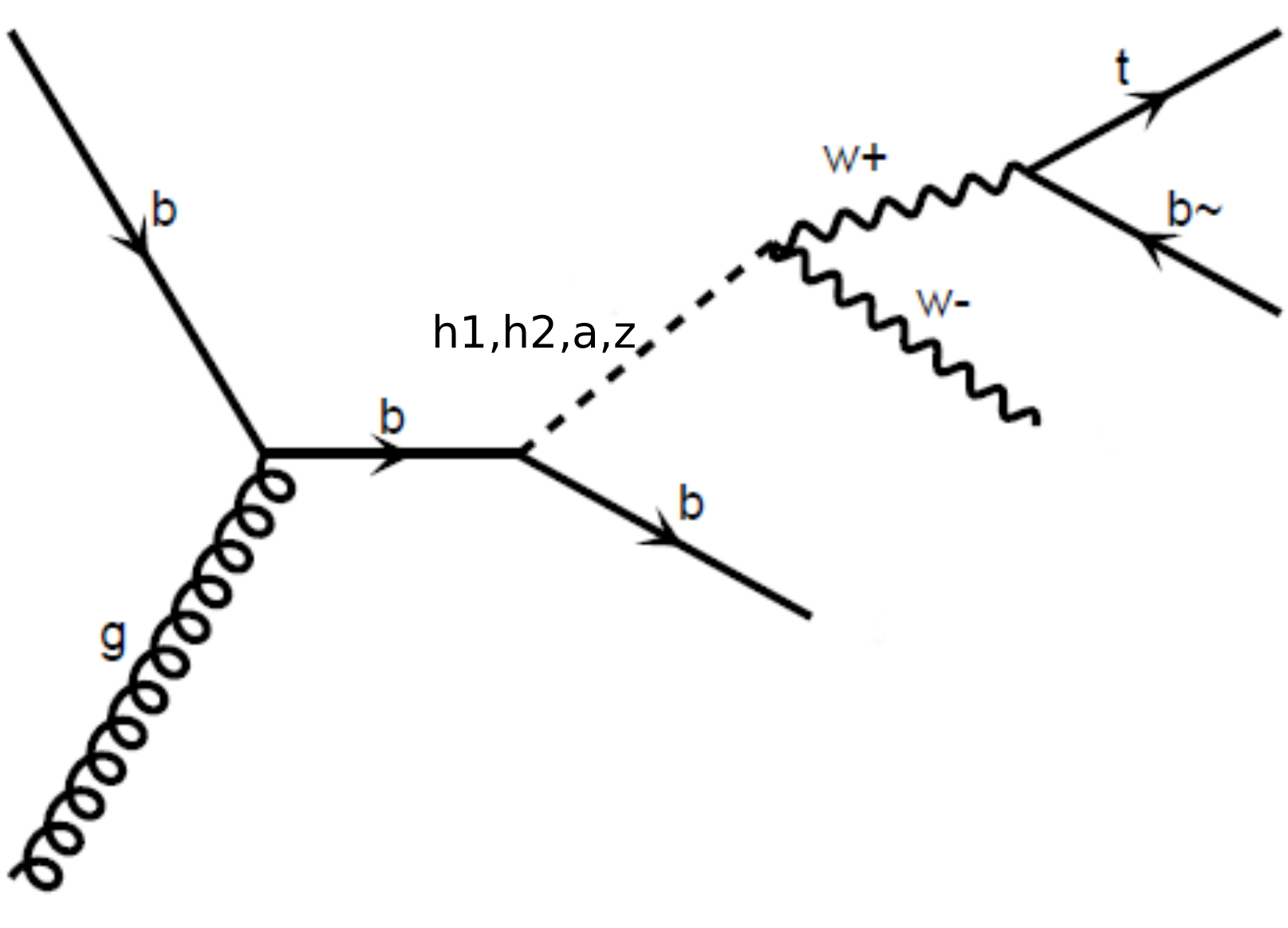} 
 \caption{}
 \end{subfigure}
 \begin{subfigure}[t]{0.3\textwidth}
 \centering
 \includegraphics[scale=0.325]{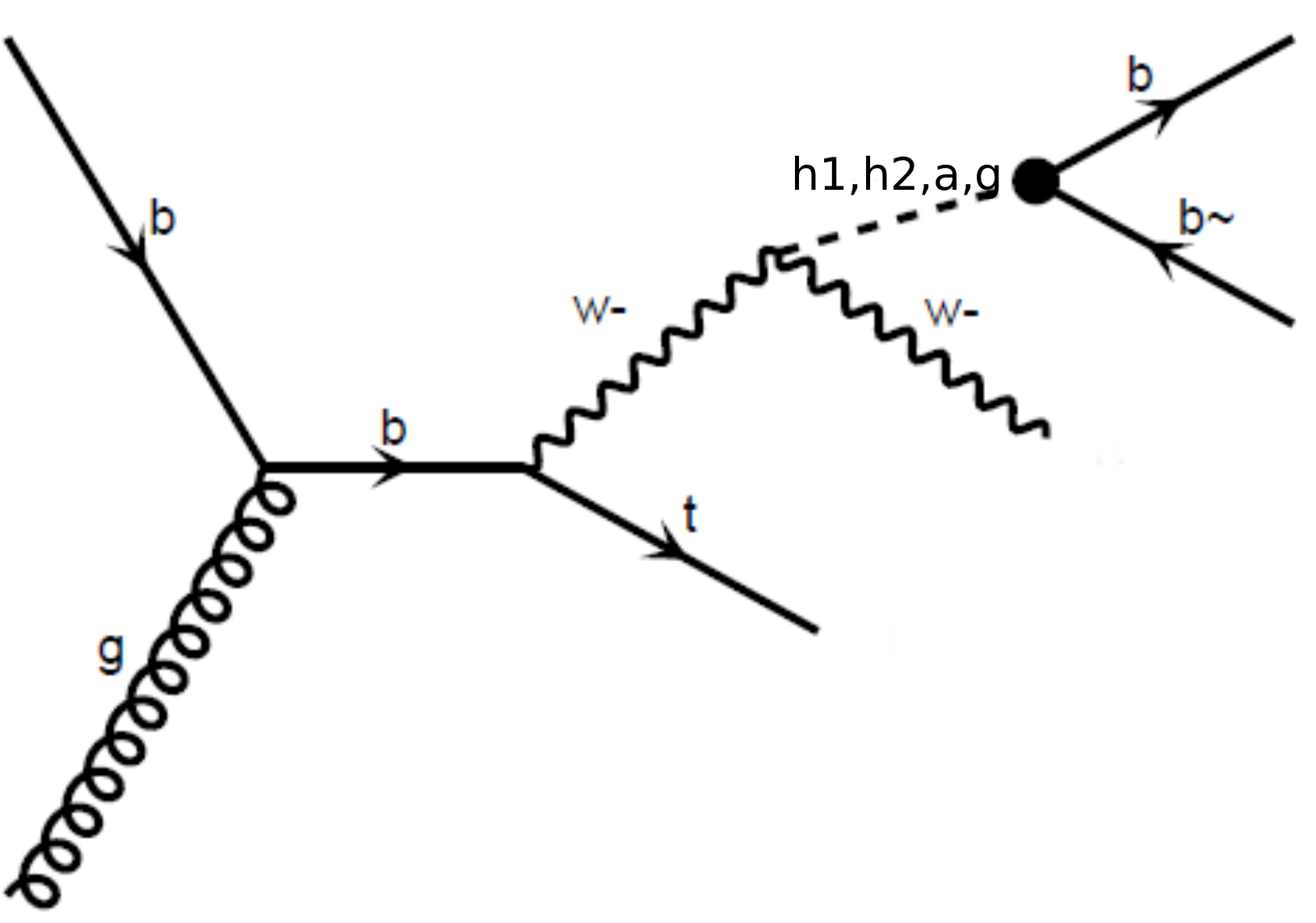} 
 \caption{}
 \end{subfigure}
 \begin{subfigure}[t]{0.3\textwidth}
 \centering
 \includegraphics[scale=0.325]{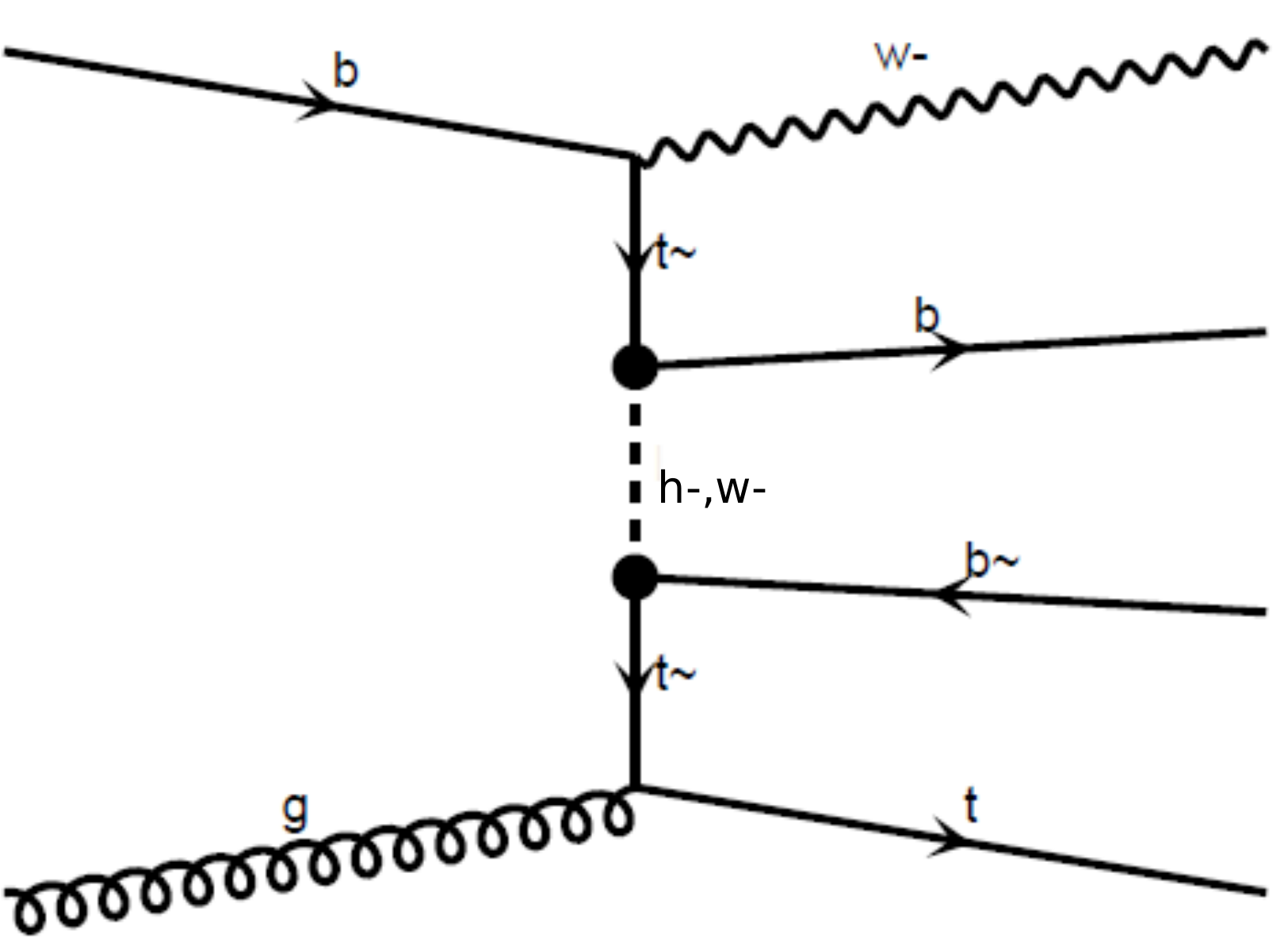} 
 \caption{}
 \end{subfigure}
 \begin{subfigure}[t]{0.3\textwidth}
 \centering
 \includegraphics[scale=0.325]{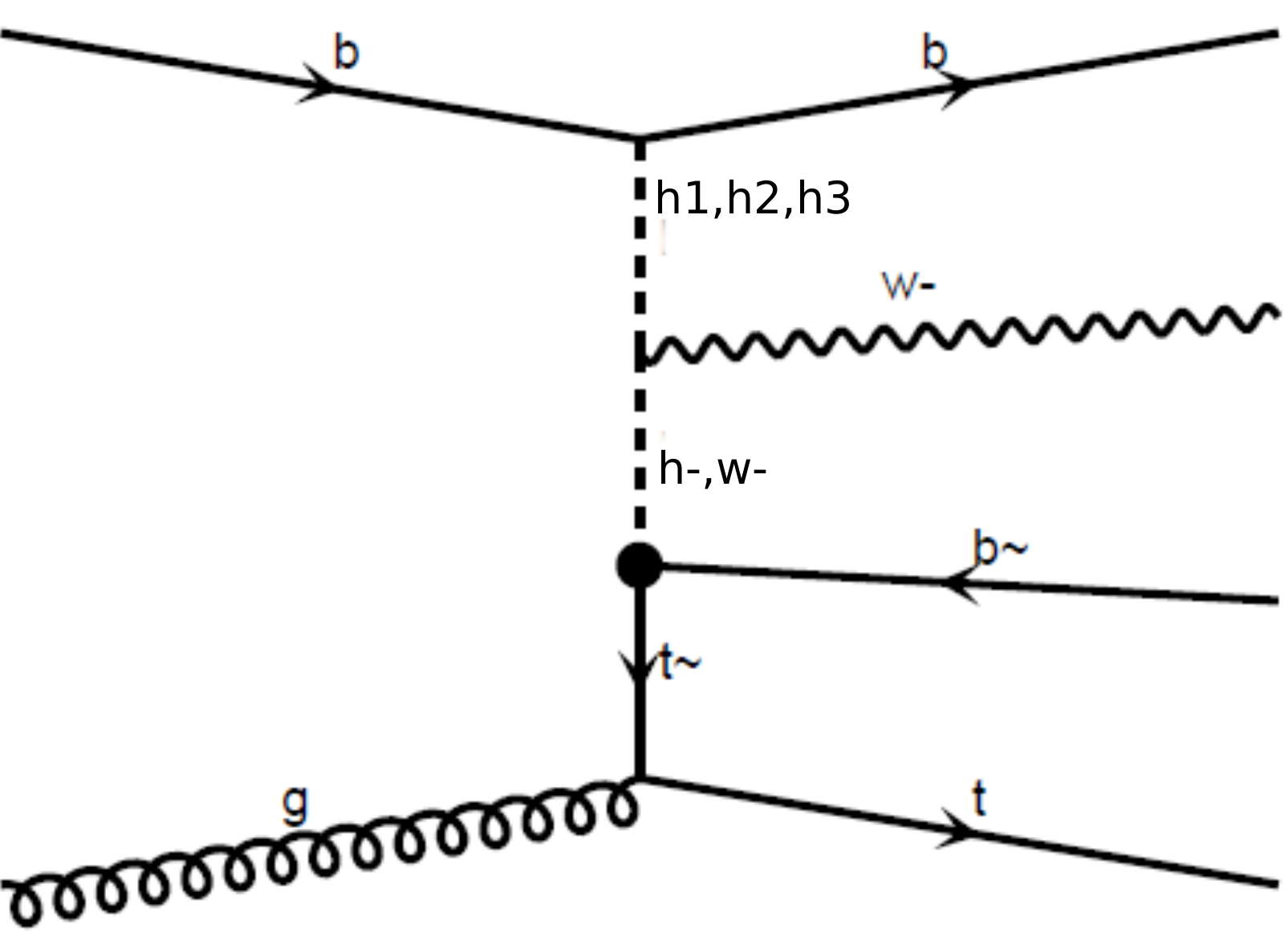} 
 \caption{}
 \end{subfigure}
 \begin{subfigure}[t]{0.3\textwidth}
 \centering
 \includegraphics[scale=0.325]{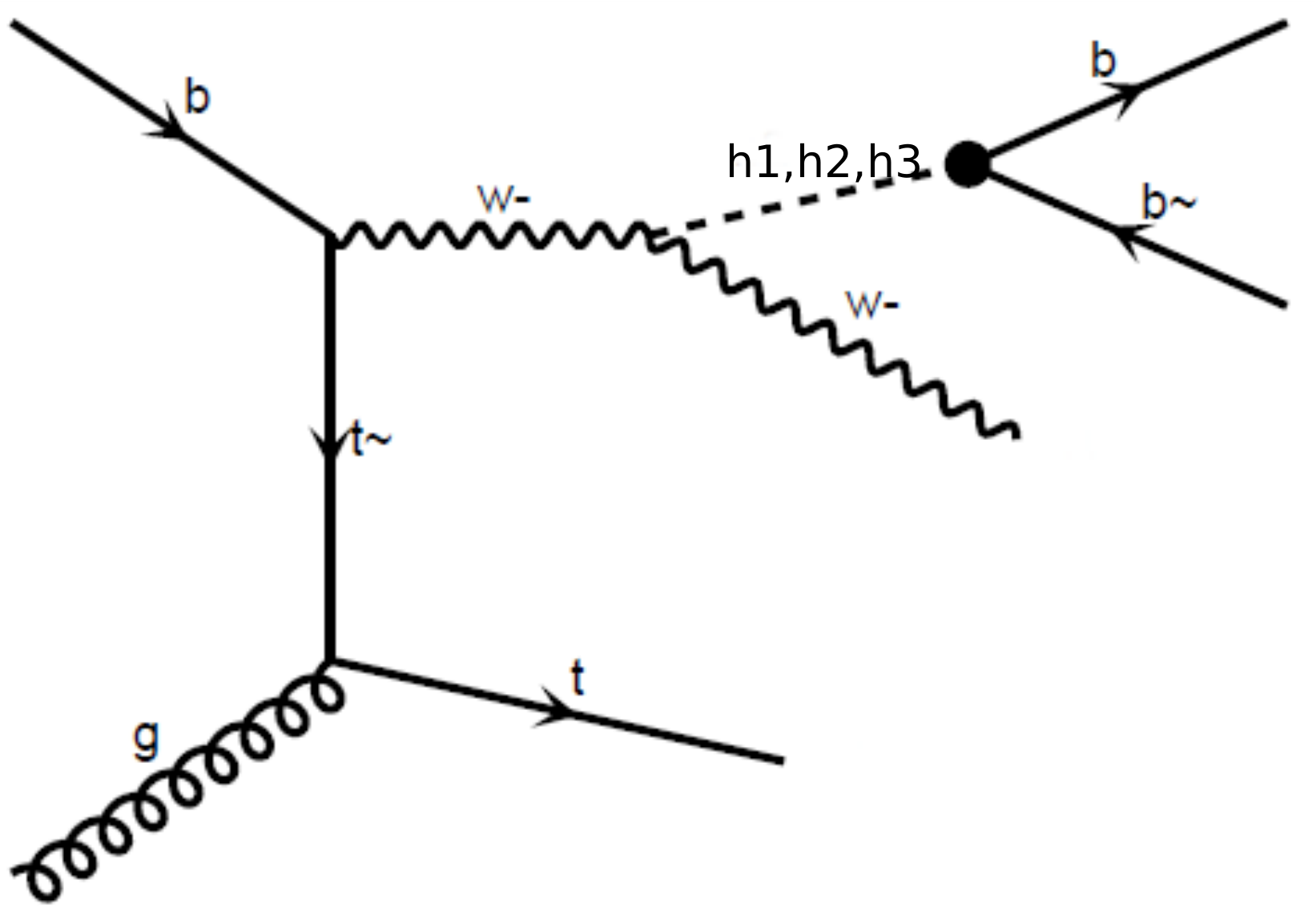} 
 \caption{}
 \end{subfigure}
 \begin{subfigure}[t]{0.3\textwidth}
 \centering
 \includegraphics[scale=0.325]{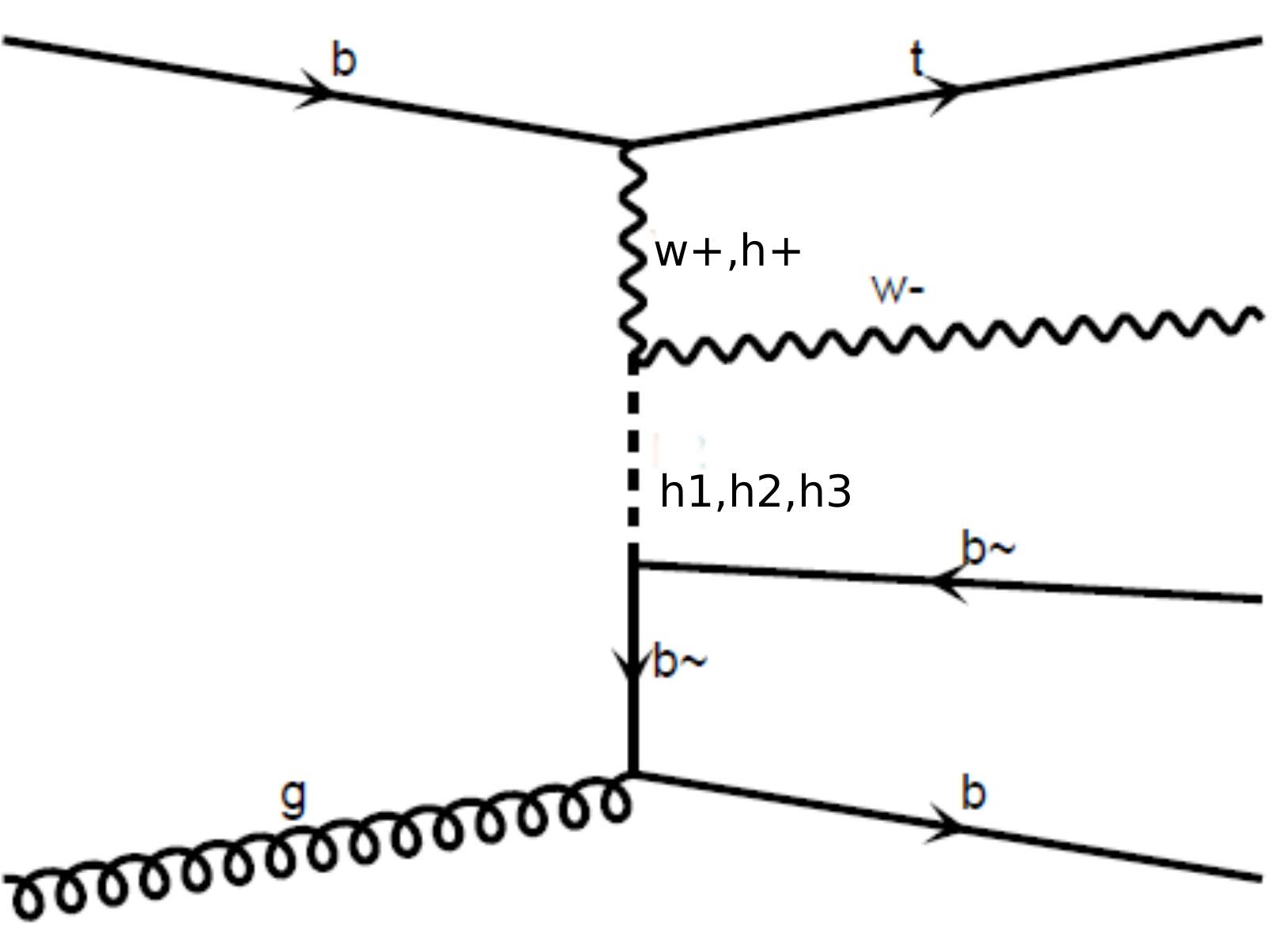} 
 \caption{}
 \end{subfigure}
 \begin{subfigure}[t]{0.3\textwidth}
 \centering
 \includegraphics[scale=0.325]{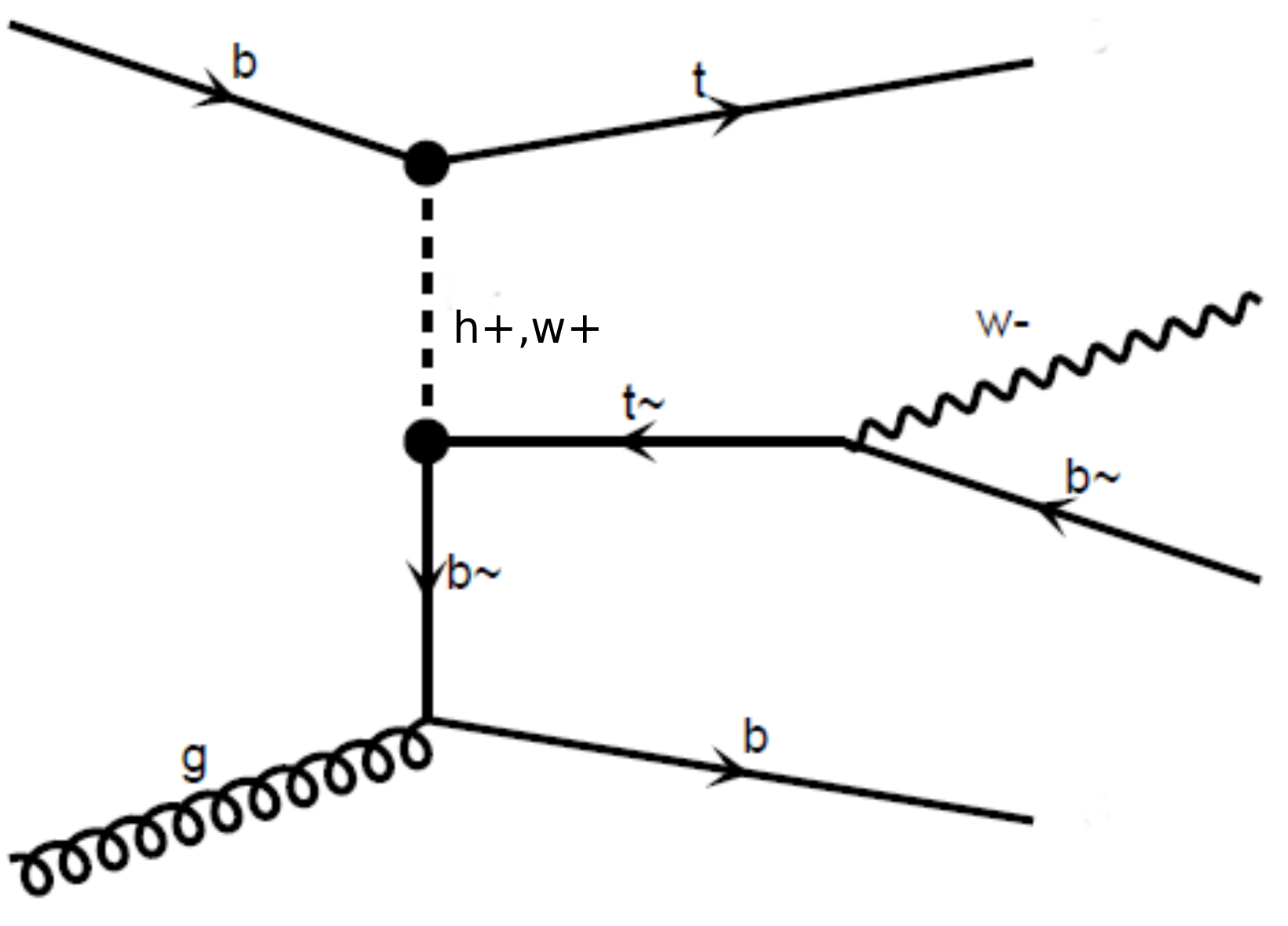} 
 \caption{}
 \end{subfigure}
 \begin{subfigure}[t]{0.3\textwidth}
 \centering
 \includegraphics[scale=0.325]{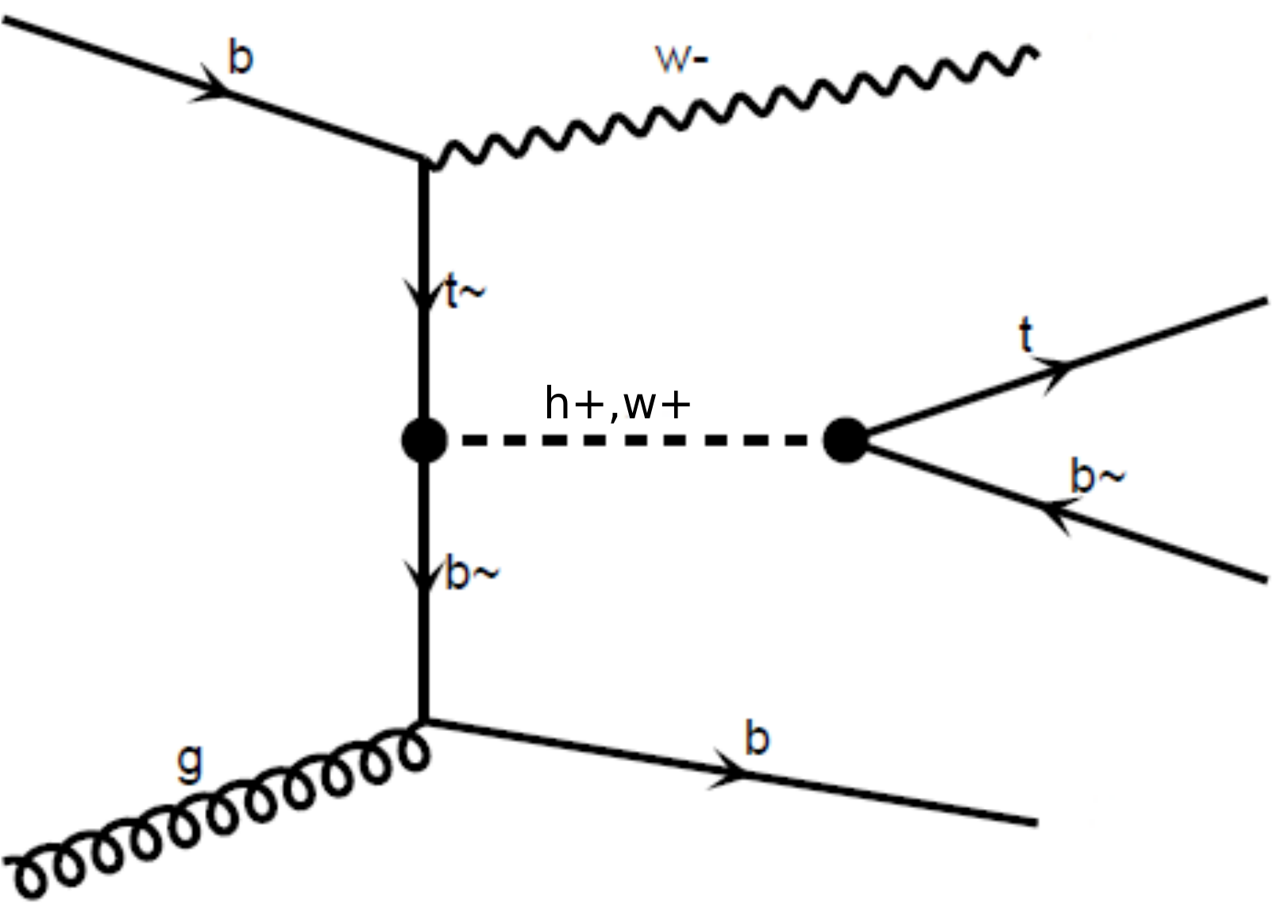} 
 \caption{}
 \end{subfigure}
 \begin{subfigure}[t]{0.3\textwidth}
 \centering
 \includegraphics[scale=0.325]{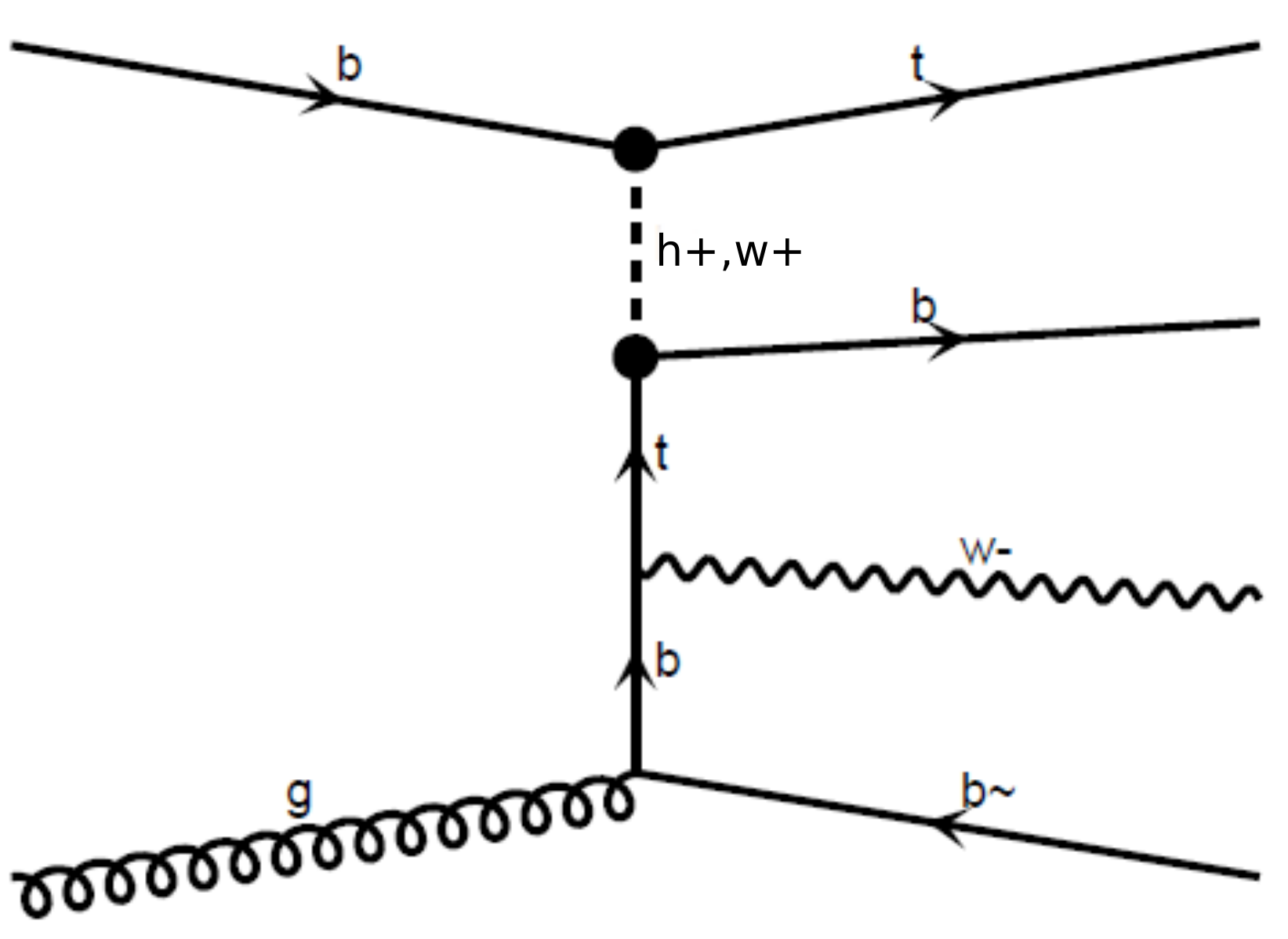} 
 \caption{}
 \end{subfigure}
 \begin{subfigure}[t]{0.3\textwidth}
 \centering
 \includegraphics[scale=0.325]{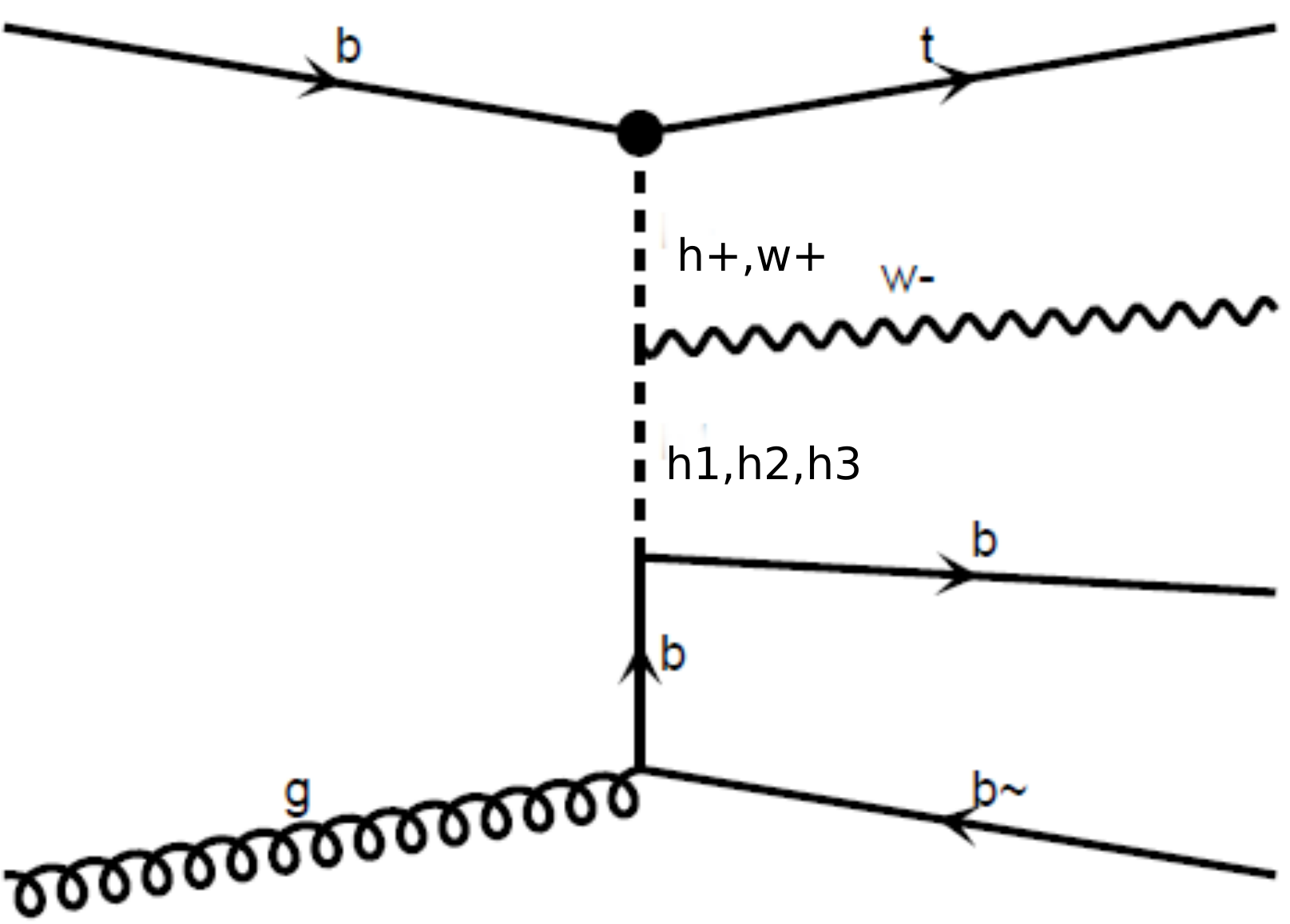} 
 \caption{}
 \end{subfigure}
 \begin{subfigure}[t]{0.3\textwidth}
 \centering
 \includegraphics[scale=0.325]{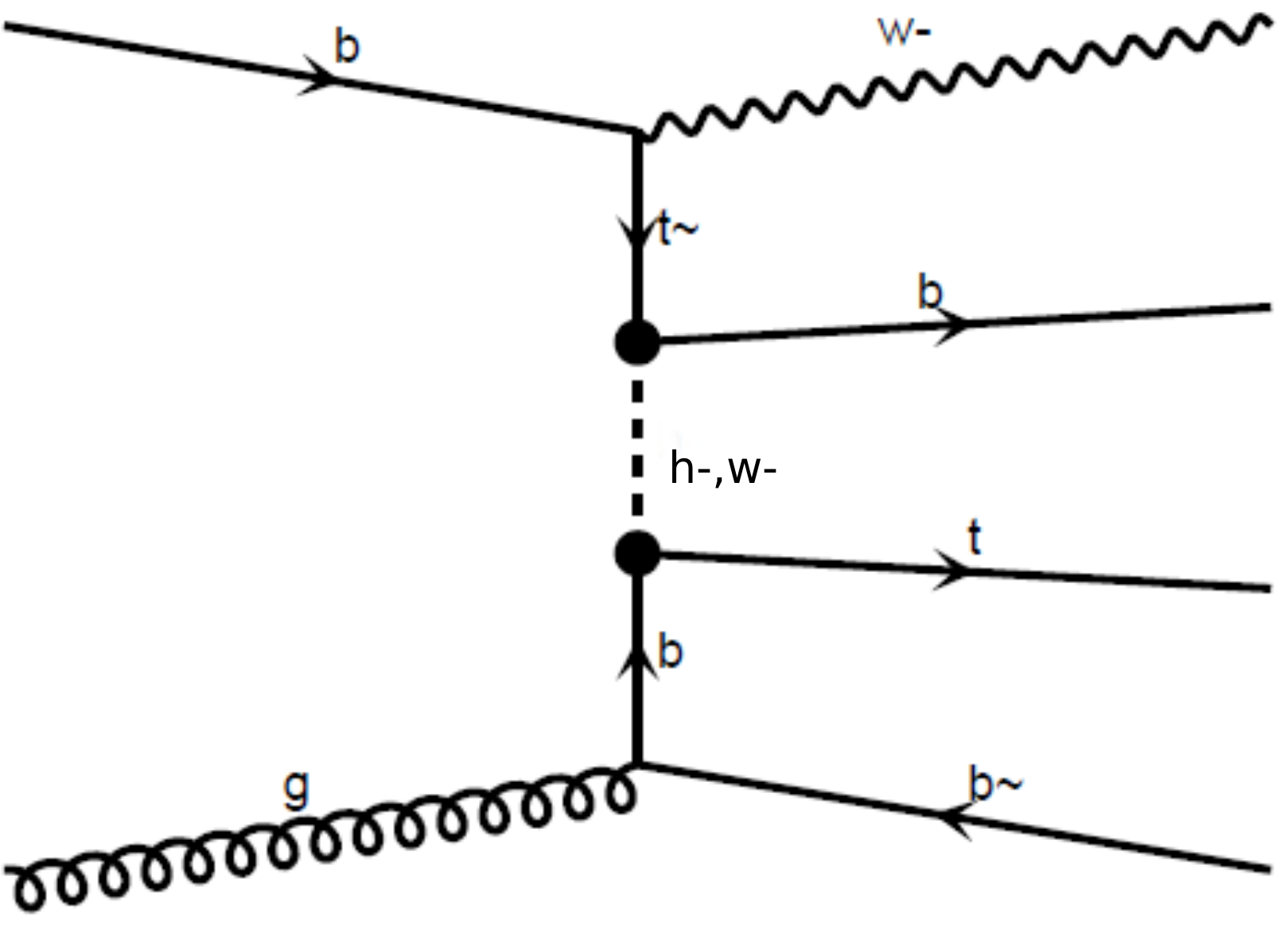} 
 \caption{}
 \end{subfigure}
 \begin{subfigure}[t]{0.3\textwidth}
 \centering
 \includegraphics[scale=0.325]{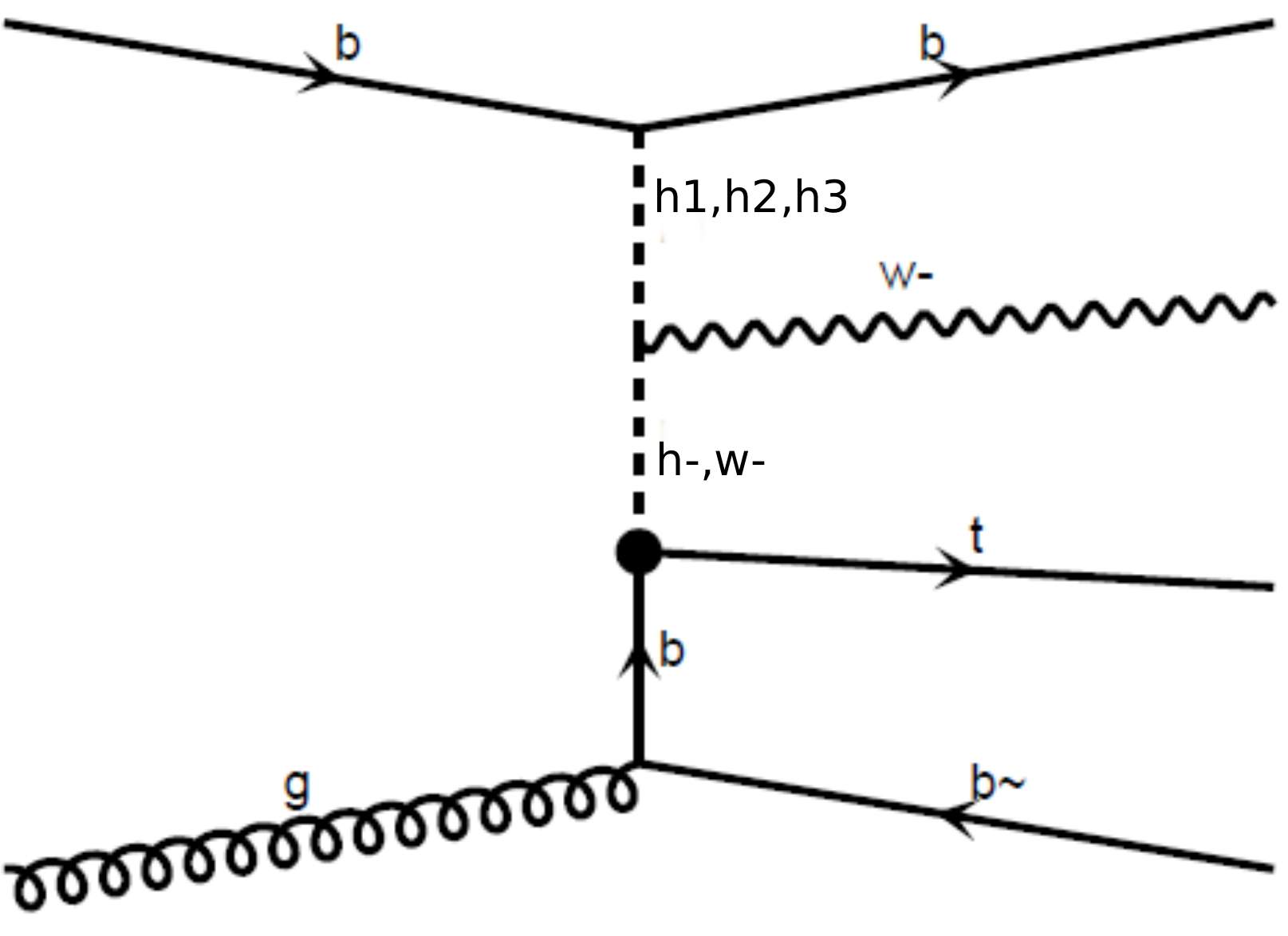} 
 \caption{}
 \end{subfigure}
\caption{\label{figs:bckg_b} Non-resonant Feynman diagrams contributing to the background for the process $pp\to tW^- b \bar b$ with h- $\equiv H^-$, h1 $\equiv h$, h2 $\equiv H$, h3 $\equiv A$ and a$\equiv \gamma$
 (as appropriate).}
\end{figure}

\begin{figure}
 \begin{subfigure}[t]{0.3\textwidth}
 \centering
 \includegraphics[scale=0.325]{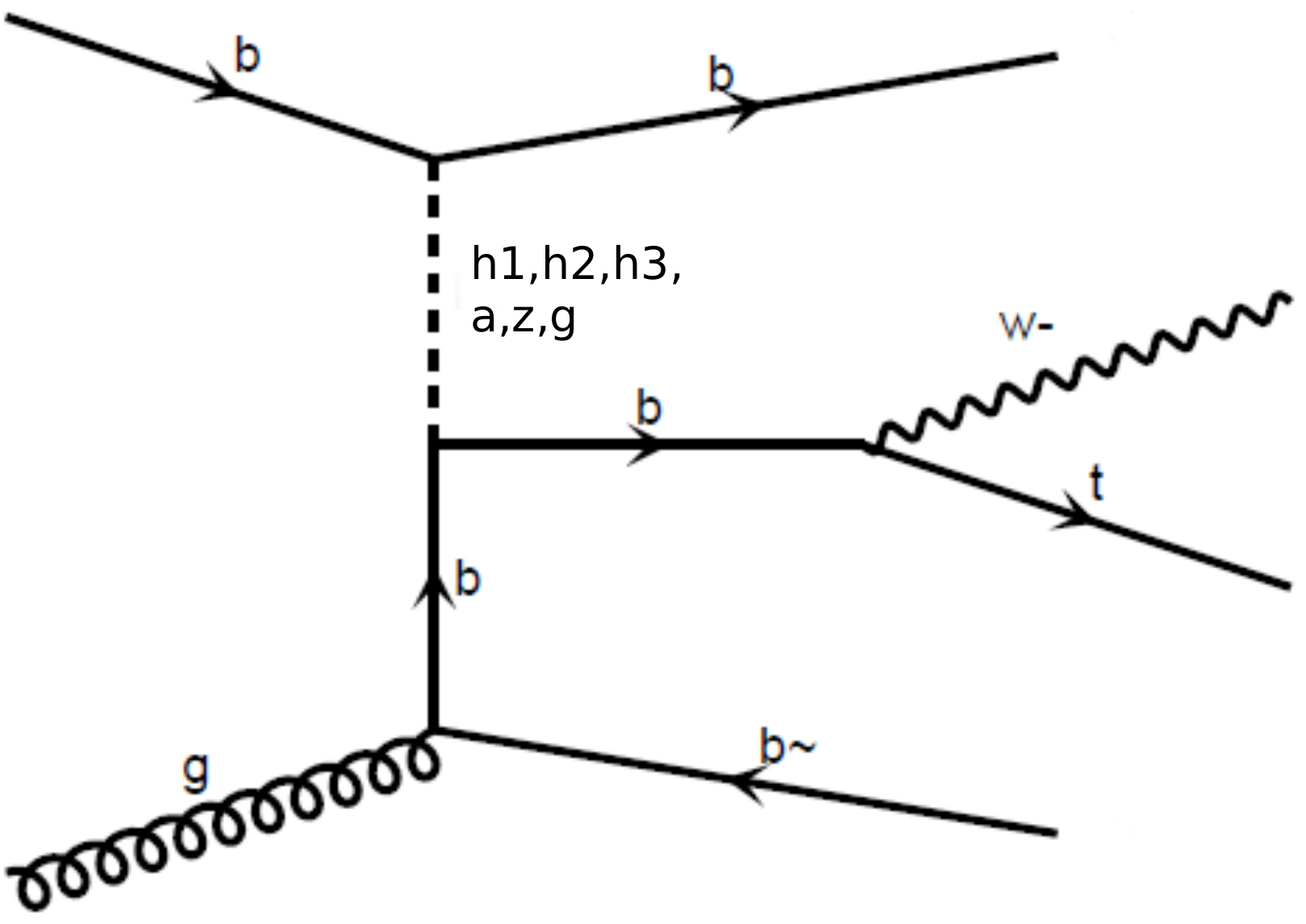} 
 \caption{}
 \end{subfigure}
 \begin{subfigure}[t]{0.3\textwidth}
 \centering
 \includegraphics[scale=0.325]{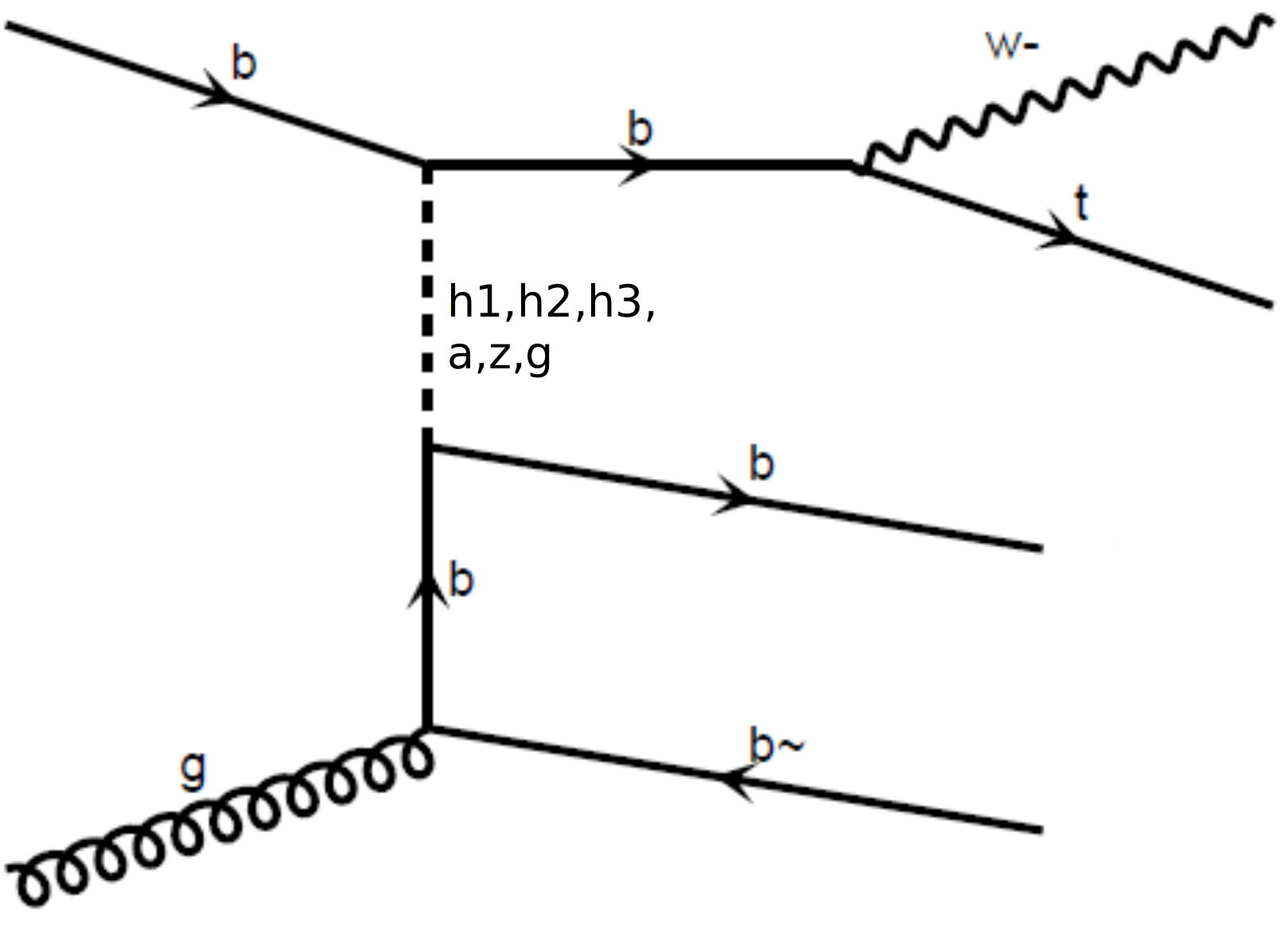} 
 \caption{}
 \end{subfigure}
 \begin{subfigure}[t]{0.3\textwidth}
 \centering
 \includegraphics[scale=0.325]{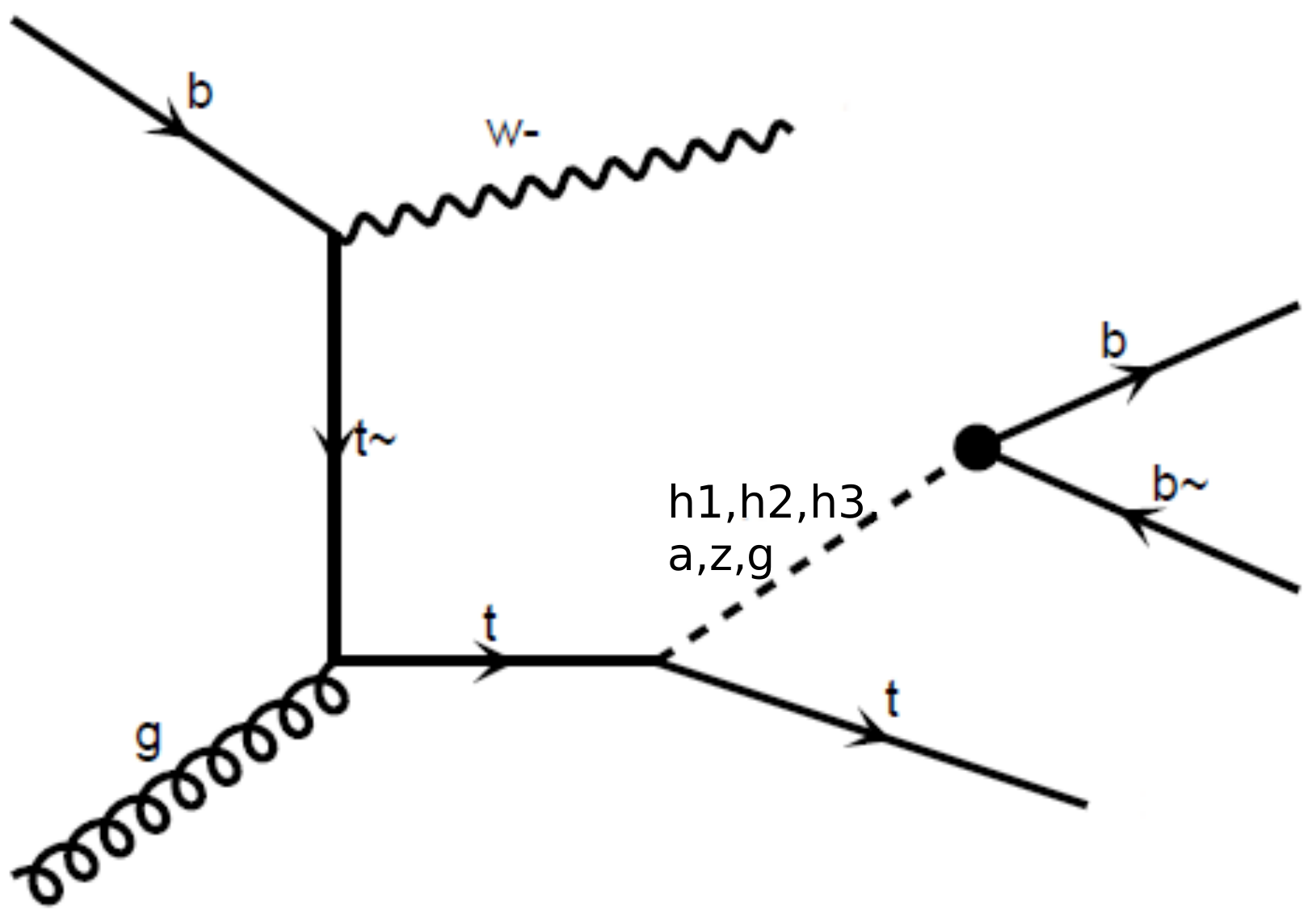} 
 \caption{}
 \end{subfigure}
 \begin{subfigure}[t]{0.3\textwidth}
 \centering
 \includegraphics[scale=0.325]{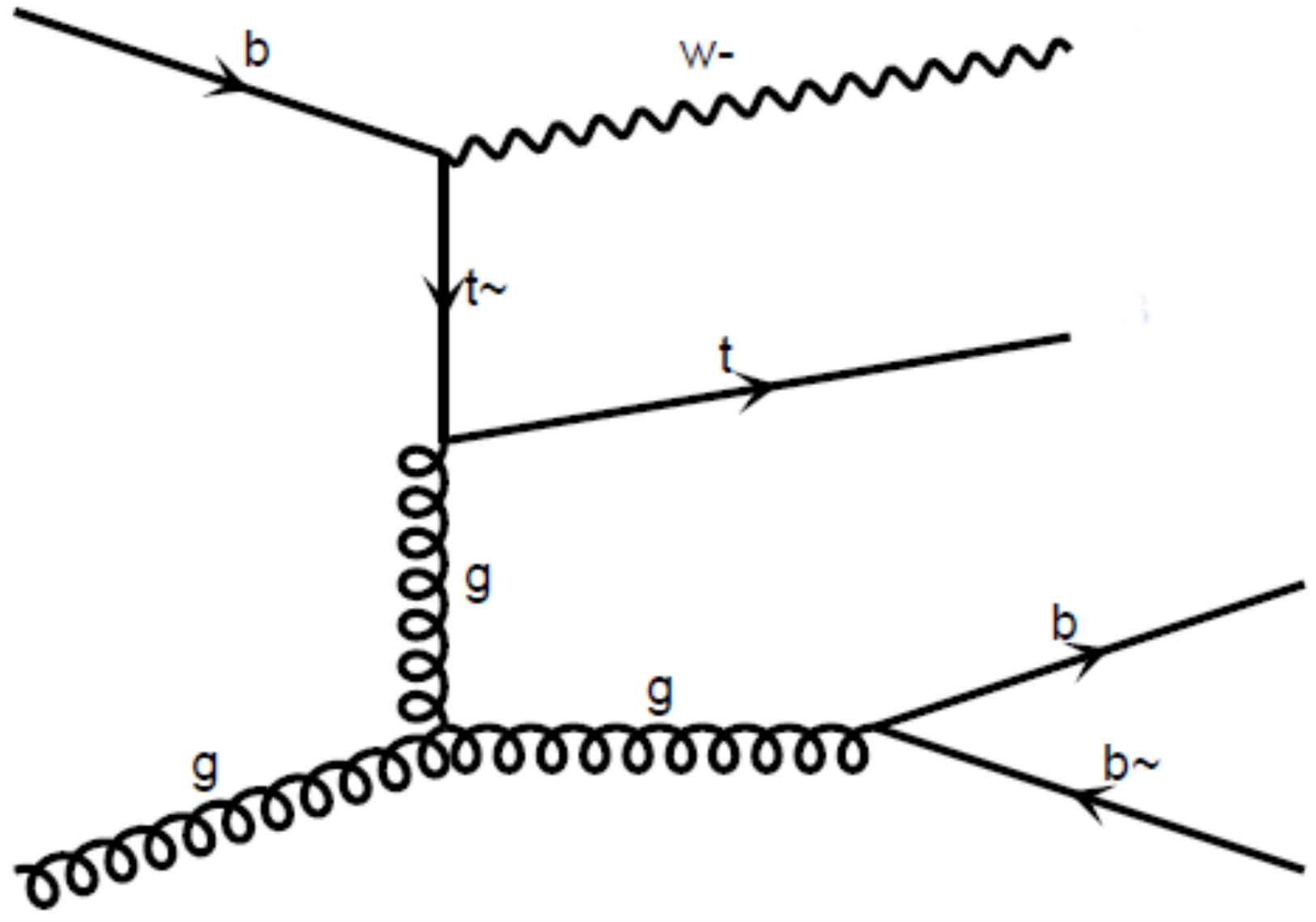} 
 \caption{}
 \end{subfigure}
 \begin{subfigure}[t]{0.3\textwidth}
 \centering
 \includegraphics[scale=0.325]{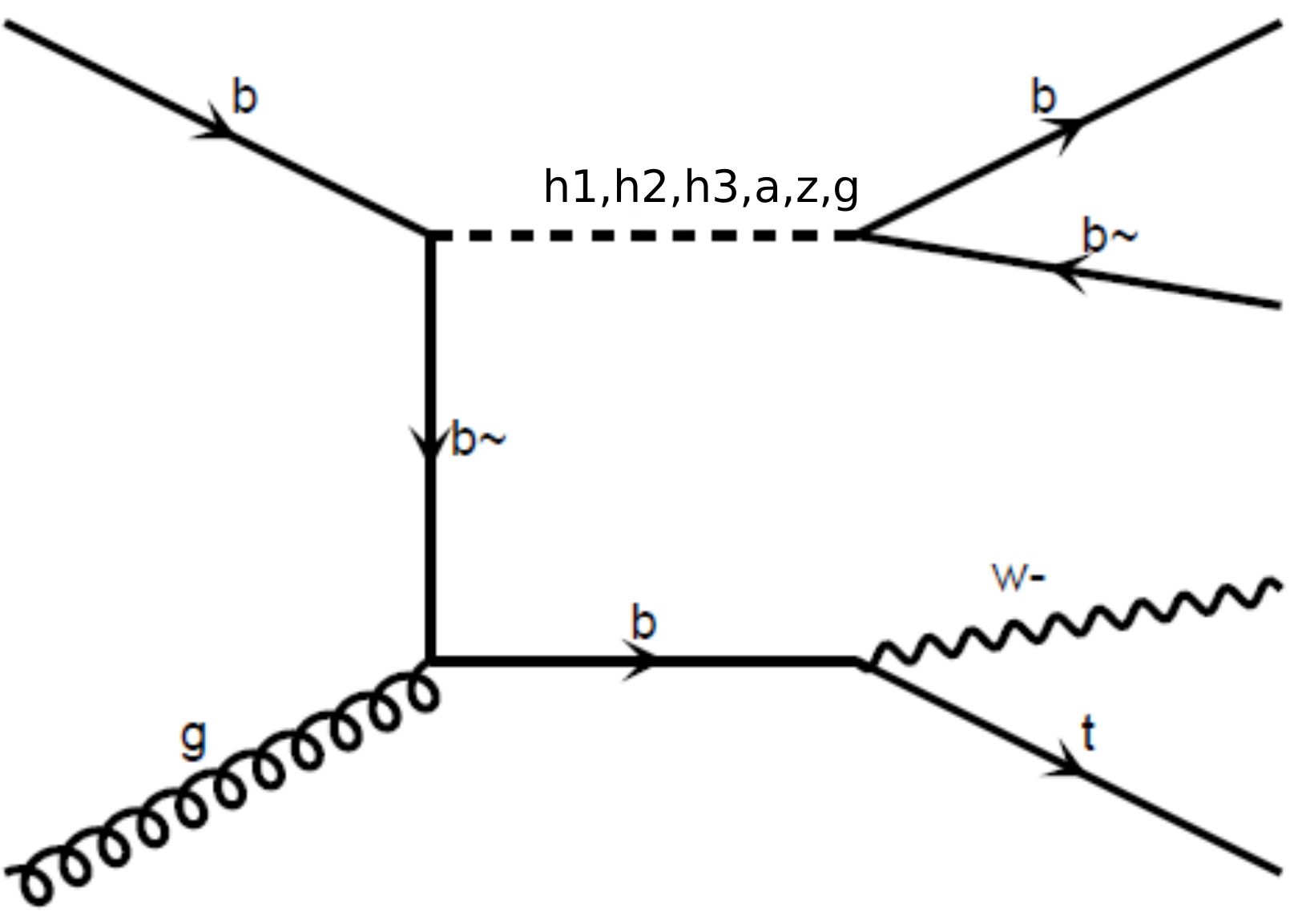} 
 \caption{}
 \end{subfigure} 
 \begin{subfigure}[t]{0.3\textwidth}
 \centering
 \includegraphics[scale=0.325]{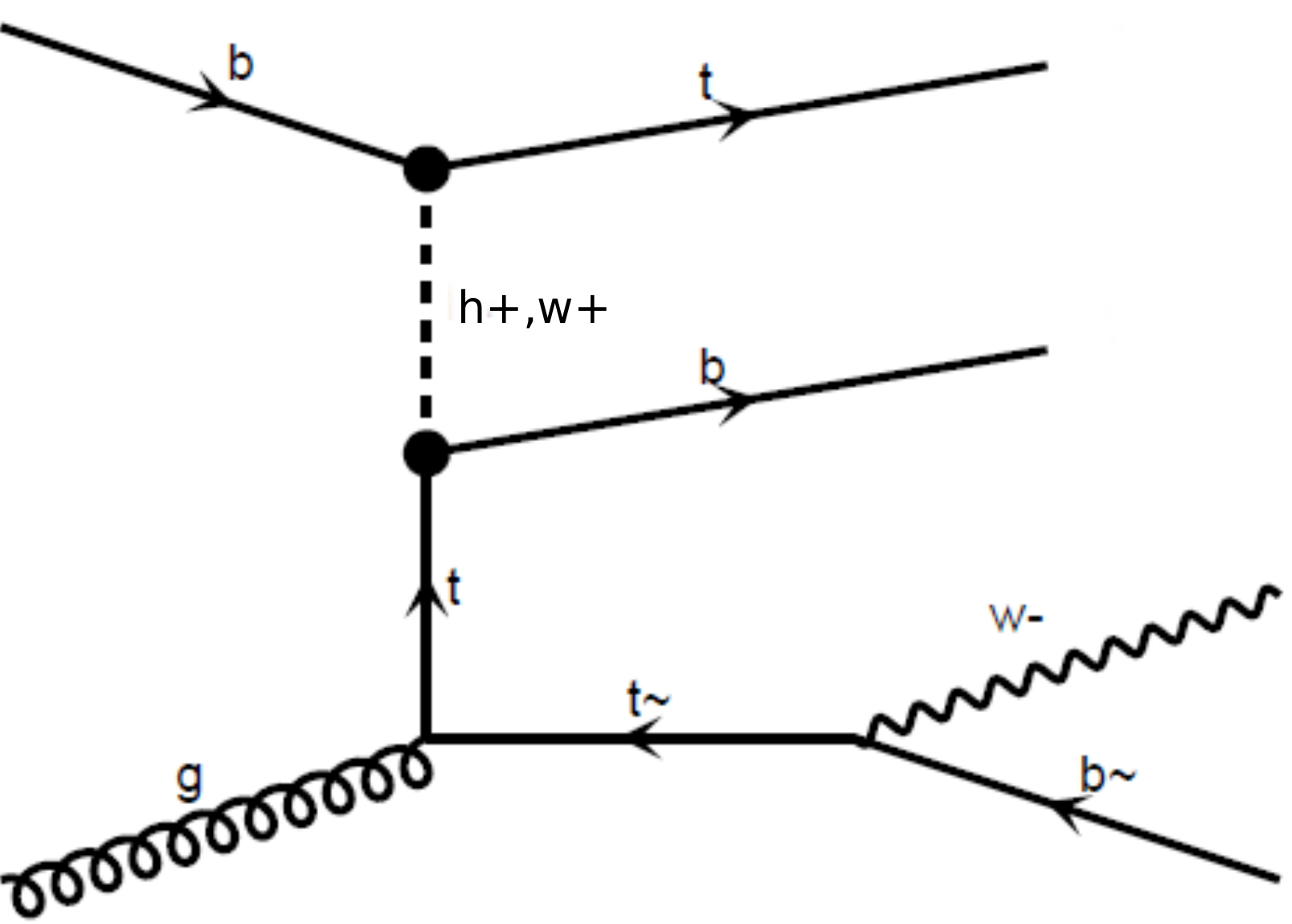} 
 \caption{}
 \end{subfigure}
 \begin{subfigure}[t]{0.3\textwidth}
 \centering
 \includegraphics[scale=0.325]{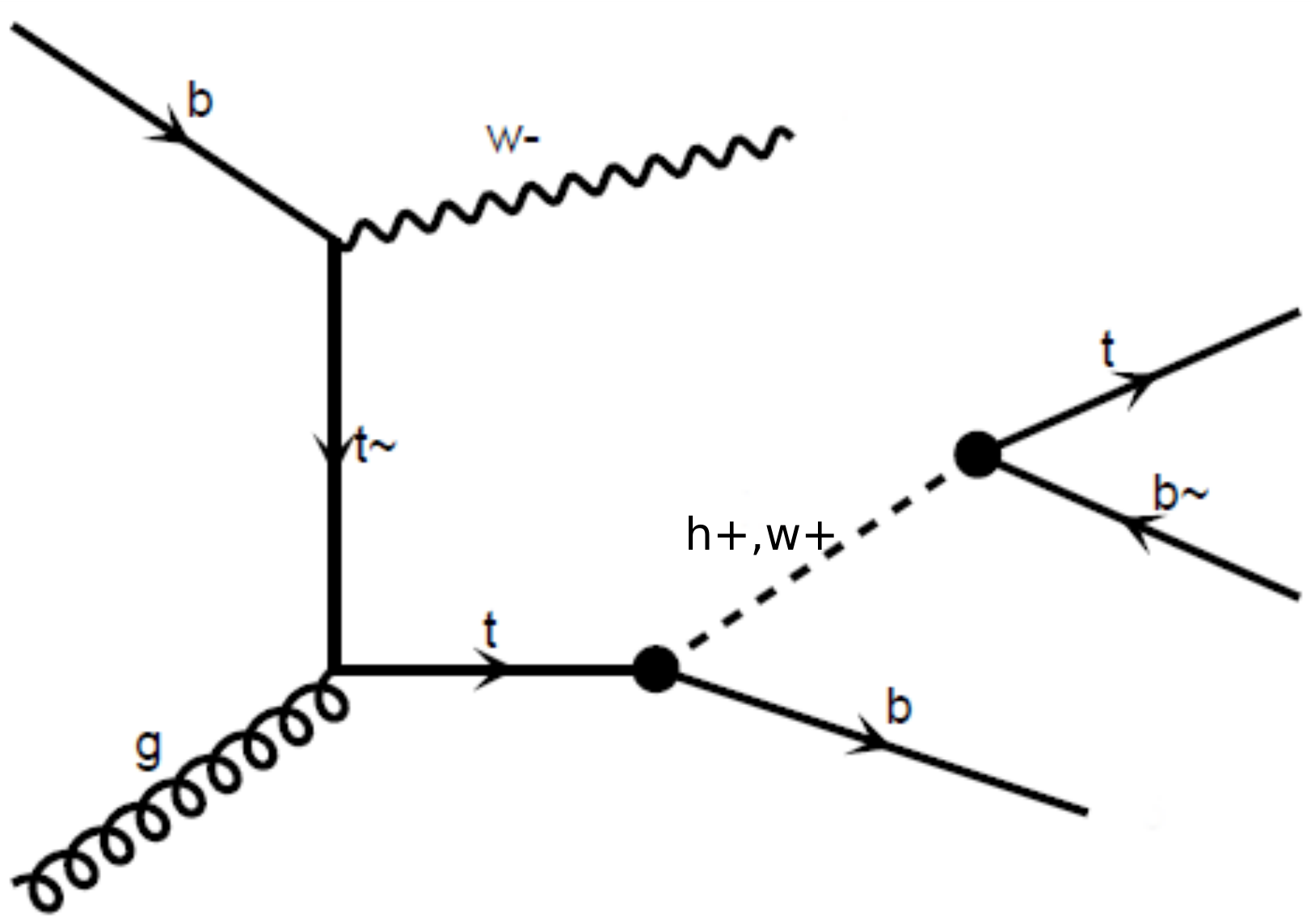} 
 \caption{}
 \end{subfigure}
\caption{\label{figs:bckg_c} Non-resonant Feynman diagrams contributing to the background for the process $pp\to tW^- b \bar b$ with h- $\equiv H^-$, h1 $\equiv h$, h2 $\equiv H$, h3 $\equiv A$ and a $\equiv \gamma$ (as appropriate).}
\end{figure}

We study the process $pp\to tW^- b\bar b$ wherein the interference effects between the charged Higgs resonant diagrams (shown in figure~\ref{figs:sig}) and the non-resonant background graphs (presented in figures~\ref{figs:bckg_a}, \ref{figs:bckg_b} and \ref{figs:bckg_c}) are found to be substantial. Non-resonant diagrams include all possible contributions coming from SM background as well as from 2HDM contributions. In total, there are 394 diagrams for the background contributing to the process $pp\to tW^- b \bar b$. The SM top-pair production associated with a $b$ quark is the dominant component of the latter. We have calculated the interference effects between the resonant diagrams (from figure~\ref{figs:sig}) and the diagrams which come from SM QCD interactions at the $\alpha_{\rm S}^3 \alpha_{\rm EW}$ order (Feynman diagrams with gluon contributions in figure~\ref{figs:bckg_a} and diagrams 1--5  in figure~\ref{figs:bckg_c}) as well as with the ones that come from SM EW interactions at the $\alpha_{\rm S} \alpha_{\rm EW}^3$ order. We found that, in most cases, the EW contributions produce a small but positive interference with the signal while the QCD contributions a large negative one. Thus, the net result is generally an overall negative interference for the total cross section of the process $pp\to t W^- b\bar b$ and the magnitude of this interference is determined by the width of the intermediate Higgs particles, i.e.,  $A$ and $H^\pm$. However, for a minority of the Benchmark Points (BPs) to be studied, the overall effect can be positive.

As far as the signal is concerned, we focus on the dominant production mode of a heavy charged Higgs, i.e., $pp\to tH^-$, followed by its decay via $H^-\to W^-h\to W^- b\bar b$, $H^-\to W^-H\to W^- b\bar b$, $H^-\to W^-A \to W^- b\bar b$ and $H^-\to  \bar t b\to W^- b\bar b$. Thus, all such decays lead to the same final state, facilitating interference effects  amongst the different signal amplitudes. However, as previously shown in~\cite{previous}, interference effects amongst the signal contributions are generally negligible. Moreover, the BPs chosen in this study are such that the $H^-\to W^- A\to W^-b\bar b$ decay mode dominates over all other charged Higgs boson decays.

For the BPs of the models that we consider, we will focus on the mass of the scalars involved in the process, $A$ and $H^\pm$, the BRs of the decays $H^\pm\to W^\pm A$ and $A \to b \bar{b}$ as well as the total width of the two scalars. As previously discussed, in order to have large interference effects between signal and background, the total width of the charged Higgs boson has to be quite large. 
However, to have a large interference, a sufficiently large width of the pseudoscalar is also essential. Otherwise the decays of the $A$ would be extremely narrow and would not overlap with any background processes. In this analysis we first consider a 2HDM Type II/Flipped where the pseudoscalar couplings to down-type fermions are proportional to $\tan\beta$ so that, for a large values of it, the width of the pseudoscalar can be made significantly large. Taking into account the latest searches on $pp \to \Phi \to \tau^+ \tau^-$~\cite{Aaboud:2017sjh}, $\Phi$ being any heavy spin-0 object, very large values of $\tan \beta$ are disallowed. In table~\ref{tab:inputs1} we present
our input parameters for the Type II and Flipped models for five chosen BPs. The five points have passed
all the constraints described before plus they are all valid for the Flipped model as well since searches for  
$pp \to \Phi \to \tau^+ \tau^-$ are negligible in the Flipped case owing to the very small coupling to $\tau$ leptons for high
$\tan \beta$. Therefore the five points are valid in the Flipped model but only the first three are valid in Type II.
The latest searches for charged Higgs bosons by ATLAS~\cite{Aad:2015typ, Aaboud:2016dig} and CMS~\cite{Khachatryan:2015qxa, CMScharged} are indeed in agreement with the values of the charged Higgs mass and the corresponding value of $\tan \beta$.

%%%%%%%%%%%%%%%%%%%%%%%%%%%%%%%%%%%%%%%%
%Input for type II and Flipped
%%%%%%%%%%%%%%%%%%%%%%%%%%%%%%%%%%%%%%%%
   \begin{table}[hpbt]
 \begin{ruledtabular}
\begin{tabular}{cccccc} 
  & $\tan\beta$ & $\sin(\beta-\alpha)$ & $m_{H^\pm}$ (GeV) & $m_{A}$ (GeV) & $m^2_{12}$ (GeV$^{2})$   \\ \hline
 BP1 (II)  & 10.25 & 0.98 & 509.14 & 248.27 &  52287.83 \\ \hline
 BP2 (II)  & 16.75  & 0.99 & 545.82 & 268.41 &  33622.43 \\ \hline
 BP3 (II)  & 18.80 & 0.99 & 457.71 & 247.22 &  16427.97 \\ \hline
 BP4  (F)  & 37.21 & 0.99 & 469.45 & 258.03 &  9800.68 \\ \hline
 BP5  (F)  & 44.10 & 1.00    & 519.45 & 288.32 &  10200.34 \\ \hline
\end{tabular}
\caption{Type II and Flipped input parameters for the BPs. }
\label{tab:inputs1}
\end{ruledtabular}
\end{table}
%%%%%%%%%%%%%%%%%%%%%%%%%%%%%%%%%%%%%%%%

%%%%%%%%%%%%%%%%%%%%%%%%%%%%%%%%%%%%%%%%
% widths and BRs for type II and Flipped 
%%%%%%%%%%%%%%%%%%%%%%%%%%%%%%%%%%%%%%%%
   \begin{table}[hpbt]
 \begin{ruledtabular}
\begin{tabular}{cccccc} 
  & $\Gamma(A)$ & $\Gamma(H^\pm)$ & ${\rm BR}(A\to b \bar b)$ & ${\rm BR}(H^+ \to b \bar t)$ & $ {\rm BR}(H^+ \to W^+ A)$   \\ \hline
 BP1 (II)  & 0.47 & 72.85 & 0.83 & 0.01 & 0.29  \\ \hline
 BP2 (II)  & 1.29 & 91.97 & 0.86 & 0.02 &  0.29 \\ \hline
 BP3 (II)  & 1.50 & 34.83 & 0.87 & 0.05 &  0.17 \\ \hline
 BP4  (F) & 5.45 & 50.45 & 0.99  & 0.13 &  0.16 \\ \hline
 BP5  (F) & 10.46 & 85.45 & 1.00  & 0.18 &  0.26 \\ \hline
\end{tabular}
\caption{Partial widths (in GeV) and BRs in Type II and Flipped for the BPs. }
\label{tab:inputs2}
\end{ruledtabular}
\end{table}
%%%%%%%%%%%%%%%%%%%%%%%%%%%%%%%%%%%%%%%%

In table~\ref{tab:inputs2} we present the partial widths for $\Gamma(A)$ and $\Gamma(H^\pm)$ and 
the ${\rm BR}(A\to b \bar b)$, ${\rm BR}(H^+ \to b \bar t)$ and $ {\rm BR}(H^+ \to W^+ A)$ for
the five BPs. Note that the major difference between the models is the column
for ${\rm BR}(A\to b \bar b)$ that is always larger in the Flipped model because the decays to $\tau$ leptons
become negligible in this model. For all other columns the differences are extremely small. This 
in turn means that the results are slightly better for the Flipped model. We choose for the detailed
analysis BP5 for of  Flipped model. Cross sections for signal, background and total (including interference) for BP5 are 0.74 pb, 10.43 pb and 10.72 pb respectively. This results into an  interference cross section of 0.45 pb which is around 60\% of the signal cross section. The results for the cross sections for all BPs for the Type II and Flipped
models are presented in table~\ref{tab:inputsx}.
%%%%%%%%%%%%%%%%%%%%%%%%%%%%%%%%%%%%%%%%
\begin{table}[h]\centering
 \begin{tabular}{|c|c|c|c|c|}\hline
 BP & Signal (pb) & Background (pb) & Total (pb) & Interference (pb) \\\hline
 BP1 & 0.031 & 10.03 &  9.96 & -0.101\\\hline
 BP2 & 0.052 &  9.96 & 10.02 & -0.008\\\hline
 BP3 & 0.144 & 10.07 & 10.18 &  0.034\\\hline
 BP4 & 0.469 &  9.94 & 10.31 & -0.102\\\hline
 BP5 & 0.742 & 10.43 & 10.72 & -0.452\\\hline
   \end{tabular}
\caption{Cross sections (in pb) for signal, background, total and interference for the BPs 
of Type II and Flipped.}
\label{tab:inputsx}
\end{table}
%%%%%%%%%%%%%%%%%%%%%%%%%%%%%%%%%%%%%%%%

In the case of the Type III model we choose three BPs which are in agreement with all constraints.
These are shown in table~\ref{tab:inputs3} and the corresponding widths and BRs are presented
in table~\ref{tab:inputs4}.

%%%%%%%%%%%%%%%%%%%%%%%%%%%%%%%%%%%%%%%%
%   Input parameters for type III
%%%%%%%%%%%%%%%%%%%%%%%%%%%%%%%%%%%%%%%%
   \begin{table}[hpbt]
 \begin{ruledtabular}
\begin{tabular}{ccccccc} 
  & $\tan\beta$ & $\sin(\beta-\alpha)$ & $m_{H^\pm}$ (GeV) & $m_{A}$ (GeV) & $m^2_{12}$ (GeV$^{-2})$  & $\chi$ \\ \hline
 BP1  & 15.84 & 0.99 & 480.75 & 369.89 &  27463.94 & 0.21\\ \hline
 BP2  & 19.41 & 0.99 & 307.23 & 225.46 & 6045.62  & -0.34 \\ \hline
% BP3  & 27.78 & 0.99 & 526.49 & 359.85 &  17695.67 & -0.45\\ \hline
 BP3  & 38.11 & 0.99 & 447.45 & 258.33 & 9833.68 & 0.71 \\ \hline
% BP5  & 41.54 & 0.99 & 457.45 & 345.14 & 9193.03  & -1\\ \hline
\end{tabular}
\caption{Type III input parameters for the BPs. }
\label{tab:inputs3}
\end{ruledtabular}
\end{table}

%%%%%%%%%%%%%%%%%%%%%%%%%%%%%%%%%%%%%%%%
%%     Type III
%%%%%%%%%%%%%%%%%%%%%%%%%%%%%%%%%%%%%%%%
   \begin{table}[hpbt]
 \begin{ruledtabular}
\begin{tabular}{cccccc} 
  & $\Gamma(A)$ & $\Gamma(H^\pm)$ & ${\rm BR}(A\to b \bar b)$ & ${\rm BR}(H^+ \to b \bar t)$ & $ {\rm BR}(H^+ \to W^+ A)$   \\ \hline
 BP1  & 2.79 & 60.72 & 0.47 & 0.02 &  0.27 \\ \hline
 BP2  & 1.69 & 12.78 &  0.88 & 0.05 & 0.21  \\ \hline
% BP3  & 5.87 & 71.91 & 0.68  & 0.09 &  0.10 \\ \hline
 BP3  & 6.10 & 52.77 & 0.87 & 0.10 & 0.17  \\ \hline
% BP5  & 9.74 & 55.77 & 0.87 & 0.02 & 0.19  \\ \hline
\end{tabular}
\caption{Partial widths in units of GeV and BRs in Type III. }
\label{tab:inputs4}
\end{ruledtabular}
\end{table}
%%%%%%%%%%%%%%%%%%%%%%%%%%%%%%%%%%%%%%%%

For the Type III, we choose the benchmark point BP2 which has the lightest $H^\pm$ mass and lowest mass 
splitting between $H^\pm$ and $W^\pm$ in order to demonstrate the interference effect in a wide range 
of mass spectra. For this benchmark point the cross sections for the signal, background 
and the total are 0.978 pb, 9.95 pb and 10.92 pb, respectively. Thus, the resulting cross section turns out to be -0.008 pb. The small interference effect in this case can be attributed to the small width of both $H^\pm$ and $A$.
The results for the cross sections for all BPs for the Type III
model are presented in table~\ref{tab:inputsyy}.

%%%%%%%%%%%%%%%%%%%%%%%%%%%%%%%%%%%%%%%%
\begin{table}[h]\centering
 \begin{tabular}{|c|c|c|c|c|}\hline
 BP & Signal (pb) & Background (pb) & Total (pb) & Interference (pb) \\\hline
 BP1 & 0.059 & 10.35  & 10.27 & -0.139\\\hline
 BP2 & 0.978 &  9.95  & 10.92 & -0.008\\\hline
% BP3 & 0.462 & 11.77  & 12.11 & -0.122\\\hline
 BP3 & 0.291 & 10.02  & 10.34 &  0.029\\\hline
% BP5 & 2.640 & 17.89  & 20.60 &  0.070\\\hline  
 \end{tabular}
\caption{Cross sections (in pb) for signal, background, total and interference for the BPs in the 
2HDM Type III.}
 \label{tab:inputsyy}
\end{table}
%%%%%%%%%%%%%%%%%%%%%%%%%%%%%%%%%%%%%%%%

All the numbers presented above are at parton level. Next we perform a detector level analysis and study if these interference effects survive even after all  acceptance and selection cuts. For this purpose, we generate the events using {\tt MadGraph} \cite{madgraph} and then we pass these  to {\tt Pythia} \cite{pythia8} for parton showering and hadronisation. After that all the events are finally passed through {\tt Delphes} \cite{delphes} for a realistic detector level analysis.   

Below we list the basic detector acceptance cuts.

\begin{itemize}
\item {\bf Acceptance cuts}
 \begin{enumerate}
  \item Events must have at least one lepton ($e$ or $\mu$) and at least 5 jets. 
  \item Leptons must have transverse momentum $p_T>20$ GeV and rapidity $|\eta|<2.5$.
  \item All jets must satisfy the following $p_T$ and $\eta$ requirements:
  $$p_{Tj}>20~ \mbox{GeV}, |\eta_j|<2.5.$$
  \item All pairs of objects must be well separated from each other,
  $$\Delta R_{jj,jb,bb,\ell j,\ell b}\geq 0.4~~ \mbox{where}~~\Delta R=\sqrt{(\Delta \phi)^2+(\Delta \eta)^2}.$$
 \end{enumerate}
\end{itemize}
\subsection{Event reconstruction}
In this section, we describe the procedure which we employ to reconstruct the masses of top quark, charged Higgs $H^\pm$, pseudoscalar $A$ and the two $W^\pm$ bosons in each event. For this purpose, we make use of a method based on a $\chi^2$ template. We then discuss the efficiency of the reconstruction. Each event in the analysis is assumed to be a $tH^-$ event decaying to $W^+W^- j j j$ and one of the $W^\pm$ is considered to decay hadronically and the other leptonically. Thus, each single event is considered to have at least one lepton, 5 jets and missing transverse energy. 

The $\chi^2$ fit takes as input the four vectors of the five leading jets,  lepton and  neutrino. The treatment of the neutrino four-vector is as follows. The transverse momentum of the neutrino is determine through balancing the initial and final particle momenta in an event. The longitudinal component of the neutrino momentum is instead determined by imposing the invariant mass constraint $M_{l\nu}^2 = M_{W^\pm}^2$. Since this condition leads to a quadratic equation, there are in general two solutions for $p_\nu^z$:	
\begin{equation}
 p_{\nu}^z=\frac{1}{2p_{\ell T}^2}\left(A_W p_{\ell}^z \pm E_\ell \sqrt{A_W^2\pm 4 p_{\ell T}^2 E_{\nu T}^2}\right),
\end{equation}
where $A_W=M_{W^\pm}^2+2p_T\cdot E_{\nu T}$. A separate $\chi^2$ is evaluated for each of the $p_\nu^z$ solutions  and the one having minimum $\chi^2$ value is retained to reconstruct the event.

We write two expressions for $\chi^2$, one corresponding to a scenario where $H^\pm$ decays fully hadronically, $\chi^2_{\rm had}$, and other where it decays semi-leptonically, $\chi^2_{\rm lep}$:
\begin{equation}\label{chi1}
\chi^2_{\rm had}=\frac{(M_{\ell \nu}-M_W)^2}{\Gamma_W^2}+\frac{(M_{jj}-M_W)^2}{\Gamma_W^2}+\frac{(M_{\ell \nu j}-M_{top})^2}{\Gamma_{top}^2}+\frac{(M_{jj}-M_A)^2}{\Gamma_A^2}+\frac{(M_{jjjj}-M_{H^\pm})^2}{\Gamma_{H^\pm}^2},
\end{equation}

\begin{equation}\label{chi2}
\chi^2_{\rm lep}=\frac{(M_{\ell \nu}-M_W)^2}{\Gamma_W^2}+\frac{(M_{jj}-M_W)^2}{\Gamma_W^2}+\frac{(M_{j j j}-M_{top})^2}{\Gamma_{top}^2}+\frac{(M_{jj}-M_A)^2}{\Gamma_A^2}+\frac{(M_{\ell \nu jj}-M_{H^\pm})^2}{\Gamma_{H^\pm}^2},
\end{equation}
where in the denominators we have the decay widths of the respective particles as calculated for the BPs in the various models.

\begin{figure}[h!]\centering
 \includegraphics[scale=0.4]{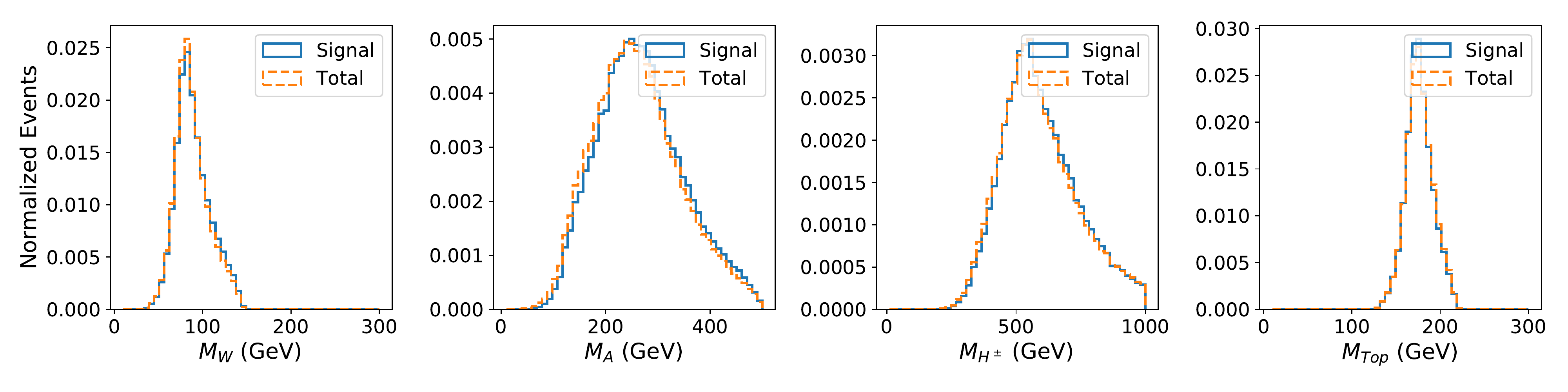}
 \caption{\label{Mass_BP5}Reconstructed masses of $W^\pm$, pseudoscalar $A$, top quark and charged Higgs $H^\pm$ for BP5 in the 2HDM Flipped.}
\end{figure}

For each event, $\chi^2$ is evaluated for each possible way of assigning the five leading jets to the reconstructed top and charged Higgs four-momenta. The number of such permutations turns out to be 15 for each of $\chi^2_{\rm had}$ and $\chi^2_{\rm lep}$. In addition, there is a twofold ambiguity in assigning the two solutions for $p_\nu^z$. Finally, there are two ways with which two of the jets can be assigned to either a $W^\pm$ boson or to the pseudoscalar. Thus, for each event, the $\chi^2$'s  are evaluated for 120 different combinations and the combination with minimum $\chi^2$ values is kept for mass reconstruction. 

\begin{figure}[h!]\centering
 \includegraphics[scale=0.4]{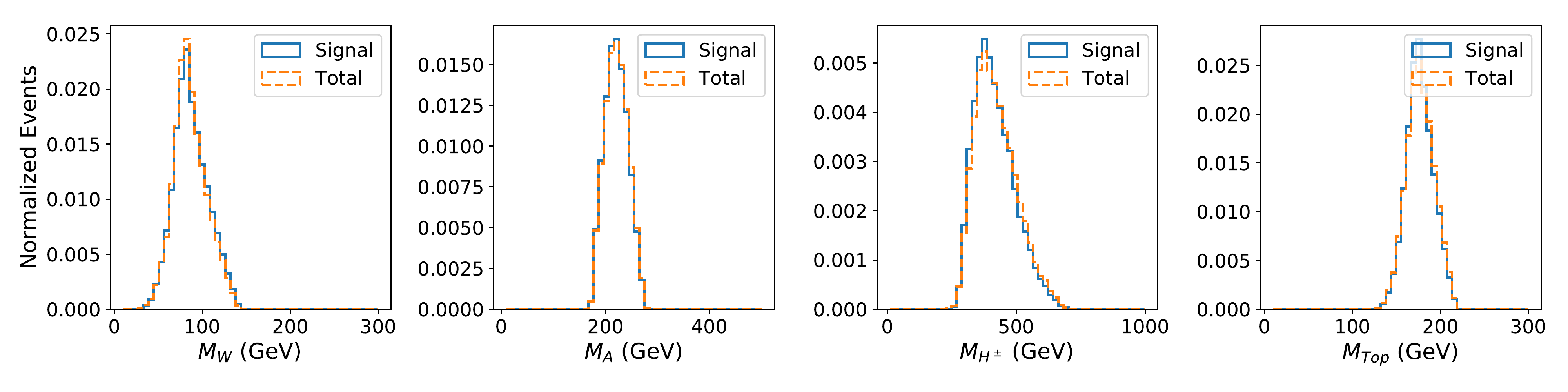}
 \caption{\label{Mass_BP2}Reconstructed masses of $W^\pm$, pseudoscalar $A$, top quark and charged Higgs $H^\pm$ for BP2 in the 2HDM Type III.}
\end{figure}

Using the procedure described above, we now proceed to reconstruct the masses of the various particles involved in the process in order to see the efficiency of it. We present the reconstructed masses of all the intermediate resonant particles in the process, {i.e.}, $W^\pm$, $A$, top and $H^\pm$ in figure~\ref{Mass_BP5} for the Flipped case (BP5) and in figure~\ref{Mass_BP2} for the Type III case (BP2). In each plot we see that the peak is found to be at the particle masses, vouching for the effectiveness of our reconstruction procedure. In presenting the plots, we take events after applying all the acceptance cuts discussed above and selection cuts mentioned in table~\ref{tab:cutflow_BP5}.

\begin{figure}[h!]\centering
\includegraphics[scale=0.5]{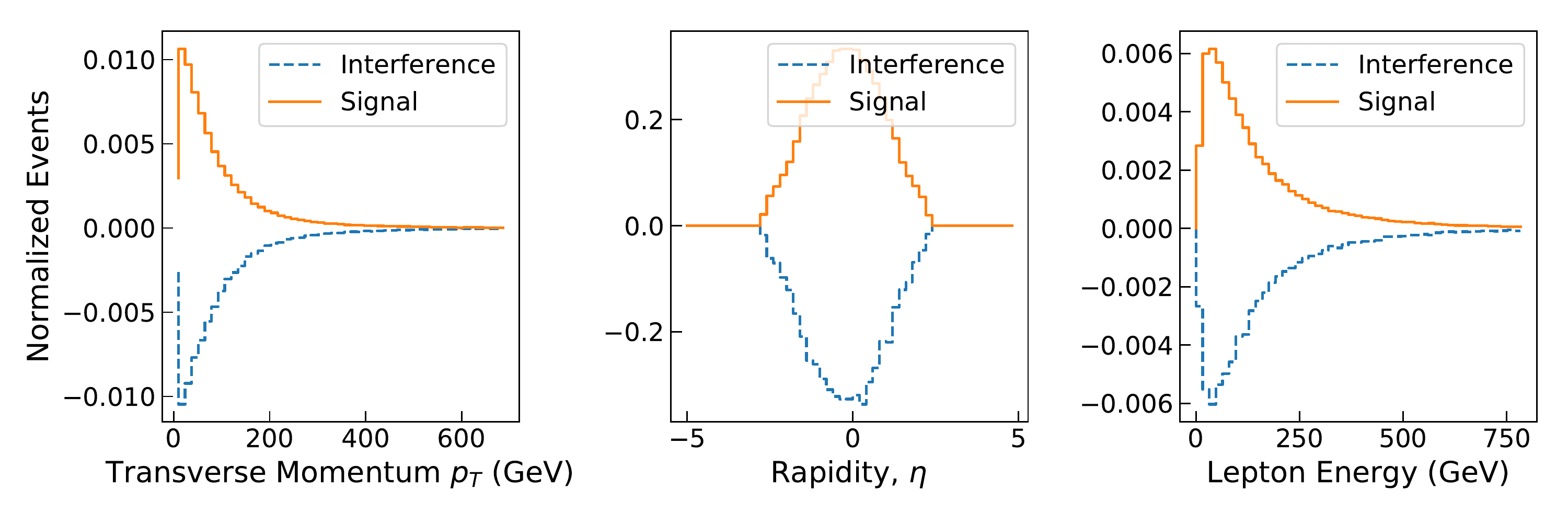}
\includegraphics[scale=0.5]{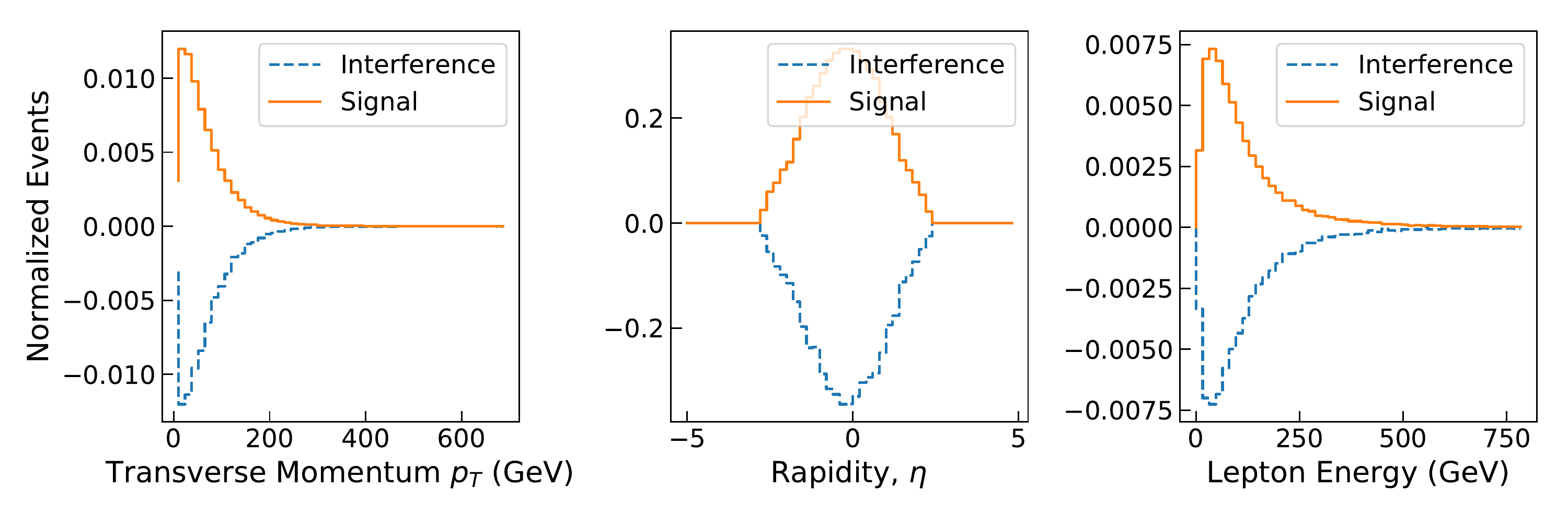}
 \caption{\label{Lep_BP5}Distributions for transverse momentum, rapidity and energy of a lepton for signal and interference for BP5 of the 2HDM Flipped (top) and BP2 of the 2HDM Type III (bottom).}
\end{figure}

In order to further investigate  interference effects, we look at various distributions, e.g., transverse momentum $p_T$, rapidity $\eta$ and energy $E$ of the lepton, for both the signal and interference contributions. The distributions for the interference are obtained by subtracting those of the signal and background processes (separately) from the total ones. The distributions for  BP5 of the 2HDM Flipped (top) and BP2 of the 2HDM Type III (bottom) are shown in figure~\ref{Lep_BP5}. We can clearly see that the shape of all  distributions for signal alone and interference are almost the same but with opposite signs, the latter being expected, as we found the overall interference between signal and irreducible background to be destructive for the used BP. It is instead remarkable the similarity found between the two contributors
to the total signal cross section. Notice that in figure~\ref{Lep_BP5} we have only shown the lepton distributions though it has been verified for all the jets involved in the process that their distributions present  the same behaviour. 

Finally, we present in table~\ref{tab:cutflow_BP5} the flow of cross section values after each cut for (Flipped) BP5 and in table~\ref{tab:cutflow_BP2} for  (Type III) BP2. We observe that the relative ratio of the signal-to-interference cross section increases with each cut for both BPs. For (Flipped) BP5, we see that the ratio rises from 60\% to almost larger than 100\% while, for (Type III) BP2, the increment is from 0.1\% to 17\%. The reason for the smaller interference cross section for the latter with respect to the former is a smaller width for both $A$ and $H^\pm$: this well illustrates the correlation between interference effects and off-shellness of the Higgs bosons involved. 

\begin{table}[h!]
% \fontsize{7pt}{8.4pt}\selectfont
\begin{center}
{\renewcommand{\arraystretch}{1.5}}
 \newcolumntype{C}[1]{>{\centering\let\newline\\\arraybackslash\hspace{0pt}}m{#1}} %need \usepackage{array}
\begin{tabular}{ ||C{0.85cm}C{2.5cm}C{1.75cm}|| C{1.5cm} | C{2.25cm} | C{1.5cm} |C{2.5cm} ||}
\hline \hline
\multicolumn{3}{||c||}{\multirow{2}{*}{Cuts}}&\multicolumn{4}{c||}{$\sigma$ [fb] } \\\cline{4-7}
&&&Signal & Background  & Total  & Interference    \\\cline{1-7}
C0:& \multicolumn{2}{c||}{No Cuts}			& 740		& 10430		& 10720		& -450 	\\ \hline
C1:& \multicolumn{2}{c||}{Only one lepton} 		& 115.0 	& 1116.2 	& 1151.2  	& -80.1 	\\ \hline
C2:& \multicolumn{2}{c||}{At least 5 light jets} 	& 91.9		& 680.8		& 703.5		& -69.2 	\\ \hline
C3:& \multicolumn{2}{c||}{Cut on $H_T>500$ GeV}		& 70.8		& 173.8		& 173.6		& -71.1		\\ \hline
% C4:& \multicolumn{2}{c||}{Cut on $\chi^2<1000$}		& 		& 		& 		& 		\\ \hline
\hline
% \multicolumn{3}{||c||}{$S/B$ }   &	&	&	& &	2.4\%							\\\hline
% \multicolumn{3}{||c||}{$S/\sqrt{B}$ with 100 fb$^{-1}$}   &  	&	& & 	& 6.1\\
% \hline
\end{tabular}
\caption{Cut flow of the cross sections for signal (BP5 in the 2HDM Flipped) and irreducible background at the 14 TeV LHC. Conjugate processes are included here. \label{tab:cutflow_BP5}}
\end{center}
\end{table}

\begin{table}[h!]
% \fontsize{7pt}{8.4pt}\selectfont
\begin{center}
{\renewcommand{\arraystretch}{1.5}}
 \newcolumntype{C}[1]{>{\centering\let\newline\\\arraybackslash\hspace{0pt}}m{#1}} %need \usepackage{array}
\begin{tabular}{ ||C{0.85cm}C{2.5cm}C{1.75cm}|| C{1.5cm} | C{2.25cm} | C{1.5cm} |C{2.5cm} ||}
\hline \hline
\multicolumn{3}{||c||}{\multirow{2}{*}{Cuts}}&\multicolumn{4}{c||}{$\sigma$ [fb] } \\\cline{4-7}
&&&Signal & Background  & Total  & Interference    \\\cline{1-7}
C0:& \multicolumn{2}{c||}{No Cuts}			& 978		& 9950		& 10920		& -8 	\\ \hline
C1:& \multicolumn{2}{c||}{Only one lepton} 		& 243.6 	& 2040.8 	& 1151.2  	& -6.4 	\\ \hline
C2:& \multicolumn{2}{c||}{At least 5 light jets} 	& 180.3		& 1221.4	& 1398.1	& -3.6 	\\ \hline
C3:& \multicolumn{2}{c||}{Cut on $H_T>500$ GeV}		& 89.8		& 491.2		& 566.9		& -14.1		\\ \hline
% C4:& \multicolumn{2}{c||}{Cut on $\chi^2<1000$}		& 		& 		& 		& 		\\ \hline
\hline
% \multicolumn{3}{||c||}{$S/B$ }   &	&	&	& &	2.4\%							\\\hline
% \multicolumn{3}{||c||}{$S/\sqrt{B}$ with 100 fb$^{-1}$}   &  	&	& & 	& 6.1\\
% \hline
\end{tabular}
\caption{Cut flow of the cross sections for signal (BP2 in the 2HDM Type III) and irreducible background at the 14 TeV LHC. Conjugate processes are included here. \label{tab:cutflow_BP2}}
\end{center}
\end{table}

\section{Conclusions}

In this paper, we have assessed whether interference effects involving heavy charged Higgs signals appearing via $W^\pm b\bar b$ final states  at the LHC, both amongst themselves and in relation to irreducible background, can be sizable and thus affect ongoing experimental searches. We have taken as reference models to perform our analysis two $Z_2$ symmetric 2HDMs, the Type
II and Flipped versions, as well as  the Type III one. We have then prepared the corresponding parameter space regions amenable to phenomenological investigation by enforcing both  theoretical (i.e., unitarity, perturbativity, vacuum stability, triviality)  and
experimental (i.e., from flavour physics, void and successful Higgs boson searches at the Tevatron and LHC, EW precisions observables from LEP and SLC) constraints. We have finally proceeded to simulate the relevant signal processes via  $bg\to tH^-$ (+ c.c.) scattering with the charged Higgs state decaying via $H^-\to W^- h,A,H\to W^- b\bar b$ or $H^-\to \bar t b\to W^- b\bar b$ (+ c.c. in all cases)  and the irreducible background given by $bg\to t W^- b\bar b$ topologies. The motivation for this is that signals and background  are treated separately in current approaches. Indeed, these may be invalidated by the fact that, on the one hand,  a heavy charged Higgs state can have a large width and, on the other hand, this can also happen for (some of) the neutral Higgs states emerging from its decays. Clearly, a prerequisite for such interference effects to onset is that such widths are large enough, say, 10\% or so, which we have verified here to be the case.
While the phenomenology we have investigated could well occur in the other decay chains in suitable regions of the parameter space, we have chosen to single out here   $H^-\to W^- A\to W^- b\bar b$, as it is the one that is most subject to 
interference effects with the irreducible background, at least in the 2HDM Type II, Flipped and  Type III setups adopted. In fact, the latter are generally predominant over interference effects amongst the different decay patterns of the $H^\pm$ signal.

After performing a sophisticated MC simulation, we have seen  that such interference effects can be very large, even of ${\cal O}(100\%)$,  both before and after $H^\pm$ selection cuts are enforced, and mostly negative. This appears to be the case for all masses tested, from 300 to 500 GeV or so, in both the 2HDM II and Flipped as well as Type III, the more so the larger the $H^\pm$ and $A$ masses (and, consequently, their widths).  Remarkably, after all  cuts are applied, the shapes of the analysed signal and interference (with the irreducible background) are essentially identical
in all kinematical observables relevant to the signal extraction, as the  selection drives these two components of the total cross section to be very similar. These findings therefore imply that current and, especially, future LHC sensitivities to heavy charged Higgs bosons signals in $W^- b\bar b$ final states
require an `inclusive' rescaling of the event yields, as the the `exclusive' shape of the signal is roughly unchanged after such interference effects are accounted for.

%%%%% Acknowledgments

\acknowledgments

RB was supported in part by Chinese Academy of Sciences (CAS) President's International Fellowship Initiative (PIFI) program (Grant No. 2017VMB0021). The work of AA, RB, SM and RS is funded through the grant H2020-MSCA-RISE-2014 No. 645722 (NonMinimalHiggs). SM is supported in part through the NExT Institute and the STFC Consolidated Grant ST/P000711/1.  PS is supported by the Australian Research Council through the ARC Center of Excellence for Particle Physics (CoEPP) at the Terascale (grant no. \ CE110001004).
RS is also supported in part by the National Science Centre, Poland, the HARMONIA project under contract UMO-2015/18/M/ST2/00518.
\clearpage
%%%%%%%%%%%%%%%%%%%%%%%%%%%%%%%%%%%%%%%%%%%%%%%%%%%

%%%%%%%%%%%%%%%%%%%%%%%%%%%%%%%%%%%%%%%%%%%%%%%%%


\begin{thebibliography}{99}

%\cite{Aoki:2011wd}
  \bibitem{Aoki:2011wd}
  M.~Aoki, R.~Guedes, S.~Kanemura, S.~Moretti, R.~Santos and K.~Yagyu,
  %``Light Charged Higgs bosons at the LHC in 2HDMs,''
  Phys.\ Rev.\ D {\bf 84} (2011) 055028
  %doi:10.1103/PhysRevD.84.055028
  [arXiv:1104.3178 [hep-ph]].
  %%CITATION = doi:10.1103/PhysRevD.84.055028;%%
  %46 citations counted in INSPIRE as of 10 Dec 2017

%\cite{Arnison:1983rp}
  \bibitem{Arnison:1983rp}
  G.~Arnison {\it et al.} [UA1 Collaboration],
  %``Experimental Observation of Isolated Large Transverse Energy Electrons with Associated Missing Energy at s**(1/2) = 540-GeV,''
  Phys.\ Lett.\  {\bf 122B} (1983) 103.
  %doi:10.1016/0370-2693(83)91177-2
  %%CITATION = doi:10.1016/0370-2693(83)91177-2;%%
  %2315 citations counted in INSPIRE as of 10 Dec 2017


%\cite{Bagnaia:1983zx}
  \bibitem{Bagnaia:1983zx}
  P.~Bagnaia {\it et al.} [UA2 Collaboration],
  %``Evidence for Z0 ---> e+ e- at the CERN anti-p p Collider,''
  Phys.\ Lett.\  {\bf 129B} (1983) 130.
  %doi:10.1016/0370-2693(83)90744-X
  %%CITATION = doi:10.1016/0370-2693(83)90744-X;%%
  %1702 citations counted in INSPIRE as of 10 Dec 2017


%\cite{Aad:2012tfa}
\bibitem{Aad:2012tfa}
  G.~Aad {\it et al.} [ATLAS Collaboration],
  %``Observation of a new particle in the search for the Standard Model Higgs boson with the ATLAS detector at the LHC,''
  Phys.\ Lett.\ B {\bf 716} (2012) 1
  %doi:10.1016/j.physletb.2012.08.020
  [arXiv:1207.7214 [hep-ex]].
  %%CITATION = doi:10.1016/j.physletb.2012.08.020;%%
  %7940 citations counted in INSPIRE as of 10 Dec 2017
  
%\cite{Chatrchyan:2012xdj}
\bibitem{Chatrchyan:2012xdj}
  S.~Chatrchyan {\it et al.} [CMS Collaboration],
  %``Observation of a new boson at a mass of 125 GeV with the CMS experiment at the LHC,''
  Phys.\ Lett.\ B {\bf 716} (2012) 30
  %doi:10.1016/j.physletb.2012.08.021
  [arXiv:1207.7235 [hep-ex]].
  %%CITATION = doi:10.1016/j.physletb.2012.08.021;%%
  %7771 citations counted in INSPIRE as of 10 Dec 2017

\bibitem{Branco}
%\cite{Branco:2011iw}
%\bibitem{Branco:2011iw} 
  G.~C.~Branco, P.~M.~Ferreira, L.~Lavoura, M.~N.~Rebelo, M.~Sher and J.~P.~Silva,
  %``Theory and phenomenology of two-Higgs-doublet models,''
  Phys.\ Rept.\  {\bf 516} (2012) 1
  %doi:10.1016/j.physrep.2012.02.002
  [arXiv:1106.0034 [hep-ph]].
  %%CITATION = doi:10.1016/j.physrep.2012.02.002;%%
  %1110 citations counted in INSPIRE as of 10 Dec 2017
  \bibitem{Ivanov:2017dad}
  I.~P.~Ivanov,
  %``Building and testing models with extended Higgs sectors,''
  Prog.\ Part.\ Nucl.\ Phys.\  {\bf 95} (2017) 160
  %doi:10.1016/j.ppnp.2017.03.001
  [arXiv:1702.03776 [hep-ph]].
  %%CITATION = doi:10.1016/j.ppnp.2017.03.001;%%
  %24 citations counted in INSPIRE as of 10 Dec 2017

%\cite{Misiak:2015xwa}
\bibitem{Misiak:2015xwa} 
  M.~Misiak {\it et al.},
  %``Updated NNLO QCD predictions for the weak radiative B-meson decays,''
  Phys.\ Rev.\ Lett.\  {\bf 114} (2015) no.22,  221801
  %doi:10.1103/PhysRevLett.114.221801
  [arXiv:1503.01789 [hep-ph]].
  %%CITATION = doi:10.1103/PhysRevLett.114.221801;%%
  %194 citations counted in INSPIRE as of 10 Dec 2017
  
\bibitem{Misiak:2017bgg} 
  M.~Misiak and M.~Steinhauser,
  %``Weak radiative decays of the B meson and bounds on $M_{H^\pm }$ in the Two-Higgs-Doublet Model,''
  Eur.\ Phys.\ J.\ C {\bf 77} (2017) no.3,  201
  %doi:10.1140/epjc/s10052-017-4776-y
  [arXiv:1702.04571 [hep-ph]].
  %%CITATION = doi:10.1140/epjc/s10052-017-4776-y;%%
  %49 citations counted in INSPIRE as of 10 Dec 2017


\bibitem{Arhrib:2017yby} 
  A.~Arhrib, R.~Benbrik, C.~H.~Chen, J.~K.~Parry, L.~Rahili, S.~Semlali and Q.~S.~Yan,
  %``$R_{K^{(*)}}$ anomaly in type-III 2HDM,''
  arXiv:1710.05898 [hep-ph].
  %%CITATION = ARXIV:1710.05898;%%

\bibitem{bg} J.~F.~Gunion, H.~E.~Haber, F.~E.~Paige, W.~K.~Tung and S.~S.~D.~Willenbrock,
  %``Neutral and Charged Higgs Detection: Heavy Quark Fusion, Top Quark Mass Dependence and Rare Decays,''
  Nucl.\ Phys.\ B {\bf 294} (1987) 621.
  %doi:10.1016/0550-3213(87)90600-6
  %%CITATION = doi:10.1016/0550-3213(87)90600-6;%%
  %193 citations counted in INSPIRE as of 10 Dec 2017

\bibitem{bq} S.~Moretti and K.~Odagiri,
  %``Production of charged Higgs bosons of the minimal supersymmetric standard model in b quark initiated processes at the large hadron collider,''
  Phys.\ Rev.\ D {\bf 55} (1997) 5627
 % doi:10.1103/PhysRevD.55.5627
  [hep-ph/9611374].
  %%CITATION = doi:10.1103/PhysRevD.55.5627;%%
  %83 citations counted in INSPIRE as of 10 Dec 2017

\bibitem{BBK} A.~A.~Barrientos Bendezu and B.~A.~Kniehl,
  %``W+- H-+ associated production at the large hadron collider,''
  Phys.\ Rev.\ D {\bf 59} (1999) 015009
  %doi:10.1103/PhysRevD.59.015009
  [hep-ph/9807480].
  %%CITATION = doi:10.1103/PhysRevD.59.015009;%%
  %121 citations counted in INSPIRE as of 10 Dec 2017

\bibitem{ioekosuke} S.~Moretti and K.~Odagiri,
  %``The Phenomenology of W+- H-+ production at the large hadron collider,''
  Phys.\ Rev.\ D {\bf 59} (1999) 055008
  %doi:10.1103/PhysRevD.59.055008
  [hep-ph/9809244].
  %%CITATION = doi:10.1103/PhysRevD.59.055008;%%
  %79 citations counted in INSPIRE as of 10 Dec 2017

\bibitem{roger} V.~D.~Barger, R.~J.~N.~Phillips and D.~P.~Roy,
  %``Heavy charged Higgs signals at the LHC,''
  Phys.\ Lett.\ B {\bf 324} (1994) 236
  %doi:10.1016/0370-2693(94)90413-8
  [hep-ph/9311372].
  %%CITATION = doi:10.1016/0370-2693(94)90413-8;%%
  %144 citations counted in INSPIRE as of 10 Dec 2017


\bibitem{gunion} 
J.~F.~Gunion,
  %``Detecting the t b decays of a charged Higgs boson at a hadron supercollider,''
  Phys.\ Lett.\ B {\bf 322} (1994) 125
  %doi:10.1016/0370-2693(94)90500-2
  [hep-ph/9312201].
  %%CITATION = doi:10.1016/0370-2693(94)90500-2;%%
  %97 citations counted in INSPIRE as of 10 Dec 2017


\bibitem{roy} D.~J.~Miller, S.~Moretti, D.~P.~Roy and W.~J.~Stirling,
  %``Detecting heavy charged Higgs bosons at the CERN LHC with four $b$ quark tags,''
  Phys.\ Rev.\ D {\bf 61} (2000) 055011
  %doi:10.1103/PhysRevD.61.055011
  [hep-ph/9906230].
  %%CITATION = doi:10.1103/PhysRevD.61.055011;%%
  %110 citations counted in INSPIRE as of 10 Dec 2017

\bibitem{roy1}  bibitem{Moretti:1999bw}
  S.~Moretti and D.~P.~Roy,
  %``Detecting heavy charged Higgs bosons at the LHC with triple b tagging,''
  Phys.\ Lett.\ B {\bf 470} (1999) 209
  %doi:10.1016/S0370-2693(99)01291-5
  [hep-ph/9909435].
  %%CITATION = doi:10.1016/S0370-2693(99)01291-5;%%
  %87 citations counted in INSPIRE as of 10 Dec 2017

\bibitem{ATLAS} G.~Aad {\it et al.} [ATLAS Collaboration],
  %``Search for charged Higgs bosons in the $H^{\pm} \rightarrow tb$ decay channel in $pp$ collisions at $\sqrt{s}=8 $ TeV using the ATLAS detector,''
  JHEP {\bf 1603} (2016) 127
 % doi:10.1007/JHEP03(2016)127
  [arXiv:1512.03704 [hep-ex]].
  %%CITATION = doi:10.1007/JHEP03(2016)127;%%
  %60 citations counted in INSPIRE as of 10 Dec 2017

\bibitem{CMS} V.~Khachatryan {\it et al.} [CMS Collaboration],
  %``Search for a charged Higgs boson in pp collisions at $ \sqrt{s}=8 $ TeV,''
  JHEP {\bf 1511} (2015) 018
  %doi:10.1007/JHEP11(2015)018
  [arXiv:1508.07774 [hep-ex]].
  %%CITATION = doi:10.1007/JHEP11(2015)018;%%
  %93 citations counted in INSPIRE as of 10 Dec 2017

\bibitem{Uppsala} R.~Enberg, W.~Klemm, S.~Moretti, S.~Munir and G.~Wouda,
  %``Charged Higgs boson in the $W^\pm$ Higgs channel at the Large Hadron Collider,''
  Nucl.\ Phys.\ B {\bf 893} (2015) 420
 % doi:10.1016/j.nuclphysb.2015.02.001
  [arXiv:1412.5814 [hep-ph]].
  %%CITATION = doi:10.1016/j.nuclphysb.2015.02.001;%%
  %17 citations counted in INSPIRE as of 10 Dec 2017

\bibitem{whroy} M.~Drees, M.~Guchait and D.~P.~Roy,
  %``Signature of charged to neutral Higgs boson decay at the LHC in SUSY models,''
  Phys.\ Lett.\ B {\bf 471} (1999) 39
 % doi:10.1016/S0370-2693(99)01329-5
  [hep-ph/9909266].
  %%CITATION = doi:10.1016/S0370-2693(99)01329-5;%%
  %57 citations counted in INSPIRE as of 10 Dec 2017

\bibitem{me} S.~Moretti,
  %``The $W^\pm$ h decay channel as a probe of charged Higgs boson production at the large hadron collider,''
  Phys.\ Lett.\ B {\bf 481} (2000) 49
  %doi:10.1016/S0370-2693(00)00423-8
  [hep-ph/0003178].
  %%CITATION = doi:10.1016/S0370-2693(00)00423-8;%%
  %33 citations counted in INSPIRE as of 10 Dec 2017

\bibitem{previous} 
%\cite{Moretti:2016jkp}
%\bibitem{Moretti:2016jkp}
  S.~Moretti, R.~Santos and P.~Sharma,
  %``Optimising Charged Higgs Boson Searches at the Large Hadron Collider Across $b\bar b W^\pm$ Final States,''
  Phys.\ Lett.\ B {\bf 760} (2016) 697.
%  doi:10.1016/j.physletb.2016.07.055
  [arXiv:1604.04965 [hep-ph]].
  %%CITATION = doi:10.1016/j.physletb.2016.07.055;%%
  %9 citations counted in INSPIRE as of 19 Nov 2017
%\cite{Akeroyd:2016ymd}

\bibitem{Akeroyd:2016ymd}
  A.~G.~Akeroyd {\it et al.},
  %``Prospects for charged Higgs searches at the LHC,''
  Eur.\ Phys.\ J.\ C {\bf 77} (2017) no.5,  276
  %doi:10.1140/epjc/s10052-017-4829-2
  [arXiv:1607.01320 [hep-ph]].
  %%CITATION = doi:10.1140/epjc/s10052-017-4829-2;%%
  %32 citations counted in INSPIRE as of 10 Dec 2017


\bibitem{Gunion:2002zf} 
  J.~F.~Gunion and H.~E.~Haber,
 % "The CP conserving two Higgs doublet model: The Approach to the decoupling limit",
  Phys.\ Rev.\ D {\bf 67} (2003) 075019
  [hep-ph/0207010].
  %%CITATION = HEP-PH/0207010;%%

\bibitem{Olive:2016xmw}
  C.~Patrignani {\it et al.} [Particle Data Group],
  %``Review of Particle Physics,''
  Chin.\ Phys.\ C {\bf 40} (2016) no.10,  100001.
%  doi:10.1088/1674-1137/40/10/100001
  %%CITATION = doi:10.1088/1674-1137/40/10/100001;%%


%\cite{Cheng:1987rs}
\bibitem{Cheng:1987rs}
  T.~P.~Cheng and M.~Sher,
  %``Mass Matrix Ansatz and Flavor Nonconservation in Models with Multiple Higgs Doublets,''
  Phys.\ Rev.\ D {\bf 35} (1987) 3484.
%  doi:10.1103/PhysRevD.35.3484
  %%CITATION = doi:10.1103/PhysRevD.35.3484;%%
  %518 citations counted in INSPIRE as of 13 Dec 2017

%\cite{Atwood:1996vj}
\bibitem{Atwood:1996vj}
  D.~Atwood, L.~Reina and A.~Soni,
  %``Phenomenology of two Higgs doublet models with flavor changing neutral currents,''
  Phys.\ Rev.\ D {\bf 55} (1997) 3156
  [hep-ph/9609279];
  %%CITATION = doi:10.1103/PhysRevD.55.3156;%%
  %390 citations counted in INSPIRE as of 13 Dec 2017
%\bibitem{Chen:2005mka} 
  C.~-H.~Chen and C.~-Q.~Geng,
 % "Scalar interactions to the polarizations of B --> phi K*",
  Phys.\ Rev.\ D {\bf 71} (2005) 115004
  [hep-ph/0504145].
  %%CITATION = HEP-PH/0504145;%%

\bibitem{GomezBock:2005hc} 
  M.~Gomez-Bock and R.~Noriega-Papaqui,
  %"Flavor violating decays of the Higgs bosons in the THDM-III",
  J.\ Phys.\ G {\bf 32} (2006) 761 
  [hep-ph/0509353];
  %%CITATION = HEP-PH/0509353;%%
%\bibitem{GomezBock:2009xz} 
  M.~Gomez-Bock, G.~Lopez Castro, L.~Lopez-Lozano and A.~Rosado,
  %"Flavor-changing neutral current in production and decay of pseudoscalar mesons in a type III two-Higgs-doublet-model with four-texture Yukawa couplings",
  Phys.\ Rev.\ D {\bf 80} (2009) 055017
  [arXiv:0905.3351 [hep-ph]].
  %%CITATION = ARXIV:0905.3351;%%

%\cite{Ferreira:2014naa}
\bibitem{Ferreira:2014naa}
  P.~M.~Ferreira, J.~F.~Gunion, H.~E.~Haber and R.~Santos,
  %``Probing wrong-sign Yukawa couplings at the LHC and a future linear collider,''
  Phys.\ Rev.\ D {\bf 89} (2014) no.11,  115003.
%  doi:10.1103/PhysRevD.89.115003
%  [arXiv:1403.4736 [hep-ph]].
  %%CITATION = doi:10.1103/PhysRevD.89.115003;%%
  %77 citations counted in INSPIRE as of 19 Nov 2017

\bibitem{Ferreira:2014dya}
  P.~M.~Ferreira, R.~Guedes, M.~O.~P.~Sampaio and R.~Santos,
  %``Wrong sign and symmetric limits and non-decoupling in 2HDMs,''
  JHEP {\bf 1412} (2014) 067.
%  doi:10.1007/JHEP12(2014)067
%  [arXiv:1409.6723 [hep-ph]].
  %%CITATION = doi:10.1007/JHEP12(2014)067;%%
  %41 citations counted in INSPIRE as of 19 Nov 2017

%\cite{Deshpande:1977rw}
\bibitem{Deshpande:1977rw}
  N.~G.~Deshpande and E.~Ma,
  %``Pattern of Symmetry Breaking with Two Higgs Doublets,''
  Phys.\ Rev.\ D {\bf 18} (1978) 2574.
%  doi:10.1103/PhysRevD.18.2574
  %%CITATION = doi:10.1103/PhysRevD.18.2574;%%
  %561 citations counted in INSPIRE as of 19 Nov 2017}
%

%\cite{Ferreira:2004yd}
\bibitem{Ferreira:2004yd}
  P.~M.~Ferreira, R.~Santos and A.~Barroso,
  %``Stability of the tree-level vacuum in two Higgs doublet models against charge or CP spontaneous violation,''
  Phys.\ Lett.\ B {\bf 603} (2004) 219 and 
   Erratum: [Phys.\ Lett.\ B {\bf 629} (2005) 114].
%  doi:10.1016/j.physletb.2004.10.022, 10.1016/j.physletb.2005.09.074
%  [hep-ph/0406231].
  %%CITATION = doi:10.1016/j.physletb.2004.10.022, 10.1016/j.physletb.2005.09.074;%%
  %119 citations counted in INSPIRE as of 19 Nov 2017


%\cite{Barroso:2013awa}
\bibitem{Barroso:2013awa} 
  A.~Barroso, P.~M.~Ferreira, I.~P.~Ivanov and R.~Santos,
  %``Metastability bounds on the two Higgs doublet model,''
  JHEP {\bf 1306}, 045 (2013)
%  doi:10.1007/JHEP06(2013)045
%  [arXiv:1303.5098 [hep-ph]].
  %%CITATION = doi:10.1007/JHEP06(2013)045;%%
  

%\cite{Kanemura:1993hm}
\bibitem{Kanemura:1993hm} 
A.~G.~Akeroyd, A.~Arhrib and E.~M.~Naimi,
  %``Note on tree level unitarity in the general two Higgs doublet model,''
  Phys.\ Lett.\ B {\bf 490} (2000) 119
  %doi:10.1016/S0370-2693(00)00962-X
  [hep-ph/0006035]. A.~Arhrib,
  %``Unitarity constraints on scalar parameters of the standard and two Higgs doublets model,''
  hep-ph/0012353.
  S.~Kanemura, T.~Kubota and E.~Takasugi,
  %``Lee-Quigg-Thacker bounds for Higgs boson masses in a two doublet model,''
  Phys.\ Lett.\ B {\bf 313} (1993) 155
%  doi:10.1016/0370-2693(93)91205-2
  [hep-ph/9303263].

%\cite{Barbieri:2006bg}
\bibitem{Barbieri:2006bg} 
  R.~Barbieri, L.~J.~Hall, Y.~Nomura and V.~S.~Rychkov,
  %``Supersymmetry without a Light Higgs Boson,''
  Phys.\ Rev.\ D {\bf 75} (2007) 035007
%  doi:10.1103/PhysRevD.75.035007
  [hep-ph/0607332].
  %%CITATION = doi:10.1103/PhysRevD.75.035007;%%

%\cite{Baak:2014ora}
\bibitem{Baak:2014ora} 
  M.~Baak {\it et al.} [Gfitter Group],
  %``The global electroweak fit at NNLO and prospects for the LHC and ILC,''
  Eur.\ Phys.\ J.\ C {\bf 74} (2014) 3046
%  doi:10.1140/epjc/s10052-014-3046-5
  [arXiv:1407.3792 [hep-ph]].
  %%CITATION = doi:10.1140/epjc/s10052-014-3046-5;%%
  
%\cite{Corbett:2015ksa}
\bibitem{Corbett:2015ksa} 
  T.~Corbett, O.~J.~P.~Eboli, D.~Goncalves, J.~Gonzalez-Fraile, T.~Plehn and M.~Rauch,
  %``The Higgs Legacy of the LHC Run I,''
  JHEP {\bf 1508} (2015) 156
%  doi:10.1007/JHEP08(2015)156
  [arXiv:1505.05516 [hep-ph]].
  %%CITATION = doi:10.1007/JHEP08(2015)156;%%

\bibitem{run2}
ATLAS collaboration,
  %``Combined measurements of the Higgs boson production and decay rates in $H\to ZZ^*\to 4\ell$ and $H\to\gamma\gamma$ final states using $pp$ collision data at $\sqrt{s}=$ 13 TeV in the ATLAS experiment,''
  ATLAS-CONF-2016-081;
  %%CITATION = ATLAS-CONF-2016-081;%%
  %28 citations counted in INSPIRE as of 10 Dec 2017
ATLAS collaboration, 
  %``Search for Higgs boson production via weak boson fusion and decaying to $b \bar b$ in association with a high-energy photon in the ATLAS detector,''
  ATLAS-CONF-2016-063;
  %%CITATION = ATLAS-CONF-2016-063;%%
ATLAS collaboration,
  %``Measurement of the $t\bar{t}Z$ and $t\bar{t}W$ production cross sections in multilepton final states using 3.2 fb$^{-1}$ of $pp$ collisions at 13 TeV at the LHC,''
  ATLAS-CONF-2016-003;
  %%CITATION = ATLAS-CONF-2016-003;%%
  %19 citations counted in INSPIRE as of 10 Dec 2017
CMS Collaboration,
  %``Updated measurements of Higgs boson production in the diphoton decay channel at $\sqrt{s}=13~\textrm{TeV}$ in pp collisions at CMS.,''
  CMS-PAS-HIG-16-020;
  %%CITATION = CMS-PAS-HIG-16-020;%%
  %49 citations counted in INSPIRE as of 10 Dec 2017
CMS Collaboration,
  %``Search for $\mathrm{t\overline{t}H}$ production in the $\mathrm{H}\rightarrow \mathrm{b\overline{b}}$ decay channel with 2016 pp collision data at $\sqrt{s}=13~\mathrm{TeV}$,''
  CMS-PAS-HIG-16-038;
  %%CITATION = CMS-PAS-HIG-16-038;%%
  %22 citations counted in INSPIRE as of 10 Dec 2017
CMS Collaboration,
  %``Updated measurements of Higgs boson production in the diphoton decay channel at $\sqrt{s}=13~\textrm{TeV}$ in pp collisions at CMS.,''
  CMS-PAS-HIG-16-020.
  %%CITATION = CMS-PAS-HIG-16-020;%%
  %49 citations counted in INSPIRE as of 10 Dec 2017

%\cite{Aaboud:2017sjh}
\bibitem{Aaboud:2017sjh}
  M.~Aaboud {\it et al.} [ATLAS Collaboration],
  %``Search for additional heavy neutral Higgs and gauge bosons in the ditau final state produced in 36 fb$^{-1}$ of $pp$ collisions at $\sqrt{s}$ = 13 TeV with the ATLAS detector,''
  arXiv:1709.07242 [hep-ex];
  %%CITATION = ARXIV:1709.07242;%%
  %11 citations counted in INSPIRE as of 08 Dec 2017
  CMS collaboration, CMS-PAS-HIG-17-020.
  
%\cite{Aad:2015typ}
\bibitem{Aad:2015typ}
  G.~Aad {\it et al.} [ATLAS Collaboration],
  %``Search for charged Higgs bosons in the $H^{\pm} \rightarrow tb$ decay channel in $pp$ collisions at $\sqrt{s}=8 $ TeV using the ATLAS detector,''
  JHEP {\bf 1603} (2016) 127
  %doi:10.1007/JHEP03(2016)127
  [arXiv:1512.03704 [hep-ex]].
  %%CITATION = doi:10.1007/JHEP03(2016)127;%%
  %60 citations counted in INSPIRE as of 09 Dec 2017  
  
%\cite{Aaboud:2016dig}
\bibitem{Aaboud:2016dig}
  M.~Aaboud {\it et al.} [ATLAS Collaboration],
  %``Search for charged Higgs bosons produced in association with a top quark and decaying via $H^{\pm} \rightarrow \tau\nu$ using $pp$ collision data recorded at $\sqrt{s} = 13$ TeV by the ATLAS detector,''
  Phys.\ Lett.\ B {\bf 759} (2016) 555
  %doi:10.1016/j.physletb.2016.06.017
  [arXiv:1603.09203 [hep-ex]].
  %%CITATION = doi:10.1016/j.physletb.2016.06.017;%%
  %41 citations counted in INSPIRE as of 09 Dec 2017
  
  %\cite{Khachatryan:2015qxa}
\bibitem{Khachatryan:2015qxa}
  V.~Khachatryan {\it et al.} [CMS Collaboration],
  %``Search for a charged Higgs boson in pp collisions at $ \sqrt{s}=8 $ TeV,''
  JHEP {\bf 1511} (2015) 018
  %doi:10.1007/JHEP11(2015)018
  [arXiv:1508.07774 [hep-ex]].
  %%CITATION = doi:10.1007/JHEP11(2015)018;%%
  %93 citations counted in INSPIRE as of 09 Dec 2017

\bibitem{CMScharged}
CMS collaboration, CMS-PAS-HIG-16-027.

\bibitem{madgraph}
  J.~Alwall, R.~Frederix, S.~Frixione, V.~Hirschi, F.~Maltoni, O.~Mattelaer, H.-S.~Shao and T.~Stelzer {\it et al.},
  %``The automated computation of tree-level and next-to-leading order differential cross sections, and their matching to parton shower simulations,''
  JHEP {\bf 1407} (2014) 079
[arXiv:1405.0301 [hep-ph]].
  %%CITATION = ARXIV:1405.0301;%%
  %154 citations counted in INSPIRE as of 24 Dec 2014
  
%\cite{Sjostrand:2014zea}
\bibitem{pythia8} 
  T.~Sjöstrand {\it et al.},
  %``An Introduction to PYTHIA 8.2,''
  Comput.\ Phys.\ Commun.\  {\bf 191} (2015) 159
  %doi:10.1016/j.cpc.2015.01.024
  [arXiv:1410.3012 [hep-ph]].
  %%CITATION = doi:10.1016/j.cpc.2015.01.024;%%
  %744 citations counted in INSPIRE as of 10 Dec 2017
 

%\cite{Ovyn:2009tx}
\bibitem{delphes} 
  S.~Ovyn, X.~Rouby and V.~Lemaitre,
  %``DELPHES, a framework for fast simulation of a generic collider experiment,''
  arXiv:0903.2225 [hep-ph].
  %%CITATION = ARXIV:0903.2225;%%
  %244 citations counted in INSPIRE as of 13 Jan 2015  


%%%%%%%%%%%%%%%%%%%%%%%%%%%%%%%%%%%%%%%%%%%%%%%%%%
\end{thebibliography}
\end{document}